\documentclass[a4paper,11pt]{article}
\pdfoutput=1 % if your are submitting a pdflatex (i.e. if you have
\usepackage{ae,aecompl}
\usepackage{jheppub} % for details on the use of the package, please
\usepackage[T1]{fontenc}
\usepackage[italian,english]{babel}
\usepackage{hyperref}
\usepackage{ifpdf}
\usepackage{subfigure}
\usepackage{gensymb}
\usepackage{changepage}
\usepackage{amssymb}
\usepackage{amsfonts}
\usepackage{epsf}
\usepackage{rotating}
\usepackage{graphicx}
\usepackage{amsmath}
\usepackage{fancyhdr}
\usepackage{lineno}
\usepackage{dutchcal}
\usepackage{babel}
\usepackage{graphics}
\usepackage{pstricks}
\usepackage{color}
\usepackage{multirow}
\usepackage{nicefrac}
\usepackage{bm}
\usepackage{slashed}
\usepackage{float}
\usepackage{array}
\usepackage{boldline}
\usepackage{array,booktabs, makecell}
\usepackage{hhline}

\usepackage{mathrsfs}
\usepackage{accents}
\usepackage{pifont}
\usepackage{diagbox}
\usepackage{physics}
\usepackage[toc,page]{appendix}
\usepackage[lmargin=1.9in,bmargin=0.2in,footskip=0.5in,total={6.5in,9.5in}]{geometry}

\usepackage{tikz}
\usepackage{tikz-feynman}
\tikzfeynmanset{compat=1.0.0}
\usepackage{braket}
\definecolor{pantoneCB}{rgb}{0.0588235, 0.298039, 0.505882}

\hypersetup{bookmarks=true,unicode=true,pdftoolbar=true,pdfmenubar=true,
	pdffitwindow=false,
	pdfnewwindow=true,colorlinks=true,allcolors=pantoneCB}

\newcolumntype{?}{!{\vrule width 3pt}}

\newcommand{\lsim}{\mathrel{\mathop{\kern 0pt \rlap
			{\raise.2ex\hbox{$<$}}}
		\lower.9ex\hbox{\kern-.190em $\sim$}}}
\newcommand{\gsim}{\mathrel{\mathop{\kern 0pt \rlap
			{\raise.2ex\hbox{$>$}}}
		\lower.9ex\hbox{\kern-.190em $\sim$}}}

\newcommand{\be}{\begin{equation}}
\newcommand{\ee}{\end{equation}}
\newcommand{\bea}{\begin{eqnarray}}
\newcommand{\eea}{\end{eqnarray}}

\def\ptmiss{\not\!\!{p_T}}
\newcommand{\bl}{\textcolor{blue}}

\newcommand{\sarah}{\texttt{SARAH} }
\newcommand{\chep}{\texttt{CalcHEP} } 

\newcommand{\py}{\texttt{PYTHIA 8.2.45} }
\newcommand{\fj}{\texttt{FastJet-3.3.3} }

\newcommand{\fbi}{fb$^{-1}$ }

\newcolumntype{?}{!{\vrule width 3pt}}

\newcommand*\xbar[1]{%
	\hbox{%
		\vbox{%
			\hrule height 0.65pt % The actual bar
			\kern0.4ex%         % Distance between bar and symbol
			\hbox{%
				\kern-0.05em%      % Shortening on the left side
				\ensuremath{#1}%
				\kern0.0em%      % Shortening on the right side
			}%
		}%
	}%
}

\newcolumntype{M}[1]{>{\centering\arraybackslash}m{#1}}

\DeclareMathSymbol{\widetildesym}{\mathord}{largesymbols}{"65}

\DeclareFontFamily{OT1}{pzc}{}
\DeclareFontShape{OT1}{pzc}{m}{it}{<-> s * [1.2] pzcmi7t}{}
\DeclareMathAlphabet{\mathpzc}{OT1}{pzc}{m}{it}

\newcommand{\y}{\mathpzc{Y}}

\newcommand{\Q}{\bm{Q}}
\newcommand{\LL}{\bm{L}}
\newcommand{\cl}{\mathpzc{l}}
\newcommand{\uq}{\mathpzc{u}}
\newcommand{\dq}{\mathpzc{d}}
\newcommand{\rtt}{\widetilde R_2^{+2/3}}

\newcommand{\half}{\frac{1}{2}}

\newcommand{\tlq}{\theta_{LQ}}
\newcommand{\X}{X}
\newcommand{\yt}{\y_2}
\newcommand{\yl}{ \y_1^L}
\newcommand{\yr}{ \y_1^R}
\newcommand{\xo}{X_{1}^{-1/3}}
\newcommand{\xt}{X_{2}^{-1/3}}
\newcommand{\xot}{X_{1,2}^{-1/3}}

\newcommand{\cmmnt}[1]{\ignorespaces}

\title{\boldmath Phenomenology of Scalar Leptoquarks at the LHC in Explaining the Radiative Neutrino Mass, Muon $g-2$ and Lepton Flavour Violating Observables}
\preprint{ IITH-PH-0004/22\\ \hspace*{13.3cm} IP/BBSR/2022-05}
\author[a]{Snehashis Parashar,}
\author[b]{Anirban Karan,}
\author[c,d]{Avnish,}
\author[a]{Priyotosh Bandyopadhyay,}
\author[c,d]{Kirtiman Ghosh}

\affiliation[a]{
	Indian Institute of Technology Hyderabad, Kandi,  Sangareddy-502284, Telangana, India.}

\affiliation[b]{Instituto de F\'isica Corpuscular (CSIC – Universitat de Val\`encia), Apt. Correus 22085, E-46071 Val\`encia, Spain}

\affiliation[c]{Institute of Physics, Bhubaneswar, Odisha-751005, India.}

\affiliation[d]{Homi Bhabha National Institute, Training School Complex, Anushakti Nagar, Mumbai 400094, India}

\emailAdd {ph20resch11006@iith.ac.in}
\emailAdd{kanirban@ific.uv.es}
\emailAdd{avnish@iopb.res.in}
\emailAdd{bpriyo@phy.iith.ac.in}
\emailAdd{kirtiman.ghosh@iopb.res.in}

\abstract{We study the phenomenology of a particular leptoquark extension of the Standard Model (SM), namely the doublet-singlet scalar leptoquark extension of the SM (DSL-SM). Besides generating Majorana mass for neutrinos, these leptoquarks contribute to muon and electron $(g-2)$ and various lepton flavour violating processes. Collider signatures of the benchmark points (BPs), consistent with the neutrino oscillation data, anomalous muon/electron magnetic moments, experimental bounds on the charged lepton flavour violation observables, etc., are studied at the LHC/FCC with centre-of-mass energies of 14, 27 and 100 TeV.  While the two $-1/3$ charged colored scalars from singlet and doublet leptoquark mix with each other, the charge $2/3$ colored scalar from the doublet leptoquark remains pure. With a near-degenerate mass spectrum, the pure and mixed leptoquark states are shown to be distinguishable from multiple finalstates, while discerning between the two mixed states remain very challenging.

}


\begin{document}
	\maketitle
	\raggedbottom
	\section{Introduction}
	
Though the Standard Model (SM) provides a beautiful theoretical explanation of the non-gravitational interaction of the elementary particles in terms of $SU(3)_C\otimes SU(2)_L \otimes U(1)_Y$ gauge group, there is ample evidence that it is not a complete theory. Neutrino oscillation indicating the non-zero masses of neutrinos and flavour mixing of leptons is one of such phenomena advocating the incompleteness of the SM which predicts massless neutrinos without any flavour mixing. While the simplest way to generate neutrino masses is to add right-handed neutrino fields to the SM particle content, it is hard to explain their extreme smallness. Such a small mass could be understood if neutrinos are Majorana particles and Majorana masses for neutrinos are generated from higher dimensional operators who violate the lepton number by two units. The most studied example of such an operator is the dimension-5 Weinberg operator \cite{Weinberg:1979sa}: ${\cal O}_5~=~\frac{c_{\alpha\beta}}{\Lambda}\left(\overline{L^C_{\alpha L}}\tilde H^*\right)\left(\tilde H^\dagger L_{\beta L}\right)$ where $\alpha,~\beta$ are the generation indices, $L_L~=~(\nu_L, l_L)^T$ is the left-handed lepton doublet of the SM, $H~=~(H^+, H^0)^T$ is the Higgs doublet and $\tilde H=i\sigma_2 H^*$. $\Lambda$ is the scale of new physics, and $c_{\alpha\beta}$ is a model-dependent coefficient. Weinberg operator gives rise to Majorana masses (suppressed by $\Lambda$) for the neutrinos after electroweak symmetry breaking (EWSB). At tree level, there are only three ways to generate the  Weinberg operator, namely, the type-I \cite{Minkowski:1977sc,Yanagida:1979as,GellMann:1980vs,Mohapatra:1979ia}, the type-II \cite{Magg:1980ut,Schechter:1980gr,Wetterich:1981bx,Lazarides:1980nt,Mohapatra:1980yp,Cheng:1980qt,Ashanujjaman:2022tdn,Ashanujjaman:2022cso,Ashanujjaman:2021txz} and the type-III \cite{Foot:1988aq,Ashanujjaman:2021zrh,Ashanujjaman:2021jhi}  seesaw mechanisms. In the framework of tree-level seesaw models, the smallness of neutrino masses ($m_\nu$) is explained via new physics at a very high scale of $\Lambda$.  Though the seesaw models are naturally motivated to have very high scale masses to explain the tinyness of the neutrino masses, if balanced with appropriate Yukawa couplings, nothing precludes them from having mass at the TeV scale and hence testable at the collider experiments. Note that the simplest TeV scale seesaw models are tightly constrained from the stringent cosmological upper bound of $\lsim 0.09$ eV on the total mass ($\sum m_\nu$) of light neutrinos \cite{DiValentino:2021hoh}, charged lepton flavour violating (CLFV) observables, electroweak (EW) precision observables, vacuum stability and perturbativity of the scalar potential (in the context of type-II seesaw only)  \cite{Jangid:2020dqh,Jangid:2020qgo,Bandyopadhyay:2020djh}, and collider experiments. However, one can construct an alternative class of models in which ${\cal O}_5$ is forbidden at the tree level, and neutrino masses are generated radiatively \cite{Bonnet:2012kz,Ma:1998dn,Babu:1988ki,FileviezPerez:2009ud,Babu:2002uu,Krauss:2002px,Cheung:2004xm,Ma:2006km,Ma:2007yx,Aoki:2008av,Aoki:2009vf,AristizabalSierra:2006gb,Avnish:2020rhx} or from a tree-level effective operator \cite{Bonnet:2009ej,Kanemura:2010bq,Anamiati:2018cuq,Kumericki:2012bh,Kumericki:2011hf,Ashanujjaman:2020tuv} with mass dimension $d>5$. The additional suppression\footnote{Note that typical scale of a new physics where neutrino mass is generated via a d-dimensional operator at n-loop level, can be estimated from the following neutrino mass formula: $m_\nu~\propto ~ \left(\frac{1}{16 \pi^2}\right)^n \times \frac{{\rm VEV}^{d-3}}{\Lambda^{d-4}}$.} to the neutrino masses arises from the loop integrals (in case of former) or higher powers of $\Lambda$ in the denominator (in case of later), bring down the new physics scale $\Lambda$ to TeV scale and hence, makes these models testable at the LHC.

Apart from neutrino mass generation, the anomalous magnetic moment of muon is another long standing puzzle in particle physics. The combined result from Fermi National Accelerator Laboratory \cite{Muong-2:2021ojo} and Brookhaven National Laboratory \cite{Muong-2:2006rrc} suggests a $4.2\sigma$ deviation in the measurement of $a_\mu\,\big(\equiv \frac{g_\mu}{2}-1\big)$ from its theoretical estimate under SM \footnote{It is worth mentioning that the combined experimental result is compared against the data-driven SM prediction from \cite{Aoyama:2020ynm} which leads to the $4.2\sigma$ deviation, not taking into account the lattice QCD prediction from \cite{Borsanyi:2020mff}, which are in fact closer to the experimental average.} \cite{Aoyama:2012wk,Czarnecki:2002nt,Gnendiger:2013pva,Davier:2017zfy,Keshavarzi:2018mgv,Hoferichter:2018kwz,Bijnens:2019ghy,Colangelo:2019uex,Borsanyi:2020mff,Aoyama:2020ynm}: $\Delta a_\mu=a_\mu^{Ex}-a_\mu^{SM}=(2.51\pm0.59)\times 10^{-9}$. Similar anomaly exists for electron $(g-2)$ also. But, while experiment with Cesium shows $\Delta a_e^{Cs}=a_e^{Ex}(Cs)-a_e^{SM}=(-8.8\pm3.6)\times 10^{-13}$ \cite{Parker:2018vye}, the same with Rubidium suggests $\Delta a_e^{Rb}=a_e^{Ex}(Rb)-a_e^{SM}=(4.8\pm3.0)\times 10^{-13}$ \cite{Morel:2020dww}. Though the expectation values of $\Delta a_e$ in the two experiments are large and opposite in sign, the significances of the measurements are reduced due to large experimental error-bars.

It is important to note that the current experimental data (from neutrino oscillation as well as scattering experiments) is inconclusive in determining the actual mechanism of neutrino mass generation. However, it is clear that the massive neutrinos and anomalous magnetic moment of the charged leptons are both experimentally facts, and hence, should be incorporated into the extensions of the SM. It would be particularly exciting if a single mechanism resolves two of these important outstanding puzzles in particle physics.  It has been shown in the literature \cite{Biggio:2008in,Zhou:2021vnf} that the minimal tree-level seesaw models are not very efficient in incorporating these $(g-2)$ anomalies. This necessitates searching for some other mechanism to provide a combined explanation for neutrino mass generation as well as muon $(g-2)$.  The SM particle spectrum extended by TeV scale leptoquarks could be a possible answer to this puzzle. However, attempts to construct a unified theory that explains non-zero neutrino masses/mixings and anomalous magnetic moments of the SM charged leptons suffer, typically, from roadblocks in the form of unacceptable phenomenological consequences such as large contributions to the SM charged lepton flavour violating processes. In particular, any ultraviolet complete theory designed to explain the anomalous magnetic moments would necessarily contain additional fields. If we want these fields to play a role in generating neutrino masses and mixings, there is always the danger of enhancing the SM charged lepton flavour-violating processes such as $\mu \to e\gamma,~\mu \to 3e$, $\mu$-$e$ conversion in nuclei, etc. which are tightly constrained from different charged lepton flavour violation experiments. In other words, the introduction of new fields and their interactions cannot be arbitrary, for not only must low-energy observables remain consistent with measurements, but the failure of collider experiments to observe such particles must be explained. Note that models comprising of doublet-singlet or doublet-triplet leptoquarks \cite{Dorsner:2016wpm,AristizabalSierra:2007nf,Dorsner:2017wwn,Babu:2019mfe,Pas:2015hca,Chua:1999si,Mahanta:1999xd,Babu:2010vp,Zhang:2021dgl} with Yukawa interactions involving the SM lepton, quark multiplets, and the leptoquarks are known to generate Majorana masses for the neutrinos at the one-loop level and also contribute to the $g-2$ of the SM charged leptons. A comprehensive study of such a framework in the context of neutrino oscillation data, $g-2$ anomalies, bounds on the CLFV observables, and collider experiments is missing in the literature.  This article intends to fill this gap.

Leptoquark models where the SM field content is enlarged by introducing colored scalars (spin-0) or vector (spin-1) fields have been there in literature for quite a few decades \cite{Georgi:1974sy,Georgi:1974my}. They  emerge naturally in higher gauge theories unifying matters \cite{Georgi:1974sy,Georgi:1974my,Fritzsch:1974nn,Dimopoulos:1979es,Farhi:1980xs,Schrempp:1984nj,Wudka:1985ef,Nilles:1983ge,Haber:1984rc,Assad:2017iib,Perez:2021ddi,Murgui:2021bdy}.  Being charged under the $SU{(3)}_C$, the Yukawa interactions of the scalar leptoquarks simultaneously involve a quark and a lepton and hence provide an elegant explanation for the recent observation of the lepton flavour non-universality in the B-meson decays.  The prospect of leptoquark models in resolving various flavour anomalies along with neutrino mass generation and muon $(g-2)$ make them discussable in recent times \cite{Couture:1995he,Marzocca:2018wcf,Gherardi:2020qhc,Crivellin:2017zlb,Crivellin:2019dwb,Aydemir:2019ynb,Becirevic:2018afm,Bigaran:2019bqv,Mandal:2018kau,Iguro:2020keo,Lee:2021jdr,Bordone:2020lnb,Bigaran:2021kmn,Borschensky:2021hbo,Dedes:2021abc,Browder:2021hbl,Sheng:2021iss,Cornella:2021sby,Crivellin:2021lix,Angelescu:2021lln,Angelescu:2018tyl,Arnan:2019olv,ColuccioLeskow:2016dox,Crivellin:2020mjs, Bhaskar:2022vgk,Ghosh:2022vpb}. The phenomenology of leptoquark models has been studied in literature \cite{Bandyopadhyay:2018syt,Bhaskar:2020gkk,Bhaskar:2021pml,Bhaskar:2021gsy,DaRold:2021pgn,Hiller:2021pul,Haisch:2020xjd,Chandak:2019iwj,Bhaskar:2020kdr,Alves:2018krf,Dorsner:2019vgp,Mandal:2018qpg,Padhan:2019dcp,Baker:2019sli,Nadeau:1993zv,Atag:1994hk,Atag:1994np,Buchmuller:1986zs,Hewett:1987yg,Hewett:1987bh,Cuypers:1995ax,Blumlein:1996qp,Belyaev,Kramer:1997hh,Plehn:1997az,Eboli:1997fb,Kramer:2004df,Hammett:2015sea,Mandal:2015vfa,Asadi:2021gah,Bandyopadhyay:2021pld,Davidson:1993qk,Mandal:2019gff,Dey:2017ede,Crivellin:2021egp, Kowalska:2018ulj}, specially in the context of the Large Hadron Collider (LHC) experiment. For example, leptoquarks have been searched experimentally at various colliders including the LHC over last three decades \cite{Behrend:1986jz,Bartel:1987de,Kim:1989qz,Abreu:1998fw,Collaboration:2011qaa,Abramowicz:2019uti,Alitti:1991dn,Abazov:2011qj,Aaltonen:2007rb,Aad:2020iuy,CMS:2020gru,Aad:2021rrh,CMS:2018qqq,CMS:2020wzx}; however their existence is yet to be confirmed. Discerning features of different scalar and vector leptoquarks at electron-photon \cite{Bandyopadhyay:2020klr}, electron-proton \cite{Bandyopadhyay:2020jez} and proton-proton \cite{Bandyopadhyay:2020wfv,Bandyopadhyay:2021pld,Dutta:2021wid} colliders have been investigated relying dominantly on the angular distribution. Constraints from Planck scale stability and perturbativity on different scalar leptoquarks have also been studied \cite{Bandyopadhyay:2016oif,Bandyopadhyay:2021kue}.

In this work, the field content of the SM is extended to include the following pair of scalar leptoquarks \cite{Dorsner:2016wpm} non-trivially charged under the SM gauge  symmetry ($\bf{SU(3)_C,~SU(2)_L},~U(1)_Y$): $S_1$ $(\bm{\xbar{3},1},1/3)$ and $\widetilde{R}_2$ $(\bm{3,2},1/6)$ \cite{Freitas:2022gqs,Zhang:2021dgl,Coy:2021hyr,Cata:2019wbu,Dorsner:2017wwn,Babu:2019mfe,Mahanta:1999xd, Chen:2022hle}\footnote{We also restrict ourselves to one generation of leptoquarks with off-diagonal Yukawa coupling, to avoid the constraints from perturbative unitarity \cite{Bandyopadhyay:2021kue}.}. The most general scalar potential and Yukawa interactions involving the SM fields and the leptoquarks, $S_1$ and $\widetilde{R}_2$, violate the lepton number conservation and result in non-zero Majorana masses for the neutrinos at the one-loop level. In other words, one needs to consider both $S_1$ and $\widetilde{R}_2$ simultaneously in the model to successfully generate the $d=5$ Weinberg operator for neutrino mass at one-loop \cite{Bonnet:2012kz}. Both the singlet and doublet leptoquark in this model contributes to the charged lepton $g-2$ and CLFV processes. We studied the possibility of simultaneously explaining neutrino oscillation data and muon anomalous magnetic moment in the framework of this model subjected to the experimental constraints from the CLFV experiments. We obtain benchmark points that are consistent with all the observational constraints, such as neutrino oscillation data, muon $(g-2)$ and lepton flavour violating decays, etc. The second part of the article is dedicated to the collider signatures of those benchmark points at the LHC experiment. The particle spectrum of this model includes two exotic colored scalars (denoted as mixed states) with electric charge $\frac{1}{3}$ resulting from the mixing of $\frac{1}{3}$ components of the singlet and doublet leptoquarks. Another exotic colored scalar (denoted as the pure state) with electric charge $\frac{2}{3}$ results from the doublet leptoquark. While this particular leptoquark combination has been extensively studied in literature in the context of various bounds and anomalies, to our knowledge, a detailed study of finalstates resulting from different choices of the Yukawa couplings, at the current and future iterations of the LHC, has not been explored for this model. We provide a thorough study of the production, decay, and resulting collider signatures of these exotic scalars for a carefully chosen set of benchmark points at the current LHC, as well as the future High Energy(HE)/High Luminosity(HL)-LHC \cite{Azzi:2019yne,Cepeda:2019klc}, and the Future Circular Collider (FCC)\cite{FCC:2018byv}. Along with the discovery prospect of the model at the LHC/FCC, our analysis on multiple finalstates that can help in distinguishing between the physical mass eigenstates of the leptoquarks in this model is presented, for a near-degenerate mass spectrum. Additionally, we suggest a method of probing the leptoquark mixing angle at a future high-energy and high-precision hadron collider, via asymmetric pair production with $t$-channel $W^\pm$-boson exchange.

This article is organized as follows: in the next section (\autoref{sec:desc}), we introduced the model with a discussion on neutrino mass generation, anomalous magnetic moments of charged leptons and leptoquark contributions to the CLFV processes. In \autoref{sec:bps}, the choice of our benchmark points from neutrino oscillation data, $(g-2)$ and CLFV processes is motivated. The subsequent section (\autoref{sec:coll}) deals with collider probe of the model at the LHC/FCC through pair production for the chosen benchmark points. In \autoref{sec:rtvx12}, we discuss the distinguishing signatures of the pure and mixed leptoquark states from pair production. The next section (\autoref{sec:x1vx2}) is dedicated to the challenges in differentiating between the two mixed leptoquark states in the context of our benchmark points, complemented with a discussion on probing the leptoquark mixing angle. Finally, we summarize our results in \autoref{sec:conc}.
	
	\section{The model}
	\label{sec:desc}

	In addition to all the SM particles, the model includes a scalar $SU(2)_L$ doublet leptoquark $\widetilde R_2\,(\bm{3,2},1/6)=(\widetilde R_2^{\,2/3},\, \widetilde R_2^{\,-1/3})^T$ and a singlet leptoquark $S_1\,(\bm{\xbar{3},1},1/3)$ \cite{Zhang:2021dgl}. The relevant part of the Lagrangian involving the leptoquarks is given as:
\small{
	\begin{align}
		\label{eq:Lag_SD}
		-\mathscr L \supset \bigg[&\y_1^L\; \xbar \Q_L^{\,c}\,  S_1\, (i\sigma_2)\,\LL_L +
		\y_1^R\;\xbar\uq_R^{\,c}\,S_1\,\cl_R
		+ \y_2\; \xbar \dq_R^{}\, \widetilde R_2^{\,T} (i\sigma_2)\, \LL_L\,+ \kappa\, H^\dagger \widetilde R_2\, S_1\, +h.c.\bigg]\nonumber\\
		&+m_1^2\big(S_1^\dagger S_1\big)+m_2^2 \big(\widetilde{R}_2^\dagger \widetilde{R}_2\big)+\alpha_1 \big(H^\dagger H\big) \big(S_1^\dagger S_1\big)+\alpha_2 \big(H^\dagger H\big) \big(\widetilde{R}_2^\dagger \widetilde{R}_2\big)+\alpha_2^\prime \big(H^\dagger \widetilde{R}_2\big) \big(\widetilde{R}_2^\dagger H\big),
\end{align}}
\normalsize where, $\Q_L\, (\bm{3,2},1/6)$ and $\LL_L\,(\bm{1,2},-1/2)$ represent the left-handed quark and lepton doublets with three generations, $\dq_R\, (\bm{3,1},-1/3)$ indicates the right-handed singlet for down-type quark with all the three generations and $H\,(\bm{1,2},1/2)$ is the SM Higgs doublet. Here, $\sigma_2$ denotes the second Pauli matrix, the superscript `$c$' and `$T$' signify charge conjugation and transpose in $SU(2)_L$ space, respectively. $\y_1^{L,R}$ and $\y_2$ are $3\times3$ complex matrices describing the Yukawa interactions of the leptoquarks $S_1$ and $\widetilde{R}_2$, respectively with different quarks and leptons. The terms $m_1$ and $m_2$ are mass terms for the singlet and the doublet leptoquarks, respectively. $\alpha_1$, $\alpha_2$ and $\alpha_2^\prime$ are real dimensionless couplings describing the strength of quartic interaction between the leptoquarks and the SM Higgs doublet. Whereas, the trilinear coupling $\kappa$ is in general complex with mass dimension one and plays crucial role in the phenomenology of this model. Note that in Eq.~\ref{eq:Lag_SD} simultaneous presence of the Yukawa couplings $\y_1^{L}$ and $\y_2$ in association with the trilinear scalar coupling $\kappa$ violates lepton number in the model. However, we keep only those terms in the Lagrangian that preserve baryon number($B$), which leads to the absence of the di-quark coupling terms with $S_1$, as well as the quartic term of the form $S_1^2 \widetilde{R}_2^\dagger H$. The presence of such terms can lead to fast proton decay\cite{Dorsner:2012nq}, and naturally we wish to avoid that. Assigning $B= -1/3$ to $S_1$ and $B= +1/3$ to $\widetilde{R}_2$ ensures the absence of these terms in the Lagrangian. Moreover, the trilinear coupling results into mixing among the singlet ($S_1$) and electromagnetic charge $\frac{1}{3}$ component of the doublet (${\widetilde{R}_2}^{1/3}$) leptoquark after the electroweak symmetry breaking (EWSB). The mass matrix involving $S_1$ and ${\widetilde{R}_2}^{1/3}$ is given by,
\begin{equation}
	\label{mass_matrix}
	\mathcal M_{LQ}^2 = \begin{pmatrix}
		m^2({S_1}) \,&\, \frac{\kappa v}{\sqrt{2}} \\
		\frac{\kappa v}{\sqrt{2}} &  m^2({\widetilde{R}_2}^{1/3})\\ 
	\end{pmatrix},
\end{equation} 
where, $v$ is the vacuum expectation value (VEV) of the SM Higgs doublet, $m^2({S_1})=m_1^2 +\half \alpha_1 v^2$ and $m^2({\widetilde{R}_2}^{1/3})=m_2^2 +\half( \alpha_2 + \alpha'_2) v^2~$. The physical electromagnetic (EM) charge $\frac{1}{3}$ scalars ($\X_1^{1/3}$ and $\X_2^{1/3}$) are obtained by diagonalizing the mass matrix in Eq.~\ref{mass_matrix} and given by,
\begin{align}
	\begin{split}
		\X_1^{1/3} ={}& \cos \tlq S_1^{1/3} + \sin \tlq \widetilde{R}_2^{1/3}, \quad\quad
		\X_2^{1/3} =  -\sin \tlq S_1^{1/3} + \cos \tlq \widetilde{R}_2^{1/3},\\
	\end{split}
\end{align}
where $\tlq$ is the mixing angle given by
\begin{equation}
	\tan 2\tlq = \frac{-\sqrt{2}\kappa v}{m^2(S_1)-m^2(\widetilde{R}_2^{1/3})}, \label{eq:tlq}
\end{equation}
and the mass eigenvalues of the mixed leptoquark eigenstates are obtained as
\begin{equation}
	M_{{1,2}}^2=\frac{1}{2}\bigg[m^2(S_1)+m^2(\widetilde{R}_2^{1/3})\mp\sqrt{\Big\{m^2(S_1)-m^2(\widetilde{R}_2^{1/3})\Big\}^2+2\,v^2\kappa^2}\bigg].
	\label{eq:mass}
\end{equation}
Whereas, the mass of the pure doublet leptoquark with EM charge $\frac{2}{3}$ is given by
\begin{equation}
	m^2({\widetilde{R}_2}^{2/3})=m_2^2 +\half \alpha_2 \,v^2. \label{eq:mr2}
\end{equation}

\subsection{Loop induced neutrino masses, CLFV, and $(g-2)$ }

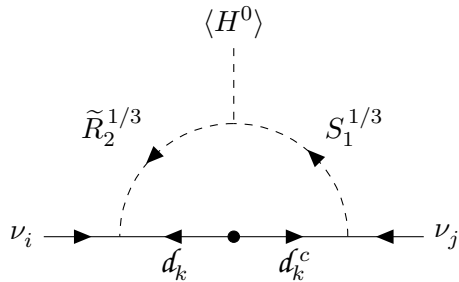
\begin{figure}[h!]
	\begin{center}
		\begin{tikzpicture}
			\begin{feynman}
				\vertex [dot] (a3){};
				\vertex [right = 1.5cm of a3] (a4);
				\vertex [right = 1.0cm of a4] (a5){\(\nu_j^{}\)};
				\vertex [left = 1.5cm of a3] (a2);
				\vertex [left = 1.0cm of a2] (a1){\(\nu_i^{}\)};
				\vertex [above = 1.5cm of a3] (b3);
				\vertex [above = 1cm of b3] (c3){\(\braket{H^0}\)};
				\diagram{(a5)--[fermion](a4),(a3)--[fermion,edge label'=\(\dq_k^c\)](a4),
					(a1)--[fermion](a2),(a3)--[fermion,edge label=\(\dq_k^{}\)](a2),
					(a4)--[charged scalar,quarter right,edge label'=\(S_1^{\,1/3}\)](b3)--[charged scalar,quarter right,edge label'=\(\widetilde R_2^{\,1/3}\)](a2),
					(c3)--[scalar](b3)};
			\end{feynman}
		\end{tikzpicture}
	\end{center}
	\caption{One-loop diagrams for generation of neutrino mass matrix through leptoquarks with $(i,j,k)$ being generation indices.}
	\label{fig:nu}
\end{figure}
In the framework of this model, Majorana mass of the light neutrinos is generated at one-loop level via the Feynman diagram depicted in \autoref{fig:nu}, where the simultaneous presence of the Yukawa couplings, $\y_1^{L}$, $\y_2$, and scalar trilinear coupling $\kappa$ leads to the lepton number violation. In the absence of any of the aforementioned couplings, this mass term vanishes. The light neutrino mass matrix from this diagram is obtained as \cite{Babu:2019mfe} \footnote{Here, Yukawa couplings of leptoquarks are considered to be real.}

\begin{align}
	\label{eq:mnu_SD}
	M_\nu\simeq\frac{3\sin{2\tlq}}{32\pi^2}\ln{\bigg(\frac{M_{1}^2}{M_{2}^2}\bigg)}\Big[\y_1^L\, \mathcal m_\dq\, \y_2^T+\y_2\, \mathcal m_\dq\,(
	\y_1^L)^T\Big],
\end{align}
where, $\mathcal m_\dq$ is the diagonal mass matrix for down-type quarks. In the above equation, it has been assumed that the leptoquarks are very heavy compared to the mass of all the down-type quarks.

\begin{figure}[h!]
	\hspace*{-7.5mm}
	\mbox{
		\begin{tikzpicture}
			\begin{feynman}
				\vertex (a0){$l_i$};
				\vertex [right=10mm of a0] (a1);
				\vertex [right=20mm of a1] (a2);
				\vertex [right=10mm of a1] (ap);
				\vertex [right=7mm of a2] (a3){$l_j$};
				\vertex [above=10mm of ap] (bp);
				\vertex [above=10mm of bp] (cp){$\gamma$};
				\diagram {(a0)--[fermion](a1)--[charged scalar,edge label=\scriptsize{${\widetilde{R}_2^{2/3}/\X_1/\X_2}$}](a2)--[fermion](a3),
					(a1)--[fermion,quarter left,edge label=${\uq/\dq}$](bp)--[fermion,quarter left](a2),
					(bp)--[boson](cp)};
			\end{feynman}
		\end{tikzpicture}
		\begin{tikzpicture}
			\begin{feynman}
				\vertex (a0){$l_i$};
				\vertex [right=10mm of a0] (a1);
				\vertex [right=20mm of a1] (a2);
				\vertex [right=10mm of a1] (ap);
				\vertex [right=7mm of a2] (a3){$l_j$};
				\vertex [above=10mm of ap] (bp);
				\vertex [above=10mm of bp] (cp){$\gamma$};
				\diagram {(a0)--[fermion](a1)--[fermion](a2)--[fermion](a3),
					(a1)--[charged scalar,quarter left](bp)--[charged scalar,quarter left](a2),
					(bp)--[boson](cp)};
			\end{feynman}
		\end{tikzpicture}
		
		\begin{tikzpicture}
			\begin{feynman}
				\vertex (a0){$l_i$};
				\vertex [right=12mm of a0] (a1);
				\vertex [right=20mm of a1] (a2);
				\vertex [right=10mm of a1] (ap);
				\vertex [right=7mm of a2] (a3){$l_j$};
				\vertex at ($(a0)!0.7!(a1)$) (xp);
				\vertex [above=10mm of ap] (bp);
				\vertex [above=15mm of xp] (cp){$\gamma$};
				\diagram {(a0)--[fermion](a1)--[fermion](a2)--[fermion](a3),
					(a1)--[charged scalar,quarter left](bp)--[charged scalar,quarter left](a2),
					(xp)--[boson](cp)};
			\end{feynman}
		\end{tikzpicture}
		\begin{tikzpicture}
			\begin{feynman}
				\vertex (a0){$l_i$};
				\vertex [right=10mm of a0] (a1);
				\vertex [right=20mm of a1] (a2);
				\vertex [right=10mm of a1] (ap);
				\vertex [right=10mm of a2] (a3){$l_j$};
				\vertex at ($(a2)!0.3!(a3)$) (xp);
				\vertex [above=10mm of ap] (bp);
				\vertex [above=15mm of xp] (cp){$\gamma$};
				\diagram {(a0)--[fermion](a1)--[fermion](a2)--[fermion](a3),
					(a1)--[charged scalar,quarter left](bp)--[charged scalar,quarter left](a2),
					(xp)--[boson](cp)};
			\end{feynman}
		\end{tikzpicture}
	}
	\caption{One-loop diagrams contributing to $l_i\to l_j\gamma$. The same diagrams will contribute to $g-2$ of muon and electron for $i=j$.}
	\label{fig:LFV}
\end{figure}
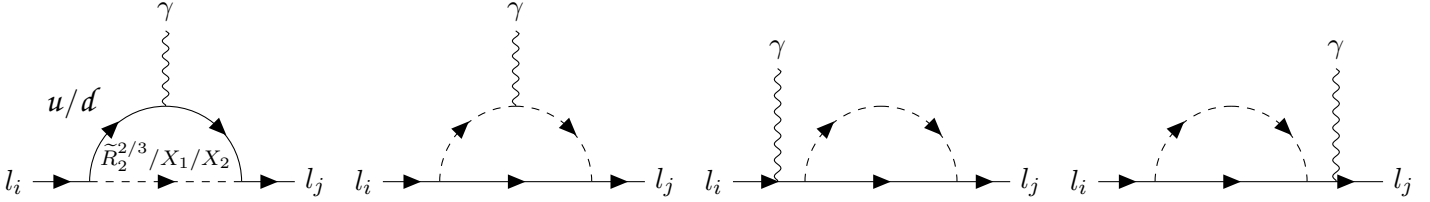

In this model, leptoquarks also contribute to CLFV processes like $l_i \to l_j \gamma $ at one-loop order, as shown in \autoref{fig:LFV}. In this work, we have implemented the model in \texttt{SARAH} \cite{Sarah} to generate model files for {\tt SPheno} \cite{Porod:2003um,Porod:2011nf} and {\tt MadGraph5} \cite{mg5} to numerically compute the mass spectrum, decays, different low energy observables including the CLFV processes, charged lepton anomalous magnetic moments, and collider signatures, respectively. However, for the sake of completeness and understanding about the dependence of CLFV processes and charged lepton anomalous magnetic moments on the Yukawa couplings approximate analytical expressions for the $\mathcal{B}(l_i \to l_j \gamma)$ and $\Delta a_l$ are summarizes in the following. 

In the framework of this model, the branching fraction for the $i^{\rm th}$ charged lepton flavour violating decay into $j^{\rm th}$ lepton and a photon, $\mathcal{B}(l_i\to l_j\gamma)$ can be expressed as \cite{Zhang:2021dgl}:
\begin{equation}
	\mathcal{B}(l_i\to l_j\gamma)=\frac{e^2(m_i^2-m_j^2)^3}{4096\pi^5 m_i^3 \Gamma_i}\bigg[\Big|\mathcal{A}_L^{\X_1}+\mathcal{A}_L^{\X_2}+\mathcal{A}_L^{\widetilde{R}_2^{2/3}}\Big|^2+\Big|\mathcal{A}_R^{\X_1}+\mathcal{A}_R^{\X_2}+\mathcal{A}_R^{\widetilde{R}_2^{2/3}}\Big|^2\bigg],
\end{equation}
where, $m_{i,j}$ are the masses of the charged leptons with flavour $(i,j)$ and $\Gamma_i$ the total decay width of $i$-th lepton. Now, the loop-functions $\mathcal A$ are given by:
{\small
	\begin{align}
		&\hspace*{-4mm}\mathcal A_L^{\X_{1,2}}=\frac{\mathcal C_{1,2}^2}{4 M_1^2}\bigg[2\,(\widetilde\y_1^L)_{ki} (\mathcal{m}_\uq^{kk})\, (\y_1^R)_{kj}\,\mathcal{F}\bigg\{\frac{(\mathcal {m}^{kk}_\uq)^2}{M_{1,2}^2}\bigg\}-\Big\{m_j(\widetilde\y_1^L)_{ki} (\widetilde\y_1^L)_{kj}
		+m_i(\y_1^R)_{ki} (\y_1^R)_{kj} \Big\}\mathcal{G}\bigg\{\frac{(\mathcal {m}_\uq^{kk})^2}{M_{1,2}^2}\bigg\}\bigg],\\
		&\hspace*{-4mm}\mathcal A_R^{\X_{1,2}}=\frac{\mathcal C_{1,2}^2}{4 M_1^2}\bigg[2\,(\y_1^R)_{ki} (\mathcal{m}_\uq^{kk})\, (\widetilde\y_1^L)_{kj}\,\mathcal{F}\bigg\{\frac{(\mathcal {m}^{kk}_\uq)^2}{M_{1,2}^2}\bigg\}-\Big\{m_i(\widetilde\y_1^L)_{ki} (\widetilde\y_1^L)_{kj}+m_j(\y_1^R)_{ki} (\y_1^R)_{kj} \Big\}\mathcal{G}\bigg\{\frac{(\mathcal {m}_\uq^{kk})^2}{M_{1,2}^2}\bigg\}\bigg],\\
		&\hspace*{-4mm}\mathcal{A}_L^{\widetilde{R}_2^{2/3}}=-\frac{m_j}{4m^2(\widetilde{R}_2^{2/3})}(\y_2)_{ki}(\y_2)_{kj}\,\mathcal I\bigg\{\frac{(\mathcal m_\dq^{kk})^2}{m^2(\widetilde{R}_2^{2/3})}\bigg\},\quad \mathcal{A}_L^{\widetilde{R}_2^{2/3}}=-\frac{m_i}{4m^2(\widetilde{R}_2^{2/3})}(\y_2)_{ki}(\y_2)_{kj}\,\mathcal I\bigg\{\frac{(\mathcal m_\dq^{kk})^2}{m^2(\widetilde{R}_2^{2/3})}\bigg\},
	\end{align}
}
where, $\widetilde\y_1^L=V \y_1^L$ ($V$ being the usual Cabibbo-Kobayashi-Maskawa (CKM) matrix) , $\mathcal C_1=\cos\tlq$, $\mathcal C_2=\sin\tlq$ along with
\begin{align}
	&\mathcal F(x)=\frac{7-8x+x^2+2(2+x)\ln x}{(1-x)^3}, \qquad \mathcal G(x)=\frac{1+4x-5x^2+2x(2+x)\ln x}{(1-x)^4},\\ &\hspace{3cm}\mathcal{I}(x)=\frac{x\big[5-4x-x^2+2(1+2x)\ln x\big]}{(1-x)^4}.
\end{align}
It is important to mention here that we use the convention of rotating the flavour eigenstates of up-type quarks by the CKM matrix to obtain their mass eigenstates\footnote{For the choice of our benchmark points, we have used {\tt SPheno} which uses the same convention.}. Therefore, the left-handed Yukawa coupling $\y_1^L$ changes to $\widetilde \y_1^L$ in the square of amplitude while considering the interaction of an up-type quark with a charged lepton through leptoquark $S_1$ (i.e. in the mass basis through the singlet component of $\xot$), while the rest of the couplings between quarks and leptons through leptoquarks remain unaltered. 

The additional contribution, resulting from the leptoquarks in the loop, to the the $j^{\rm th}$ charged lepton magnetic moment can be expressed in the following way:
\begin{equation}
	\Delta a_j=\Delta a_j^{\X_1}+\Delta a_j^{\X_2}+\Delta a_j^{\widetilde{R}_2^{2/3}},
\end{equation}
\small{
	\begin{align}
		&\Delta a_j^{\X_{1,2}}=-\frac{\mathcal C_{1,2}^2 m_j}{32\pi^2 M_1^2}\bigg[2\,(\mathcal{m}_\uq^{kk})\,(\widetilde\y_1^L)_{kj}  (\y_1^R)_{kj}\,\mathcal{F}\bigg\{\frac{(\mathcal {m}^{kk}_\uq)^2}{M_{1,2}^2}\bigg\}-m_j\Big\{ \Big|(\widetilde\y_1^L)_{kj}\Big|^2+\Big|(\y_1^R)_{kj}\Big|^2 \Big\}\mathcal{G}\bigg\{\frac{(\mathcal {m}_\uq^{kk})^2}{M_{1,2}^2}\bigg\}\bigg],\\
		&\hspace{5cm}\Delta a_j^{\widetilde{R}_2^{2/3}}=\frac{m_j^2|(\y_2)_{kj}|^2}{32\pi^2m^2(\widetilde{R}_2^{2/3})}\,\mathcal I\bigg\{\frac{(\mathcal m_\dq^{kk})^2}{m^2(\widetilde{R}_2^{2/3})}\bigg\},
	\end{align}
}
\normalsize
with all the relevant functions defined above. One can argue from the above expressions that for TeV-scale leptoquark masses, the contribution from $\rtt$ in explaining the experimental excess of muon $(g-2)$ is negligible compared to $\xot$, and assuming no mixing, $S_1$ alone can explain the excess. However, we inevitably require both these leptoquarks in the model to generate the Majorana neutrino mass as discussed, and hence we consider muon $(g-2)$ contribution from both $\rtt$ and $\xot$, for the sake of completeness. The muon-electron conversion rate with the presence different nuclei for generic leptoquark model has been discussed using effective field theory approach in Ref. \cite{Mandal:2019gff}.

\section{Benchmark points}
\label{sec:bps}

The primary goal of this work is to probe the model at the LHC/FCC for the particular set of model parameters, which are not only consistent with the neutrino mass and oscillation data, but also satisfy the current experimental values of electron and muon $g-2$, while respecting the upper bounds on CLFV decays of charged leptons. Following the brief introduction of the model including radiative neutrino mass generation, CLFV processes and $g-2$, we are now equipped enough to search for benchmark points on which we will perform the collider study. 

We diagonalize the neutrino mass matrix of \autoref{eq:mnu_SD} to obtain the Yukawa couplings $\y_1^{L}$ and $\y_2$ which are consistent with the neutrino mass sum and mass-squared difference limits. Then, we utilize {\tt SPheno} \cite{Porod:2003um,Porod:2011nf} version 4.0.4 to find out which set of such couplings can satisfy the $g-2$ data simultaneously with the neutrino mass data, with adequate choices of entries in $\y_1^{R}$. Then, with the help of {\tt SPheno}, the CLFV observables are obtained for such couplings, while keeping them consistent with the neutrino oscillation parameters as well. In the following, we list three benchmark points (BP), each with different phenomenological aspect. The parameters of the scalar potential are kept fixed across the three BPs as noted in \autoref{tab:para}. The masses of the leptoquarks $m_1$ and $m_2$ are considered in accordance with the LHC data at $2\sigma$ level\cite{CMS:2018ncu,CMS:2018lab, ATLAS:2020dsk}. For this set of parameters in \autoref{tab:para}, we obtain $\tlq=-0.618$ radians. This leads to $\X_1$ containing $\sim66.5\%$ of $S_1$ and $\sim33.5\%$ of $\widetilde{R}_2^{1/3}$ while for $\X_2$ these proportions get reversed.

\begin{table}[h!]
	\centering
	\begin{tabular}{|c|c|c|c|c|c|c|}
		\hline
		Parameters&$m_1$&$m_2$&$\alpha_1$&$\alpha_2$&$\alpha_2^\prime$&$\kappa$\\
		\hline
		Values&1.5 TeV&1.5 TeV&0.2&0.2&0.2&50 GeV\\
		\hline
	\end{tabular}
	\caption{Parameters except Yukawa couplings for all the BPs (at  5 GeV renormalization scale of {\tt SPheno}).}
	\label{tab:para}
\end{table}

It is worth mentioning here that, generic studies of the parameter space of this model exist in contemporary literature \cite{Zhang:2021dgl, Freitas:2022gqs}, taking into account different sets of bounds alongside the neutrino mass generation. Ref. \cite{Freitas:2022gqs} for example, presents a parameter scan for the $\widetilde{R}_2 + S_1$ combination considering not only the muon $g-2$ and CLFV bounds, but also the various $B$-decay anomalies and the latest CDF-II measurement of the $W^\pm$-mass\cite{CDF:2022hxs}, with complex Yukawa couplings. In our work however, the collider searches for these leptoquarks is given prime importance, which requires adherence to a set of benchmark scenarios. We wish to perform a detailed and comparative study of multiple finalstates at the LHC/FCC, and hence we have strategically chosen our BPs in such a way that the Yukawa couplings follow three different regimes of relative strength, in an attempt to draw a more comprehensive picture of the parameter space and their respective phenomenology. For our first BP, the entries of the $\yt$ are taken to be very tiny in comparison to those of $\yl$, as prescribed in \cite{Babu:2019mfe}. In the second regime i.e. BP2, two entries of $\yt$ are chosen in the same order of magnitude as those of $\yl$. And lastly, in the third benchmark scenario, $\yt$ contains an element that is larger than any of those in $\yl$, leading to more impact of the leptoquark mixing in the decay branching ratios. All these BPs are fine-tuned to respect the indirect bounds from neutrino oscillation data and CLFV decays, and if one wishes to perform a generic collider study over a large parameter space, one can certainly extrapolate these three regimes by varying the Yukawa couplings around the chosen BPs, which may lead to changes in the values corresponding to the indirect bounds.

In the subsections that follow, we will discuss the choice of our benchmark points from the perspective of each of the sets of bounds that we prioritize. The extremely sensitive nature of the bounds from neutrino mass and oscillation, anomalous magnetic moment, and other CLFV observables means we require very high precision on the Yukawa couplings to satisfy all of them simultaneously. However, from a collider perspective, it is not possible to probe Yukawa couplings up to such accuracy. Therefore, we start off with the neutrino masses as our first priority, and choose the couplings up to the least decimal point with which these sub-eV masses are satisfied within the mass sum and mass squared difference limits. The next immediate priority is given to the muon and electron $g-2$ measurements. After that, the required fine-tuning to satisfy the neutrino oscillation data and the CLFV bounds will be discussed.

\subsection{Choice of Yukawa Couplings}
\label{sec:yuk_br}
As discussed previously, the couplings stated here are kept up to the lowest possible precision that are in agreement with the resultant neutrino mass eigenvalues. The LHC/FCC is insensitive towards the higher orders of precision, which allows us the freedom to fine-tune the entries.
	
\textbf{BP1:} The choice of Yukawa couplings for BP1 are as follows:
\begin{equation*}
	\y_1^L=\begin{pmatrix}
		0.220&\,\,0.001&\,\,-0.030 \\
		0.150&\,\,-0.005&\,\,0.140 \\
		0.005&\,\,0.120&\,\,0.006
	\end{pmatrix}; \quad 
	\y_1^R=\begin{pmatrix}
		1.000&\,\,0.000&\,\,0.000\\
		0.150&\,\,0.000&\,\,0.200 \\
		0.000&\,\,0.120&\,\,0.009
	\end{pmatrix}; 
\end{equation*}
\begin{equation}
	\label{eq:BP1}
	\quad \y_2=10^{-3}\begin{pmatrix}
		0.00000&\,\,-0.03900&\,\,-0.00082 \\
		0.28200&\,\,0.00000&\,\,0.00000 \\
		0.35900&\,\,0.01810&\,\,0.00000
	\end{pmatrix}.
\end{equation}

In this scenario, the tiny values of $\y_2$ makes $\widetilde{R}_2$ almost inert and the therefore the total decay width of $\rtt$ becomes very small, i.e. $6.27\times 10^{-6}$ GeV, but large enough to create prompt decays. Due to this fact the dynamics of $\xo$ and $\xt$ are mostly dominated by the singlet component $S_1$. The mass eigenstates of the leptoquarks with charge $1/3$, i.e. $\xot$ are almost degenerate with a mass difference of $\sim7$ GeV; however, the total decay width of $\X_1$ is approximately twice of that of $\X_2$. The branching fractions of the leptoquarks in their mass basis, for different modes are presented in \autoref{bp1decay}. Among all the components of $\y_1^L$ and $\y_1^R$, the large $(\y_1^R)_{11}=1.0$ makes $ue$ the dominant decay channel of $\xot$ with 82.4\% of branching fraction. The doublet Yukawa $\y_2$ has negligible effect on their branching, and only governs the decay modes of $\rtt$.  
	\begin{table}[h!]
		\centering
		\renewcommand{\arraystretch}{1}
		\begin{tabular}{|M{0.8cm}|M{1.5cm}|M{1cm}||M{0.8cm}|M{3cm}|M{1cm}||M{0.8cm}|M{3cm}|M{1cm}|}
			\hline
			\multicolumn{9}{|c|}{Dominant Decay Modes of leptoquarks for BP1}\\
			\hline
			\multicolumn{3}{|c||}{\makecell{$\widetilde R_2^{+2/3}$ \\ $m(\widetilde R_2^{2/3})=1.502$ TeV}}&\multicolumn{3}{c||}{\makecell{$ \X_1^{-1/3}$ \\ $M_1=1.499$ TeV}}&\multicolumn{3}{c|}{\makecell{$\X_2^{-1/3}$ \\ $M_2=1.506$ TeV}}\\[2mm]
			\hline
			\multicolumn{3}{|c||}{$\Gamma(\widetilde R_2^{2/3})=6.27\times10^{-6}$ GeV}&\multicolumn{3}{c||}{$\Gamma(\X_1)=25.6$ GeV}&\multicolumn{3}{c|}{$\Gamma(\X_2)=13.0$ GeV}\\
			\hline
			Mode & Dominant Yukawa& BR(\%) & Mode & Dominant Yukawa& BR(\%) & Mode& Dominant Yukawa & BR(\%)\\
			\hline 
			$be^+$&$(\y_2)_{31}$&61.6&$ue^-$&$(\yr)_{11} + (\widetilde \y_1^L)_{11}$&82.4&$ue^-$&$(\yr)_{11} + (\widetilde \y_1^L)_{11}$&82.4\\
			$se^+$&$(\y_2)_{21}$&37.5&$c\tau^-$&$(\yr)_{23}+(\widetilde \y_{1}^L)_{23}$&4.6&$c\tau^-$&$(\yr)_{23}+(\widetilde \y_{1}^L)_{23}$&4.6\\
			&&&$d\nu_e$&$(\yl)_{11}$&3.8&$d\nu_e$&$(\yl)_{11}$&3.8\\
			%		&&$b\nu_\mu$&8.8&$c\tau$&8.1\\
			%		&&$t\mu$&8.6&$b\nu_\tau$&3.6\\
			%		&&$b\nu_e$&6.2&$t\tau$&3.5\\
			%		&&$u\mu$&3.9&&\\
			%		&&$d\nu_\mu$&3.9&&\\
			\hline
		\end{tabular}
		\caption{Dominant decay modes and branching ratios of the three leptoquark eigenstates for BP1.}
		\label{bp1decay}
\end{table}

\textbf{BP2}: For this benchmark point, the $\y_1^L$ and $\y_1^R$ Yukawa couplings are chosen in such a way that, for $\xo$ and $\xt$ leptoquarks we can obtain more than one dominant decay modes with comparable branching fractions, unlike in BP1. The structures of these couplings are given below:

	\begin{equation*}
		\y_1^L=\begin{pmatrix}
			0.0000&\,\,0.3740&\,\,-0.0100\\
			-0.0008&\,\,0.5990&\,\,0.0420\\
			0.0200&\,\,0.0223&\,\,0.0418
		\end{pmatrix}; \quad 
		\y_1^R=\begin{pmatrix}
			0.00&\,\,0.50&\,\,0.00\\
			0.00&\,\,0.50&\,\,0.05 \\
			0.02&\,\,0.00&\,\,0.03
		\end{pmatrix}; 
	\end{equation*}
	\begin{equation}
		\label{eq:BP2}
		\quad \y_2=10^{-3}\begin{pmatrix}
			5.0000&\,\,-1.1893&\,\,-1.3230 \\
			-144.0400&\,\,5.0000&\,\,-2.1195 \\
			-108.8400&\,\,5.0363&\,\,0.0000
		\end{pmatrix}.
	\end{equation}

The total decay widths and branching fractions of all the leptoquarks under this BP are listed in \autoref{bp2decay}. In this case, the total decay width of $\widetilde{R}_2^{+2/3}$ gets enhanced to $\sim 1$ GeV, due to larger values of $(\y_2)_{21}$ and $(\y_2)_{31}$, i.e. $\mathcal O(10^{-1})$, compared to the BP1 scenario, where the total decay width is $\mathcal{O}(10^{-6})$ GeV. Owing to the larger values of $(\y_2)_{21}$ and $(\y_2)_{31}$, $\rtt$ decays dominantly to $se^+$ (63.5$\%$) and $be^+$ (36.3\%). The mass splitting and the ratio of the total decay widths between $X_1$ and $X_2$ remain similar as BP1. However, for this particular choice of $\y_1^{R,L}$, $\xo$ and $\xt$ both dominantly decay to $u\mu$ and $c\mu$, with equal branching fractions of $32.7\%$ in case of $\xo$, and $31.7\%$ in case of $\xt$, respectively. This is attributed to the fact that, while we already have $(\y_{1}^R)_{12}=(\y_{1}^R)_{22}=0.5$, after the CKM rotation of $\y_1^L$ we also get $(\widetilde\y_{1}^L)_{12}=(\widetilde\y_{1}^L)_{22}\approx0.5$. Additionally, two other significant decay modes of $\xot$ open up, namely $s\nu$ ($\sim$23\%) and $d\nu$ ($\sim$9\%).
\begin{table}[h!]
	\centering
	\renewcommand{\arraystretch}{1}
	\begin{tabular}{|M{0.8cm}|M{1.5cm}|M{1cm}||M{0.8cm}|M{3cm}|M{1cm}||M{0.8cm}|M{3cm}|M{1cm}|}
		\hline
		\multicolumn{9}{|c|}{Dominant Decay Modes of leptoquarks for BP2}\\
		\hline
		\multicolumn{3}{|c||}{\makecell{$\widetilde R_2^{+2/3}$ \\ $m(\widetilde{R}_2^{2/3})=1.502$ TeV}}&\multicolumn{3}{c||}{\makecell{$ \X_1^{-1/3}$ \\ $M_1=1.499$ TeV}}&\multicolumn{3}{c|}{\makecell{$\X_2^{-1/3}$ \\ $M_2=1.506$ TeV}}\\
		\hline
		\multicolumn{3}{|c||}{$\Gamma(\widetilde R_2^{2/3})=9.76\times10^{-1}$ GeV}&\multicolumn{3}{c||}{$\Gamma(\X_1)=30.3$ GeV}&\multicolumn{3}{c|}{$\Gamma(\X_2)=15.9$ GeV}\\
		\hline
		Mode & Dominant Yukawa &BR(\%) & Mode &Dominant Yukawa &BR(\%) & Mode & Dominant Yukawa&BR(\%)\\
		\hline 
		$se^+$&$(\y_2)_{21}$&63.5&$u\mu^-$&$(\yr)_{12} + (\widetilde\y_{1}^L)_{12}$&32.7&$u\mu^-$&$(\yr)_{12}+(\widetilde\y_{1}^L)_{12}$&31.7\\
		$be^+$&$(\y_2)_{31}$&36.3&$c\mu^-$&$(\yr)_{22}+(\widetilde\y_{1}^L)_{22}$&32.7&$c\mu^-$&$(\yr)_{22}+(\widetilde\y_{1}^L)_{22}$&31.7\\
		&&&$s\nu_\mu$&$(\yl)_{22}$&23.5&$s\nu_\mu$&$(\yl)_{22}$&22.8\\
		&&&$d\nu_\mu$&$(\yl)_{12}$&9.2&$d\nu_\mu$&$(\yl)_{12}$&8.9\\
		%		&&$b\nu_\mu$&8.8&$c\tau$&8.1\\
		%		&&$t\mu$&8.6&$b\nu_\tau$&3.6\\
		%		&&$b\nu_e$&6.2&$t\tau$&3.5\\
		%		&&$u\mu$&3.9&&\\
		%		&&$d\nu_\mu$&3.9&&\\
		\hline
	\end{tabular}
	\caption{Dominant decay modes and branching ratios of the three leptoquark eigenstates for BP2.}
	\label{bp2decay}
\end{table}

\textbf{BP3}: For the third benchmark points, we wish to have significantly different branching ratios for $\xo$ and $\xt$, contrary to the two previous scenarios. Hence, the Yukawa couplings for this BP are chosen as follows:

	\begin{equation*}
		\y_1^L=\begin{pmatrix}
			-0.0070&0.0490&0.4870\\
			0.0290&0.0092&0.1124 \\
			0.0012&0.0500&0.0017
		\end{pmatrix}; \quad 
		\y_1^R=\begin{pmatrix}
			0.0&0.5&0.5\\
			0.5&0.0&0.0 \\
			0.0&0.3&0.0
		\end{pmatrix}; 
	\end{equation*}
	\begin{equation}
		\label{eq:BP3}
		\quad \y_2=10^{-3}\begin{pmatrix}
			0.00000&878.31000&-2.62350 \\
			-2.03300&200.00000&-0.48457 \\
			89.30000&3.00000&-2.67500
		\end{pmatrix}.
	\end{equation}

\begin{table}[h!]
	\centering
	\renewcommand{\arraystretch}{1}
	\begin{tabular}{|M{0.8cm}|M{1.5cm}|M{1cm}||M{0.8cm}|M{3cm}|M{1cm}||M{0.8cm}|M{3cm}|M{1cm}|}
		\hline
		\multicolumn{9}{|c|}{Dominant Decay Modes of leptoquarks for BP3}\\
		\hline
		\multicolumn{3}{|c||}{\makecell{$\widetilde R_2^{+2/3}$ \\ $m(\widetilde{R}_2^{2/3})=1.502$ TeV}}&\multicolumn{3}{c||}{\makecell{$ \X_1^{-1/3}$ \\ $M_1=1.499$ TeV}}&\multicolumn{3}{c|}{\makecell{$\X_2^{-1/3}$ \\ $M_2=1.506$ TeV}}\\
		\hline
		\multicolumn{3}{|c||}{$\Gamma(\widetilde R_2^{+2/3})=24.4$ GeV}&\multicolumn{3}{c||}{$\Gamma(\X_1)=34.9$ GeV}&\multicolumn{3}{c|}{$\Gamma(\X_2)=29.8$ GeV}\\
		\hline
		Mode & Dominant Yukawa &BR(\%) & Mode &Dominant Yukawa & BR(\%) & Mode &Dominant Yukawa & BR(\%)\\
		\hline 
		$d\mu^+$&$(\yt)_{12}$&94.1&$u\tau^-$&$(\yr)_{13}+(\widetilde\y_1^L)_{13}$&28.4&$d\bar{\nu}_\mu$&$(\yt)_{12}$&51.4\\
		$s\mu^+$&$(\yt)_{22}$&4.9&$d\bar{\nu}_\mu$&$(\yt)_{12}$&22.2&$u\tau^-$&$(\yr)_{13}+(\widetilde\y_1^L)_{13}$&16.9\\
		&&&$u\mu^-$&$(\yr)_{12}+(\widetilde\y_1^L)_{12}$&14.3&$u\mu^-$&$(\yr)_{12}+(\widetilde\y_1^L)_{12}$&8.5\\
		&&&$ce^-$&$(\yr)_{21}+(\widetilde\y_1^L)_{21}$&14.2&$ce^-$&$(\yr)_{21}+(\widetilde\y_1^L)_{21}$&8.4\\
		&&&$d\bar{\nu}_\tau$&$(\y_1^L)_{13}$&13.5&$d\bar{\nu}_\tau$&$(\y_1^L)_{13}$&8.0\\
		%		&&$b\nu_\mu$&8.8&$c\tau$&8.1\\
		%		&&$t\mu$&8.6&$b\nu_\tau$&3.6\\
		%		&&$b\nu_e$&6.2&$t\tau$&3.5\\
		%		&&$u\mu$&3.9&&\\
		%		&&$d\nu_\mu$&3.9&&\\
		\hline
	\end{tabular}
	\caption{Dominant decay modes and branching ratios of the three leptoquark eigenstates for BP3.}
	\label{bp3decay}
\end{table}

The branching fractions for the leptoquark mass eigenstates are depicted in \autoref{bp3decay}. As a result of the large $(\y_2)_{12}$ and $(\y_2)_{22}$, the total decay width of $\rtt$ in this scenario, i.e. 24.4 GeV, becomes comparable to that of $\xot$. Because of the same fact, the branching fractions of $\xo$ and $\xt$ look quite different, since $\xo$ and $\xt$ contain different proportions of $S_1$ and $\widetilde{R}_2^{1/3}$. Here, the total decay widths of $\xo$ and $\xt$ are also quite close unlike the previous two cases, as a result of both $\y_1^R$ and $\y_2$ containing entries of $\mathcal{O}(10^{-1})$. The element $(\y_2)_{12}$ = 0.876269 results in $\rtt$ decaying dominantly to $d\mu^+$ with a branching fraction of $94.1\%$. As the $\X_2$ leptoquark contains 66\% of $\widetilde{R}_2^{-1/3}$, so this same large coupling enhances the decay of $\xt \to d\nu_\mu$,  with a 51.4\% branching fraction. The other decay modes of $\xt$ are dominated by $\y_1^{L,R}$, the larger of them being $u\tau$ (16.9\%). On the other hand, as $\xo$ contains 66\% of $S_1$, the largest branching fraction is observed in the $u\tau$ decay channel (28.4\%). However, the larger value of $(\y_2)_{12}$ still affects its decay, owing to the 33.5\% of $\widetilde{R}_2^{-1/3}$ remaining in it. This results in the 22.2\% branching fraction of $\xo$ decaying to $d\nu_\mu$. However, significant difference between the BRs of $\xot \to d\nu_\mu$ is observed, which is the purpose of choosing the Yukawa couplings as described in \autoref{eq:BP3}.

\subsection{Neutrino mass from the chosen BPs}

Following the discussion on our choice of Yukawa couplings and their effect on the leptoquark decay channels, we will now look into the experimental bounds on various low-energy observables and their corresponding values under these benchmark scenarios. The first motivation of choosing this model containing a doublet and a singlet leptoquark was to generate neutrino masses at one-loop. It turns out that, up to the stated accuracy of the Yukawa coupling entries, the neutrino mass eigenvalues shape up as depicted in \autoref{tab:mnu}. though there is no bound on individual masses of active neutrinos, their total mass should be less than 0.09 eV \cite{DiValentino:2021hoh}. We observe that, in all the three benchmark points, the neutrino mass sum stays within a value of 0.075 eV.

\begin{table}[h!]
	\renewcommand{\arraystretch}{1}
	\centering
	\begin{tabular}{|c|c|c|c|c|}
		\hline
		Parameter& Expt. bound  & BP1 &BP2& BP3\\
		\hline\hline
		$m_{\nu_{1}}$ (eV)&---&$0.009$&$0.010$ &$0.009$ \\
		$m_{\nu_{2}}$ (eV)&---&$0.012$&$0.013$& $0.013$\\
		$m_{\nu_{3}}$ (eV)&---&$0.051$&$0.052$&$0.052$\\
		$\sum m_\nu$ (eV) & $<0.09$ \cite{DiValentino:2021hoh} &$0.072$&$0.075$&$0.074$\\
		\hline
	\end{tabular}
	\caption{Masses of neutrinos under different benchmark points and experimental bound.}
	\label{tab:mnu}
\end{table}

The next immediate bounds comes from the neutrino mass-squared differences. It is important to note that, for the purpose of this work, the normal hierarchy of neutrino mass is considered, where $m_{\nu_3} > m_{\nu_2} > m_{\nu_1}$ \cite{deSalas:2020pgw}. However, one can choose to work with the inverted hierarchy as well, tweaking the parameter space as required. In consideration of the normal hierarchy, the respective mass-squared differences and the experimental $3\sigma$ range are provided in \autoref{tab:nmsq}. Up to the quoted order of precision, the entries of the Yukawa coupling matrices results into values within the allowed $3\sigma$ range.

\begin{table}[h!]
	\renewcommand{\arraystretch}{1}
	\centering
	\begin{tabular}{|c|c|c|c|c|}
		\hline
		Parameter& Expt. $3\sigma$ range  & BP1 &BP2& BP3\\
		\hline
		$\Delta m_{21}^2 \,(\times 10^{-5} \, \text{eV}^2 )$&$6.94-8.14$ &7.03&7.84&8.12\\
		$\Delta m_{31}^2 \,(\times 10^{-3} \, \text{eV}^2 )$ & $2.47-2.63$&2.47&2.63&2.63\\
		\hline
	\end{tabular}
	\caption{Neutrino mass-squared differences for the three benchmark points, along with the experimental $3\sigma$ values (normal hierarchy).}
	\label{tab:nmsq}
\end{table}

However, further fine-tuning at least at the third decimal point of the chosen Yukawa couplings is unavoidable to fully satisfy the neutrino oscillation parameters within the $3\sigma$ range, which we will discuss towards the end of this section.

\subsection{Muon and electron $g-2$ from the chosen BPs}

With the help of {\tt SPheno}, it is observed that the benchmark points up to the stated accuracy also satisfy the muon and electron $g-2$ data \cite{Muong-2:2021ojo,Parker:2018vye,Morel:2020dww}. The experimental results for these observables and their values under our choice of benchmark points are listed in \autoref{tab:g-2}. It should be noticed that the values of the electron $g-2$ for our BPs are more in accordance with the experimental values for Rubidium, which is the more recent measurement.

\begin{table}[h!]
	\renewcommand{\arraystretch}{1}
	\centering
	\begin{tabular}{|c|c|c|c|c|}
		\hline
		Obs&Expt. Value&BP1&BP2&BP3\\
		\hline
		$\Delta a_\mu$
		&$(2.51\pm0.59)\times10^{-9}$ \cite{Muong-2:2021ojo}&$2.97\times 10^{-9}$&$2.44\times 10^{-9}$&$3.10\times 10^{-9}$\\
		\hline
		$\Delta a_e^{Cs}$
		&$(-8.8\pm3.6)\times10^{-13}$ \cite{Parker:2018vye}&\multirow{2}{*}{$7.71\times 10^{-13}$}&\multirow{2}{*}{$4.00\times 10^{-13}$}&\multirow{2}{*}{$7.02\times 10^{-13}$}\\
		$\Delta a_e^{Rb}$&$(4.8\pm3.0)\times10^{-13}$ \cite{Morel:2020dww}&&&\\
		\hline
	\end{tabular}
	\caption{Experimental estimates of muon and electron $(g-2)$ and their values for the three benchmark points, given by {\tt SPheno}.}
	\label{tab:g-2}
\end{table}

\subsection{Fine-tuning of the BPs}

Having discussed the bounds that are satisfied with the lowest possible precision of the Yukawa couplings, we now move towards the fine-tuning of the same, through which the neutrino oscillation data and the CLFV observables can also be addressed. The fine-tunings are required mainly for $\y_1^L$ and $\y_2$, at least at the third decimal place, as these two Yukawa couplings contribute towards the neutrino mass and mixing. For our purpose, $\y_1^R$ does not require any such fine-tuning. It is important to note that, from the perspective of analysis at the LHC/FCC, such fine-tuned precise values of the couplings do not affect the observations of the finalstates. As we have seen from the discussion on the decay branching ratios in \autoref{sec:yuk_br}, only the entries of $\mathcal{O}(10^0 - 10^{-1})$ lead to significantly observable decay channels. Therefore, fine-tuning at $\mathcal{O}(10^{-3})$ or less does not affect the collider study. Nonetheless, such minute adjustments at $\mathcal{O}(\leq 10^{-3})$ are inevitable for us to be consistent with the neutrino oscillation and CLFV bounds. Below, we list the high-precision values of the Yukawa couplings in the three aforementioned benchmark points.

\textbf{BP1:}
\begin{equation*}
	\y_1^L=\begin{pmatrix}
		0.221039&\,\,0.000714&\,\,-0.031542 \\
		0.153678&\,\,-0.004908&\,\,0.136279 \\
		0.004975&\,\,0.119897&\,\,0.005766
	\end{pmatrix}; \quad 
	\y_1^R=\begin{pmatrix}
		1.000&\,\,0.000&\,\,0.000\\
		0.150&\,\,0.000&\,\,0.200 \\
		0.000&\,\,0.120&\,\,0.009
	\end{pmatrix}; 
\end{equation*}
\begin{equation}
	\label{eq:BP1_prec}
	\quad \y_2=10^{-3}\begin{pmatrix}
		0.000000000&\,\,-0.038621100&\,\,-0.000825144 \\
		0.280505000&\,\,0.000000000&\,\,0.0000224450 \\
		0.359512000&\,\,0.018091600&\,\,0.000000000
	\end{pmatrix}.
\end{equation}

\textbf{BP2:}
	\begin{equation*}
		\y_1^L=\begin{pmatrix}
			0.000119&\,\,0.374489&\,\,-0.009901\\
			-0.000818&\,\,0.599381&\,\,0.042168 \\
			0.019983&\,\,0.022305&\,\,0.041819
		\end{pmatrix}; \quad 
		\y_1^R=\begin{pmatrix}
			0.00&\,\,0.50&\,\,0.00\\
			0.00&\,\,0.50&\,\,0.05 \\
			0.02&\,\,0.00&\,\,0.03
		\end{pmatrix}; 
	\end{equation*}
	\begin{equation}
		\label{eq:BP2_prec}
		\quad \y_2=10^{-3}\begin{pmatrix}
			5.00000&\,\,-1.17252&\,\,-1.31907 \\
			-144.02600&\,\,5.00000&\,\,-2.11232 \\
			-108.86400&\,\,5.01788&\,\,0.00000
		\end{pmatrix}.
	\end{equation}

\textbf{BP3:}
	\begin{equation*}
		\y_1^L=\begin{pmatrix}
			-0.00675900&0.04901100&0.48713800\\
			0.02920300&0.00922200&0.11267200 \\
			0.00123556&0.05012900&0.00171300
		\end{pmatrix}; \quad 
		\y_1^R=\begin{pmatrix}
			0.0&0.5&0.5\\
			0.5&0.0&0.0 \\
			0.0&0.3&0.0
		\end{pmatrix}; 
	\end{equation*}
	\begin{equation}
		\label{eq:BP3_prec}
		\quad \y_2=10^{-3}\begin{pmatrix}
			0.0000000&876.2690000&-2.6172423 \\
			-1.9016000&200.0000000&-0.4845686 \\
			87.8210000&3.0095000&-2.6750000
		\end{pmatrix}.
	\end{equation}

With the values of the Yukawa couplings modified to higher orders of precision, we obtain the neutrino oscillation parameters as depicted in \autoref{tab:osc} for the three fine-tuned BPs, along with the experimental $3\sigma$ ranges. As we are working with real Yukawa couplings, the  CP-violating phase $\delta_{CP}$ is chosen to be $180^\circ$, for convenience, so that the PMNS mixing matrix also becomes real. The $3\sigma$ range of the absolute value for each component of the 3$\times$3 PMNS matrix, denoted as $U_{ij}$, are also presented in \autoref{tab:osc}.

\begin{table}[h!]
	\renewcommand{\arraystretch}{1}
	\centering
	\begin{tabular}{|c|c|c|c|c|}
		\hline
		Parameter& Expt. $3\sigma$ range  & BP1 &BP2& BP3\\
		\hline
		$\sin^2 \theta_{12}$ &$0.27-0.37$&0.27&0.30& 0.31\\
		$\sin^2 \theta_{13}$ &$0.020-0.024$&0.023& 0.023& 0.023\\
		$\sin^2 \theta_{23}$  &$0.43-0.61$& 0.50&0.59& 0.59\\
		$\delta_{CP}$&$120^\circ -369^\circ$ &$180^\circ$&$180^\circ$&$180^\circ$\\
		\hline\hline
		$U_{11}$&$0.801<|U_{11}|<0.845$&$-0.844$&$0.830$&$-0.823$\\ 
		$U_{12}$&$0.513<|U_{12}|<0.579$&$-0.514$&$0.537$&$-0.547$\\ 
		$U_{13}$&$0.143<|U_{13}|<0.155$&$0.152$&$0.150$&$0.153$\\ 
		$U_{21}$&$0.234<|U_{21}|<0.500$&$0.276$&$-0.252$&$0.256$\\ 
		$U_{22}$&$0.471<|U_{22}|<0.689$&$-0.661$&$0.601$&$-0.598$\\ 
		$U_{23}$&$0.637<|U_{23}|<0.776$&$-0.698$&$-0.758$&$-0.759$\\ 
		$U_{31}$&$0.271<|U_{31}|<0.525$&$-0.459$&$0.498$&$-0.507$\\ 
		$U_{32}$&$0.477<|U_{32}|<0.694$&$0.547$&$-0.591$&$0.586$\\ 
		$U_{33}$&$0.613<|U_{33}|<0.756$&$-0.700$&$0.634$&$-0.632$\\ \hline
	\end{tabular}
	\caption{Neutrino oscillation data (normal ordering) \cite{deSalas:2020pgw} and values of oscillation parameters under different benchmark points.}
	\label{tab:osc}
\end{table}

Finally, we consider the constraints from various CLFV processes, for which the experimental bounds, along with their values under three benchmark points of this model, are presented in \autoref{tab:LFV}. The tightest constraint comes from the $\mu \to e \gamma$ by MEG collaboration \cite{MEG:2016leq} providing the branching fraction to be smaller than $4.2\times 10^{-13}$. Experiment Sindrum II dealing with $\mu-e$ conversion by gold atom puts another stringent bound stating that the branching ratio of this process with respect to the nuclear capture probability should be less than $7\times 10^{-13}$ \cite{SINDRUMII:2006dvw}. The other bounds are not very strong relative to these two and are automatically satisfied. For our simulation we generated all these results through {\tt SPheno}. 

\begin{table}[h!]
	\renewcommand{\arraystretch}{1}
	\centering
	\begin{tabular}{|c|c|c|c|c|c|}
		\hline
		No&CLFV Mode&Expt. Branching&BP1&BP2&BP3\\
		\hline
		1&$\mu\to e\gamma$&$<4.2\times10^{-13}$&$9.19\times10^{-14}$&$3.36\times10^{-15}$&$5.09\times10^{-15}$\\
		\hline
		2&$\tau\to e\gamma$&$<3.3\times10^{-8}$&$3.04\times10^{-8}$&$3.18\times10^{-8}$&$6.36\times10^{-17}$\\
		\hline
		3&$\tau\to \mu\gamma$&$<4.4\times10^{-8}$&$3.71\times10^{-8}$&$4.10\times10^{-8}$&$4.53\times10^{-10}$\\
		\hline
		4&$\mu\to 3e$&$<1.0\times10^{-12}$&$6.46\times10^{-16}$&$8.57\times10^{-15}$&$3.93\times10^{-16}$\\
		\hline
		5&$\tau\to 3e$&$<2.7\times10^{-8}$&$4.08\times10^{-10}$&$3.83\times10^{-10}$&$2.47\times10^{-17}$\\
		\hline
		6&$\tau\to 3\mu$&$<2.1\times10^{-8}$&$9.56\times10^{-11}$&$1.49\times10^{-10}$&$8.55\times10^{-10}$\\
		\hline
		7&$\mu-e,\text{Ti}$&$<4.3\times10^{-12}$&$2.19\times10^{-13}$&$6.34\times10^{-13}$&$5.66\times10^{-13}$\\
		\hline
		8&$\mu-e,\text{Au}$&$<7\times10^{-13}$&$2.38\times10^{-13}$&$6.31\times10^{-13}$&$6.29\times10^{-13}$\\
		\hline
		9&$\mu-e,\text{Pb}$&$<4.6\times10^{-11}$&$2.30\times10^{-13}$&$6.04\times10^{-13}$&$6.08\times10^{-13}$\\
		\hline
		10&$\tau^-\to e^-\mu^+\mu^- $&$<2.7\times10^{-8}$&$8.78\times10^{-11}$&$7.53\times10^{-11}$&$7.02\times10^{-17}$\\
		\hline
		11&$\tau^-\to \mu^-e^+e^- $&$<1.8\times10^{-8}$&$4.39\times10^{-10}$&$5.04\times10^{-10}$&$5.30\times 10^{-10}$\\
		\hline
		12&$\tau^-\to e^+\mu^-\mu^- $&$<1.7\times10^{-8}$&$2.71\times10^{-30}$&$1.84\times10^{-25}$&$1.12\times 10^{-22}$\\
		\hline
		13&$\tau^-\to \mu^+ e^- e^-$&$<1.5\times10^{-8}$&$2.92\times10^{-27}$&$2.08\times10^{-22}$&$1.07\times 10^{-24}$\\
		\hline
		14&$Z\to e\mu$&$<7.5\times10^{-7}$&$7.81\times10^{-23}$&$4.91\times10^{-17}$&$5.23\times 10^{-19}$\\
		\hline
		15&$Z\to e\tau$&$<9.8\times10^{-6}$&$3.12\times10^{-13}$&$4.87\times10^{-14}$&$1.56\times10^{-18}$\\
		\hline
		16&$Z\to \mu\tau$&$<1.2\times10^{-5}$&$5.87\times10^{-14}$&$5.36\times10^{-13}$&$1.05\times 10^{-11}$\\
		\hline
		17&$h\to e\mu$&$<6.1\times10^{-5}$&$3.16\times10^{-18}$&$1.13\times10^{-19}$&$1.12\times10^{-20}$\\
		\hline
		18&$h\to e\tau$&$<4.7\times10^{-3}$&$3.11\times10^{-12}$&$1.75\times10^{-9}$&$6.70\times10^{-21}$\\
		\hline
		19&$h\to \mu\tau$&$<2.5\times10^{-3}$&$2.02\times10^{-9}$&$4.14\times10^{-12}$&$6.01\times10^{-16}$\\
		\hline
		
	\end{tabular}
	\caption{Bounds on various LFV processes and their {\tt SPheno} generated values for different benchmark points.}
	\label{tab:LFV}
\end{table}

\section{Collider phenomenology}
\label{sec:coll}
After setting up the model and the benchmark scenarios in the previous sections, we now perform a series of simulations to probe this model at the current 14 TeV LHC, as well as the future HL/HE-LHC and the FCC. In this section, we will be studying the pair production of the three leptoquark (LQ) eigenstates of $\rtt$ and $\X_{1,2}^{-1/3}$ from proton-proton collisions at three different centre-of-mass energies ($E_{CM}$) of 14, 27 and 100 TeV, respectively.  

\subsection{Simulation at the LHC/FCC: setup}
\label{subsec:setup}
For the simulations at the LHC/FCC, the hard scattering event files are generated in the \texttt{.lhe} format from \chep\cite{Belyaev:2012qa} version 3.8.7.\texttt{SPheno} version 4.0.4 is utilized to obtain spectrum files (\texttt{.spc} format) to be read by \chep as parameter cards. The parton shower, hadronization and jet clustering are done with \py\cite{Pythia8}. To form the jets, \fj \cite{fj} is used with Cambridge-Aachen jet clustering algorithm\cite{CMS:2009lxa} with a jet radius of 0.5. the same softwares and parameters are utilized to simulate the five dominant SM backgrounds (BG) at the LHC, namely $t\bar{t}$, $VV$, $VVV$, $t\bar{t}V$, and $tVV$ (where $V$ represents the electroweak vector bosons $W/Z$), so that we can compare the signal events with the background to obtain the significance of discovery. The next-to-leading order (NLO) $K$-factors for the backgrounds are calculated with \texttt{MadGraph5\_aMC@NLO} version 3.1.0 \cite{mg5}. Additionally, the following cuts are imposed:
\begin{itemize}
	\item Calorimeter coverage, $\abs{\eta}<$ 4.5.
	\item Transverse momentum cut for jets and leptons, $p_{T,\text{min}}^{\text{jet,lep}} = 20.0$ GeV.
	\item Leptons are hadronically cleaned, with minimized hadronic activity within a cone of radius $\Delta R_{hl} = 0.3$ with the relation $\sum p_T^{\text{had}} = 0.15\, p_T^{\text{lep}}$.
	\item Leptons are isolated from jets, with a cone radius cut of $\Delta R_{lj}>0.4$.
	\item Pertaining to the high mass of the leptoquarks i.e. 1.5 TeV in the benchmark points, a cut on the total hardness of the event, defined as $p_T^H  = \sum(p_T^{\text{lep}} + p_T^{\text{jet}} + \ptmiss) \geq 1.2$ TeV is applied to both the signal and background events, at the analysis level. Additionally, for convergence of the events at the high-momentum tail, the SM background events are generated with a phase space cut of $\sqrt{\hat{s}} \geq 1.2$ TeV. 
	\item For further filtering of the backgrounds, the following dijet and dilepton invariant mass vetoes are imposed: $\abs{M_{jj}-M_Z} \geq 10\, \text{GeV},\, \abs{M_{jj}-M_W} \geq 10 \,\text{GeV},$ and $\abs{M_{\ell\ell}-M_Z} \geq 10 \,\text{GeV}$. 
\end{itemize}

Moreover, for ease of analysis and clarity of signal, we perform flavour-tagging of heavy jets. For $b$-tagging, we take the efficiency to be $\sim70\%$ following a secondary vertex reconstruction algorithm \cite{CMS:2017wtu}. For $c$-tagging we consider a conservative efficiency of $\sim56\%$ with a misidentification rate of $\sim0.12$ \cite{Bols:2020bkb}. Again, while tagging the $\tau$-jets using one- or three-prong $\pi^\pm$ tracks, we take the hadronic $\tau$-jet identification efficiency of $\sim 75\%$ for a misidentification rate of $\sim 10^{-2}$, as reported in \cite{CMS:2018jrd}. With this setup, we are now ready to move ahead with our simulations of leptoquark pair production.

\subsection{Leptoquark pair production}
\label{subsec:pairprod}

\begin{figure}[h!]
	\centering
	%		\tikzfeynmanset{every scalar@@/.style={thick, dashed}, every boson@@/.style={thick, decoration={snake,amplitude=1mm},decorate}, every plain@@/.style={thick}}
	\begin{tikzpicture}
		\begin{feynman}
			\vertex (a1);
			\vertex [above left =1.5cm of a1] (i1){\footnotesize\(q\)};
			\vertex [below left =1.5cm of a1] (i2){\footnotesize\(\xbar{q}\)};
			\vertex [right =1.2cm of a1] (a2);
			\vertex [above right =1.5cm of a2] (i3){\footnotesize\(LQ\)};
			\vertex [below right =1.5cm of a2] (i4){\footnotesize\(\xbar{LQ}\)};
			%			\vertex [above right =1cm of i3] (h1){\footnotesize\(T^0\)};
			%			\vertex [right =1cm of i3] (h2);
			%			\vertex [below =1cm of i4] (h3){\footnotesize\(\textcolor{white}{d}\)}; %dummy vertex%
			%			\vertex [above right =0.5cm of h2] (w1){\footnotesize\(l\)};
			%			\vertex [below right =0.5cm of h2] (w2){\footnotesize\(\bar{\nu}\)};
			%			\vertex [below =1cm of a1] (t){\footnotesize\(\textbf{ITM}\)};
			
			\diagram { (i1)--[fermion](a1)--[gluon](a2), (i2)--[anti fermion](a1), (a2)--[scalar](i3), (a2)--[scalar](i4)
			};
		\end{feynman}
	\end{tikzpicture}
	\hfill
	\begin{tikzpicture}
		\begin{feynman}
			\vertex (a1);
			\vertex [above left =1.5cm of a1] (i1);
			\vertex [below left =1.5cm of a1] (i2);
			\vertex [right =1.2cm of a1] (a2);
			\vertex [above right =1.5cm of a2] (i3){\footnotesize\(LQ\)};
			\vertex [below right =1.5cm of a2] (i4){\footnotesize\(\xbar{LQ}\)};
			%			\vertex [above right =1cm of i3] (h1){\footnotesize\(T^0\)};
			%			\vertex [right =1cm of i3] (h2);
			%			\vertex [below =1cm of i4] (h3){\footnotesize\(\textcolor{white}{d}\)}; %dummy vertex%
			%			\vertex [above right =0.5cm of h2] (w1){\footnotesize\(l\)};
			%			\vertex [below right =0.5cm of h2] (w2){\footnotesize\(\bar{\nu}\)};
			%			\vertex [below =1cm of a1] (t){\footnotesize\(\textbf{ITM}\)};
			
			\diagram { (i1)--[gluon](a1)--[gluon](a2), (i2)--[gluon](a1), (a2)--[scalar](i3), (a2)--[scalar](i4)
			};
		\end{feynman}
	\end{tikzpicture}
	\hfill
	\begin{tikzpicture}
		\begin{feynman}
			\vertex (i1);
			\vertex [right =1.75cm of i1] (a1);
			\vertex [below =1.5cm of a1] (a2);
			\vertex [left =1.75cm of a2] (i2);
			\vertex [right =1.5cm of a1] (i3){\footnotesize\(LQ\)};
			\vertex [right =1.5cm of a2] (i4){\footnotesize\(\xbar{LQ}\)};
			%			\vertex [above right =1cm of i3] (h1){\footnotesize\(T^0\)};
			%			\vertex [right =1cm of i3] (h2);
			\vertex [below =0.5cm of i4] (h3){\footnotesize\(\textcolor{white}{d}\)}; %dummy vertex%
			%			\vertex [above right =0.5cm of h2] (w1){\footnotesize\(l\)};
			%			\vertex [below right =0.5cm of h2] (w2){\footnotesize\(\bar{\nu}\)};
			%			\vertex [below =1cm of a1] (t){\footnotesize\(\textbf{ITM}\)};
			
			\diagram { (i1)--[gluon](a1)--[scalar, edge label ={\footnotesize\(LQ\)}](a2), (i2)--[gluon](a2), (a1)--[scalar](i3), (a2)--[scalar](i4)
			};
		\end{feynman}
		
	\end{tikzpicture}
	\hfill
	\begin{tikzpicture}
		\begin{feynman}
			\vertex (a1);
			\vertex [above left =1.5cm of a1] (i1);
			\vertex [below left =1.5cm of a1] (i2);
			%		\vertex [right =1.2cm of a1] (a2);
			\vertex [above right =1.5cm of a1] (i3){\footnotesize\(LQ\)};
			\vertex [below right =1.5cm of a1] (i4){\footnotesize\(\xbar{LQ}\)};
			%			\vertex [above right =1cm of i3] (h1){\footnotesize\(T^0\)};
			%			\vertex [right =1cm of i3] (h2);
			%			\vertex [below =1cm of i4] (h3){\footnotesize\(\textcolor{white}{d}\)}; %dummy vertex%
			%			\vertex [above right =0.5cm of h2] (w1){\footnotesize\(l\)};
			%			\vertex [below right =0.5cm of h2] (w2){\footnotesize\(\bar{\nu}\)};
			%			\vertex [below =1cm of a1] (t){\footnotesize\(\textbf{ITM}\)};
			
			\diagram { (i1)--[gluon](a1), (i2)--[gluon](a1), (a1)--[scalar](i3), (a1)--[scalar](i4)
			};
		\end{feynman}
	\end{tikzpicture}
	%	\hfill
	\hspace{2cm}
	\begin{tikzpicture}
		\begin{feynman}
			\vertex (i1){\footnotesize\(q\)};
			\vertex [right =1.75cm of i1] (a1);
			\vertex [below =1.5cm of a1] (a2);
			\vertex [left =1.5cm of a2] (i2){\footnotesize\(\xbar{q}\)};
			\vertex [right =1.5cm of a1] (i3){\footnotesize\(LQ\)};
			\vertex [right =1.5cm of a2] (i4){\footnotesize\(\xbar{LQ}\)};
			%			\vertex [above right =1cm of i3] (h1){\footnotesize\(T^0\)};
			%			\vertex [right =1cm of i3] (h2);
			\vertex [below =0.5cm of i4] (h3){\footnotesize\(\textcolor{white}{d}\)}; %dummy vertex%
			%			\vertex [above right =0.5cm of h2] (w1){\footnotesize\(l\)};
			%			\vertex [below right =0.5cm of h2] (w2){\footnotesize\(\bar{\nu}\)};
			%			\vertex [below =1cm of a1] (t){\footnotesize\(\textbf{ITM}\)};
			
			\diagram { (i1)--[fermion](a1)--[fermion, edge label ={\footnotesize\(\ell/\nu\)}](a2), (i2)--[anti fermion](a2), (a1)--[scalar](i3), (a2)--[scalar](i4)
			};
		\end{feynman}
		
	\end{tikzpicture}

\caption{Feynman diagrams for pair production of leptoquarks from $pp$ collisions at the LHC.}
\label{fig:feynpairprod}
\end{figure}
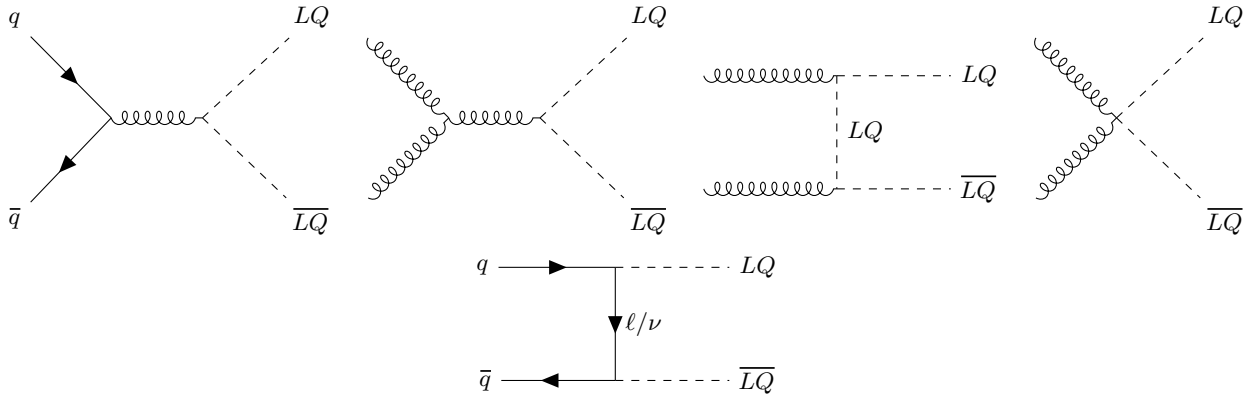

For the pair production of leptoquarks (LQ) in $pp$ collisions, we are considering the QCD processes, along with the contribution from $t$-channel lepton exchange diagrams as shown in \autoref{fig:feynpairprod}. The QCD dominated leading-order partonic cross-section for the pair production of scalar leptoquarks through gluon and quark fusion channels can be expressed as \cite{Blumlein:1996qp,Bandyopadhyay:2020wfv}:
\begin{align}
\hat\sigma_{gg}=\frac{\pi\alpha_s^2}{96\hat s}\bigg[\beta\big(41-31\beta^2\big)-\big(17-18\beta^2+\beta^4\big)\log\Big|\frac{1+\beta}{1-\beta}\Big|\bigg]\quad \text{and}\quad
\hat\sigma_{q\bar q}=\frac{2\pi\alpha_s^2}{27\hat s} \beta^3,
\end{align}
where, $\beta=\sqrt{1-4 M_{LQ}^2/\hat s}$ with $M_{LQ}$ being the mass of the leptoquark, $\hat s$ the partonic centre-of-mass energy and $\alpha_s$ the strong coupling constant. As shown in Ref. \cite{Bandyopadhyay:2020wfv}, the effect of the $t$-channel lepton exchange diagram becomes significant only when the scattering angle is very small. However, to obtain the total cross-section for the pair production of leptoquarks at hadronic collider, one has to wrap each of the partonic cross-section with parton distribution function (PDF) and sum over all the different contributions.

The leptoquark model is implemented in \sarah\cite{Sarah}, and model files for \chep\cite{Belyaev:2012qa} are generated from there. Reading the parameter cards from \texttt{SPheno}, \chep evaluates the pair production cross-sections at leading order (LO) QCD, taking the \texttt{cteq6l1} parton distribution function (PDF) \cite{Pumplin:2002vw}. However, we wish to perform our pair production analysis at the next-to-leading order (NLO) QCD. For that, we first neglect the lepton exchange diagrams to match the cross-sections with the published results from \cite{Kramer:2004df,Mandal:2015lca} and make a judicial choice of NLO QCD $K$-factor = 1.84 for $M_\text{LQ} = 1.5$ TeV. Next we include the $t$-channel lepton exchange diagrams along with the QCD processes, where the BPs with $\mathcal{O}(1)$ Yukawa couplings can have some significant effect on the total cross-section. Referring to the results published in \cite{Borschensky:2020hot, Borschensky:2021hbo}, it is observed that, for 1.5 TeV leptoquark mass, the ratio of cross-sections at NLO QCD with both QCD-mediated + $t$-channel lepton processes, compared to only QCD processes without $t$-channel, come out to be about 1.1. Therefore, our effective $K$-factor now becomes $1.84 \times 1.1 \approx 2.02$. Moreover, the cross-sections are evaluated at the renormalization scale $\mu_R$ = $M_\text{LQ}=1.5$ TeV, at which the value of strong coupling $\alpha_S$ for \texttt{cteq6l1} PDF is 0.0899. 

\begin{table}[h!]
	\centering
	\renewcommand{\arraystretch}{1}
	\begin{tabular}{|M{0.8cm}|M{1.2cm}|M{1.2cm}|M{1.4cm}|M{1.2cm}|M{1.2cm}|M{1.4cm}|M{1.2cm}|M{1.2cm}|M{1.4cm}|}
		\hline
		\multirow{3}{*}{BPs}&\multicolumn{9}{c|}{$pp\to LQ \overline{LQ}$ at different $E_{CM}$s}\\
		\cline{2-10}
		&\multicolumn{3}{c|}{$\sigma_{\widetilde{R}_2^{2/3} \widetilde{R}_2^{-2/3}}$ (fb)}&\multicolumn{3}{c|}{$\sigma_{\X_1^{1/3} \X_1^{-1/3}}$ (fb)}&\multicolumn{3}{c|}{$\sigma_{\X_2^{1/3} \X_2^{-1/3}}$ (fb)}\\
		\cline{2-10}
		&14 TeV & 27 TeV & 100 TeV 	&14 TeV & 27 TeV & 100 TeV 	&14 TeV & 27 TeV & 100 TeV \\
		\hline
		BP1 &0.39&10.91&779.3&0.61&15.21&890.8&0.35&10.53&772.8\\
		\hline
		BP2 &0.39&10.91&779.6&0.32&10.13&778.5&0.32&9.90&755.5\\
		\hline
		BP3 &0.54&14.70&886.5&0.30&9.80&770.0&0.33&10.48&780.9\\
		\hline
	\end{tabular}
	
	\caption{Production cross-sections of leptoquark pairs from $pp$ collisions at the LHC at $E_{CM} = 14, 27$ and 100 TeV, with \texttt{cteq6l1} as PDF, and the NLO $K$-factor = 2.02.}
	\label{tab:pairprod}
\end{table}

With this setup of parameters, we evaluate the cross-sections at three different centre-of-mass energies ($E_{CM}$) of 14, 27 and 100 TeV, corresponding to present and future iterations of the LHC/FCC. These cross-sections for the three benchmark points are presented in \autoref{tab:pairprod}. From this table we note that, in case of BP1, the $\X_1$ pair production cross-section is large due to $(\yr)_{11} = 1.0$, while in BP3, $\widetilde{R}_2$ pair production is enhanced by the dominance of $(\yt)_{12} = 0.876$. In BP2, the slight difference in cross-sections of $\X_{1,2}$ pair production is resulting from the small mass gap of $M_{\X_1} - M_{\X_2} \approx 7$ GeV. The effect of this mass gap is overcome by the $(\yt)_{12} = 0.876$ in BP3, which enhances the cross-section for $\xt$ slightly higher than $\xo$, despite being heavier.

\begin{table}[h]
	\centering
	\renewcommand{\arraystretch}{1}
	\begin{adjustwidth}{-0.7cm}{}
		%		\hspace{-1cm}
		\begin{tabular}{|M{1cm}|M{4cm}|M{0.8cm}|M{4cm}|M{0.8cm}|M{4cm}|M{0.8cm}|}
			\hline
			\multirow{3}{*}{BPs}&\multicolumn{6}{c|}{Observable finalstates for different leptoquark pairs with effective branching ratios}\\
			\cline{2-7}
			&\multicolumn{2}{c|}{$\widetilde{R}_2^{2/3} \widetilde{R}_2^{-2/3}$}&\multicolumn{2}{c|}{$\X_1^{1/3} \X_1^{-1/3}$}&\multicolumn{2}{c|}{$\X_2^{1/3} \X_2^{-1/3}$}\\
			\cline{2-7}
			&finalstate & BR$_\text{eff}$&finalstate & BR$_\text{eff}$&finalstate & BR$_\text{eff}$\\
			\hline
			\multirow{3}{*}{BP1}&$2\,b\text{-jets} + 2\,\text{OS}e$ &0.38&$2\,j + 2\,\text{OS}e$ &0.68&$2\,j + 2\,\text{OS}e$ &0.68\\
			&$1\,b\text{-jet} + 1j + 2\,\text{OS}e$& 0.46&&&&\\
			&$2j + 2\,\text{OS}e$ &0.14&&&&\\
			\hline
			\multirow{6}{*}{BP2}&$2j + 2\,\text{OS}e$& 0.40&$2j + 2\,\text{OS}\mu$ &0.11&$2j + 2\,\text{OS}\mu$& 0.10\\
			&$1\,b\text{-jet} + 1j + 2\,\text{OS}e$& 0.46&$1\,c\text{-jet} + 1j + 2\,\text{OS}\mu$ &0.21&$1\,c\text{-jet} + 1j + 2\,\text{OS}\mu$& 0.20\\
			&$2\,b\text{-jets} + 2\,\text{OS}e$ &0.13&$2\,c\text{-jets}+ 2\,\text{OS}\mu$ &0.11&$2\,c\text{-jets} + 2\,\text{OS}\mu$ &0.10\\
			&&&$2j + \ptmiss$ &0.11&$2j + \ptmiss$&0.10\\
			&&&$1\,c\text{-jet} + 1j + 1\mu + \ptmiss$& 0.21&$1\,c\text{-jet} + 1j + 1\mu + \ptmiss$ &0.20\\
			&&&$2j + 1\mu+\ptmiss$ &0.21&$2j + 1\mu+\ptmiss$& 0.20\\
			\hline
			\multirow{9}{*}{BP3}&$2j + 2\,\text{OS}\mu$& 1.0&$2j + 2\,\tau\text{-jets}$& 0.08&$2j + 2\,\tau\text{-jets}$ &0.03\\
			&&&$2j + 1\,\tau\text{-jet}+\ptmiss$ &0.20&$2j + 1\,\tau\text{-jet}+\ptmiss$ &0.20\\
			&&&$2j + \ptmiss$ &0.13&$2j + \ptmiss$&0.35\\
			&&&$2j + 1\,\tau\text{-jet}+1\mu/e$ &0.16&$2j + 1\,\tau\text{-jet}+1\mu/e$ &0.06\\
			&&&$2j + 2\,\text{OS}\mu e$ &0.04&$2j + 2\,\text{OS}\mu e$ &0.014\\
			&&&$2j + 2\,\text{OS}\mu$ &0.02&$2j + 2\,\text{OS}\mu$ &0.007\\
			&&&$2\,c\text{-jets}+ 2\,\text{OS}e$ &0.02&$2\,c\text{-jets}+ 2\,\text{OS}e$ &0.007\\
			&&&$1j + 1\,c\text{-jet}+1\,\tau\text{-jet}+1e$ &0.08&$1j + 1\,c\text{-jet}+1\,\tau\text{-jet}+1e$ &0.03\\
			&&&$2j+1e/\mu+\ptmiss$ &0.20&$2j+1e/\mu+\ptmiss$ &0.20\\
			\hline
		\end{tabular}
	\end{adjustwidth}
	
	\caption{List of possible finalstates for pair production of leptoquarks at the LHC. Here, light jets are represented by $j$, OS refers to ``opposite signed".}
	\label{tab:pairprodbr}
\end{table}

In \autoref{tab:pairprodbr}, we illustrate the effective branching fractions of different observable finalstates at the LHC/FCC, in accordance with the Yukawa couplings and decay channels discussed in \autoref{sec:bps}. Here, $\ptmiss$ refers to the missing transverse momentum that is carried by particles such as neutrinos which are not observable at the LHC, and OS means ``oppositely signed". The light jets i.e. non-flavour-tagged jets are represented as $j$. The finalstates and probabilities for $\rtt$ pair production look very different to those of $\xot$, due to the difference in charge, as well as for $\rtt$ being purely doublet. In BP1 and BP2, the mixed states $\xot$ have almost the same probabilities for their finalstates, which are governed mostly by $\y_1^{L,R}$. In case of BP3, the effect of $\y_2$ comes into play, leading to different probabilities for the same finalstates. We will be extensively using the \autoref{tab:pairprodbr} in the collider analysis that follows, in order to decide upon which finalstates to look for. The first objective for us is to probe the model at the LHC/FCC, and obtain a model signature with $\geq5\sigma$ significance from the pair production events of all the three leptoquark mass eigenstates. This study is discussed in detail in the following subsections.

\subsection{Model signatures}
\label{sec:modsig}

To explore the feasibility of probing the model at the LHC/FCC, we first identify finalstates that are observable across all the three benchmark points. These finalstates also need to be direct consequences of the chosen Yukawa couplings, rather than emerging only due to initial- or final-state radiation (ISR, FSR) effects. To observe a significant excess of events over the SM backgrounds, we study the differential distributions of the kinematic variables pertaining to the emergent leptons and jets, so that some advanced cuts can be applied in our analysis.

\subsubsection{Kinematics and finalstate topologies}
\begin{figure}[h]
	\makebox[\linewidth][c]{%
		\centering
		\subfigure[]{\includegraphics[width=0.33\linewidth]{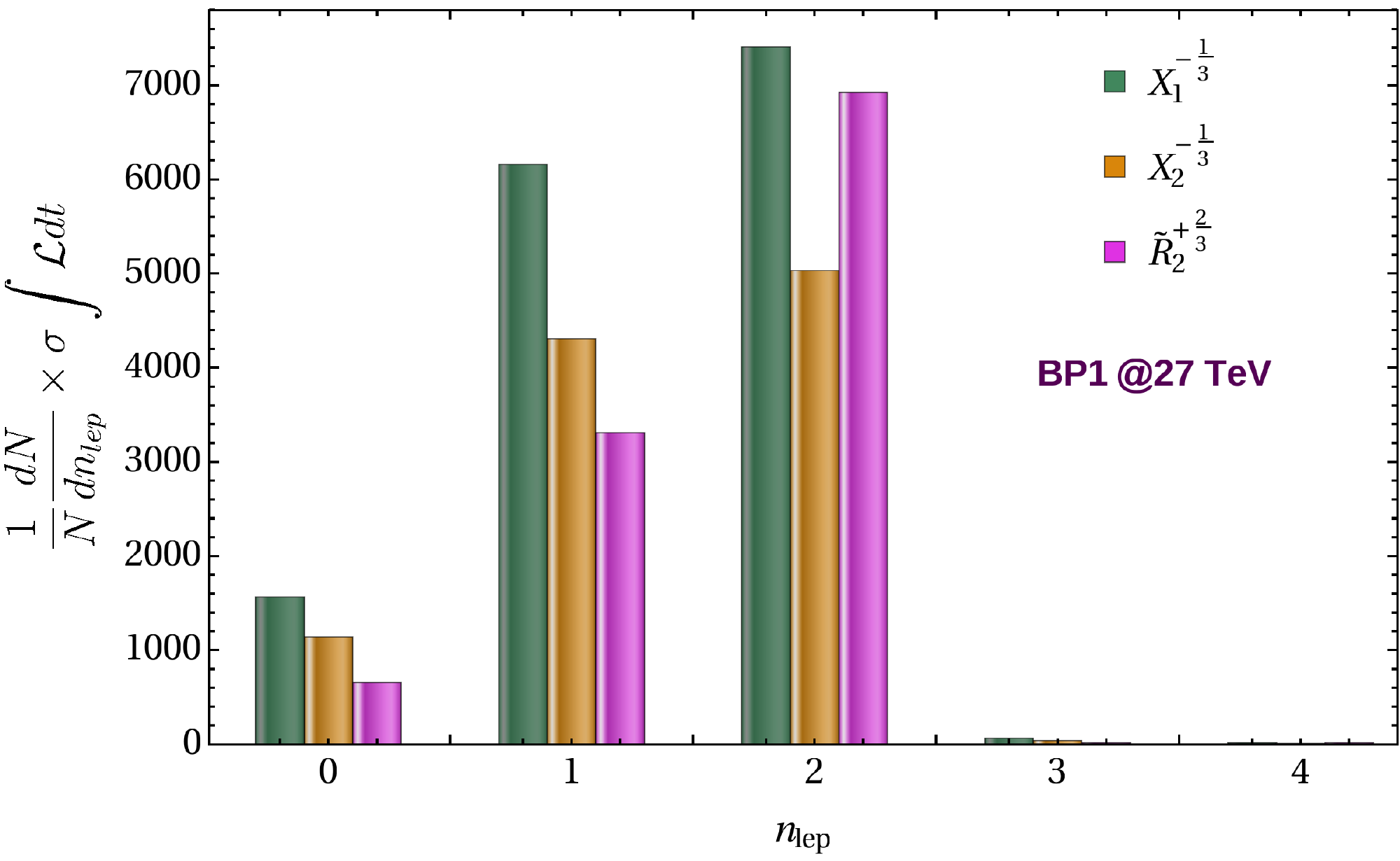}\label{lm1}}
		\subfigure[]{\includegraphics[width=0.33\linewidth]{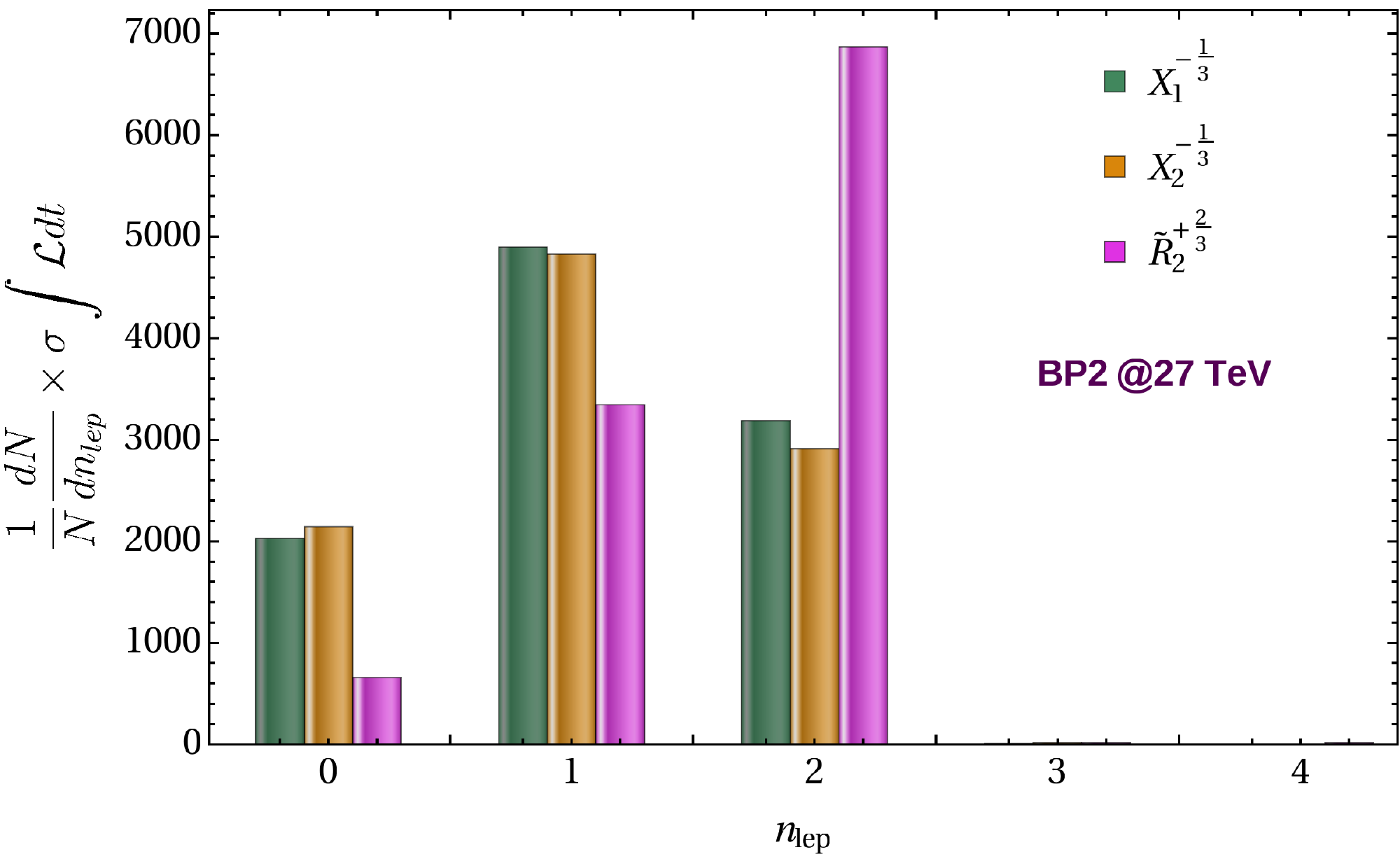}\label{lm2}}
		\subfigure[]{\includegraphics[width=0.33\linewidth]{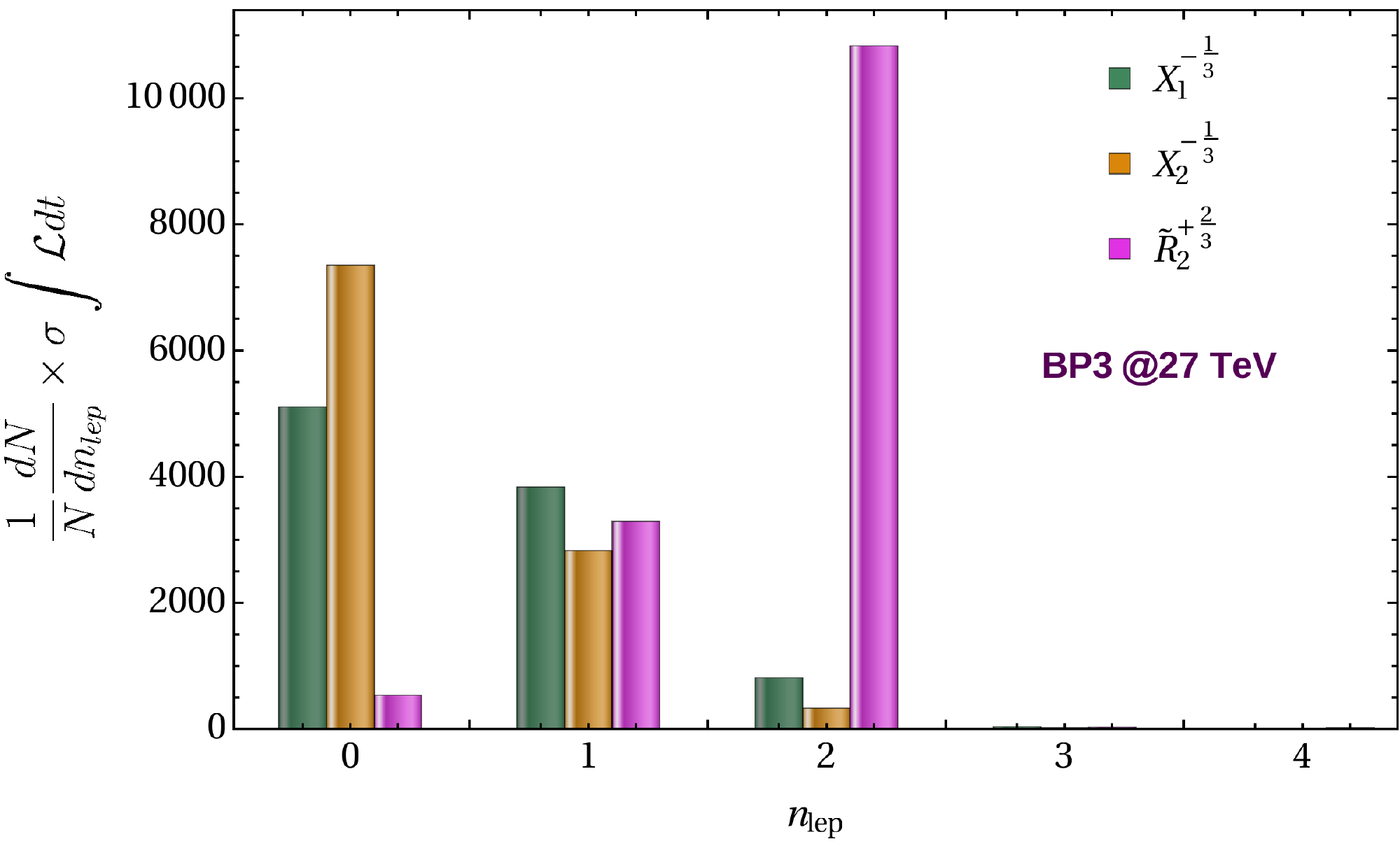}\label{lm3}}
	}
\caption{Distribution of lepton multiplicity coming from pair production of $\rtt$(pink) and $X_{1,2}^{-1/3}$ (green, yellow) for (a) BP1, (b) BP2 and (c) BP3, at the 27 TeV LHC. }
\label{fig:lepmul}
\end{figure}

In \autoref{fig:lepmul} the multiplicity distribution of isolated charged leptons emerging from the leptoquark pair production processes are presented, with panels (a)-(c) corresponding to BP1-BP3 respectively. The event numbers have been calculated with an integrated luminosity of 1000 fb$^{-1}$. Although the distributions are plotted for 27 TeV centre-of-mass energy owing to larger number of events than for 14 TeV, the pattern of distributions do not alter much at 14 TeV or 100 TeV energies. In each case, the green, yellow and pink bars correspond to $X_1^{-1/3}$, $X_2^{-1/3}$, and $\rtt$ pair productions respectively. For each BP, the distributions for $\rtt$ preserve the same shape and they peak at $n_{\text{lep}}$ = 2, which is obvious from the decay of the pair produced leptoquarks and also predicted by the effective branching fractions noted in \autoref{tab:pairprodbr}. Depending on the dominant branching fractions of $\xot$, the lepton multiplicity distribution varies for the three BPs. However, in all three cases this multiplicity becomes negligible for $n_{\text{lep}} > 2$, as there is no other possible source of leptons apart from the leptoquarks in the hard process. In BP1, the majority of events carry two leptons for both $\xot$, due to a 68\% probability of obtaining a $2j + \text{OS}e$ finalstate. In BP2 the multiplicity peaks at one lepton for both of these leptoquarks, with the numbers remaining almost the same for each. However, in BP3, even though pair production of both $\xo$ and $\xt$ leads to a majority of zero-lepton events, the ratios are not uniform unlike the other two BPs. The zero-lepton events are accounted for by the large branching ratios of $\xot$ into neutrinos and $\tau$-jets. Such events are more dominant for $\xt$, as a result of the greater probabilities of decaying into neutrinos and $\tau$-jets compared to $\xo$, shown in \autoref{tab:pairprodbr}.

\begin{figure}[h]
	\makebox[\linewidth][c]{%
		\centering
		\subfigure[]{\includegraphics[width=0.33\linewidth]{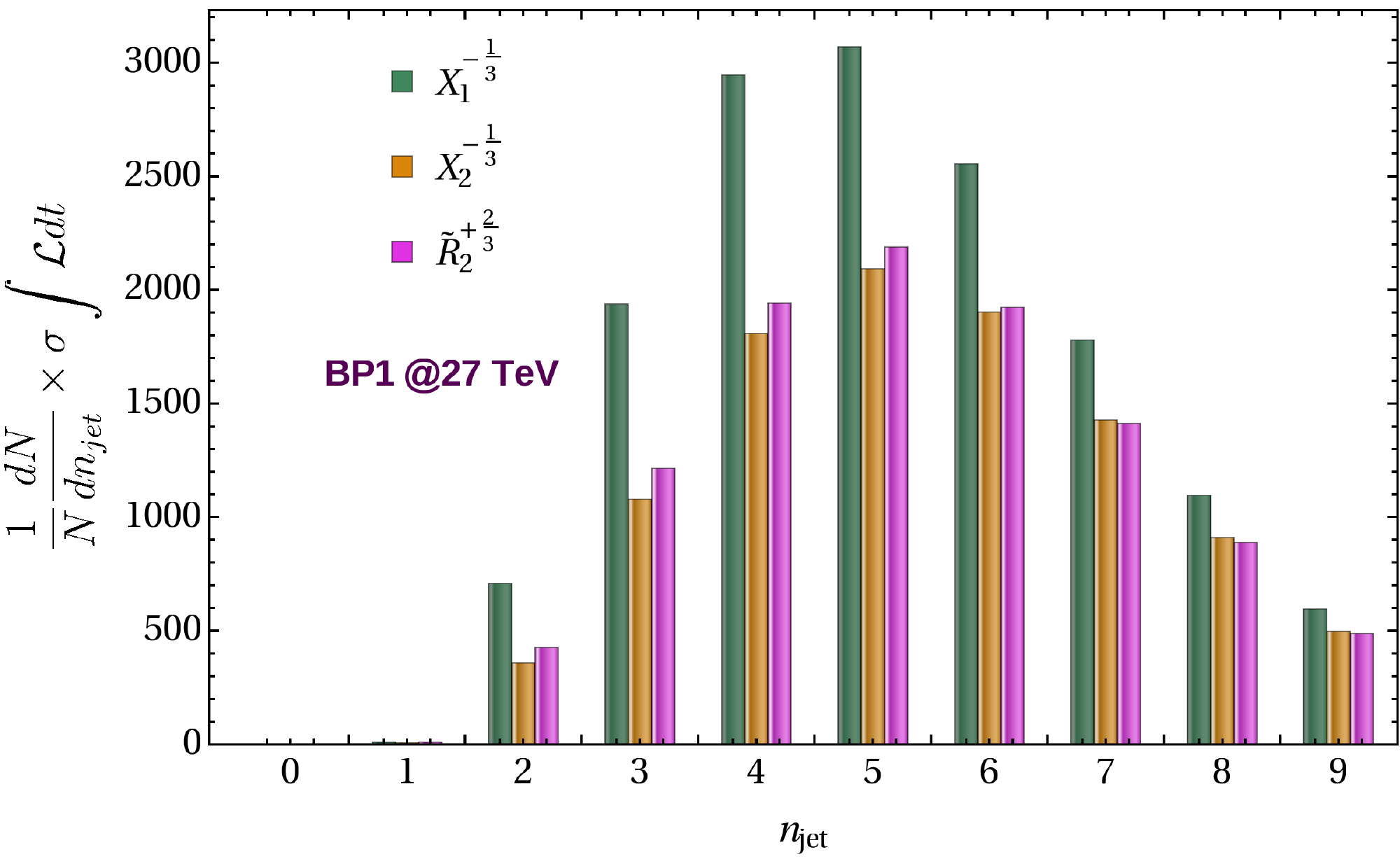}\label{jm1}}
		\subfigure[]{\includegraphics[width=0.33\linewidth]{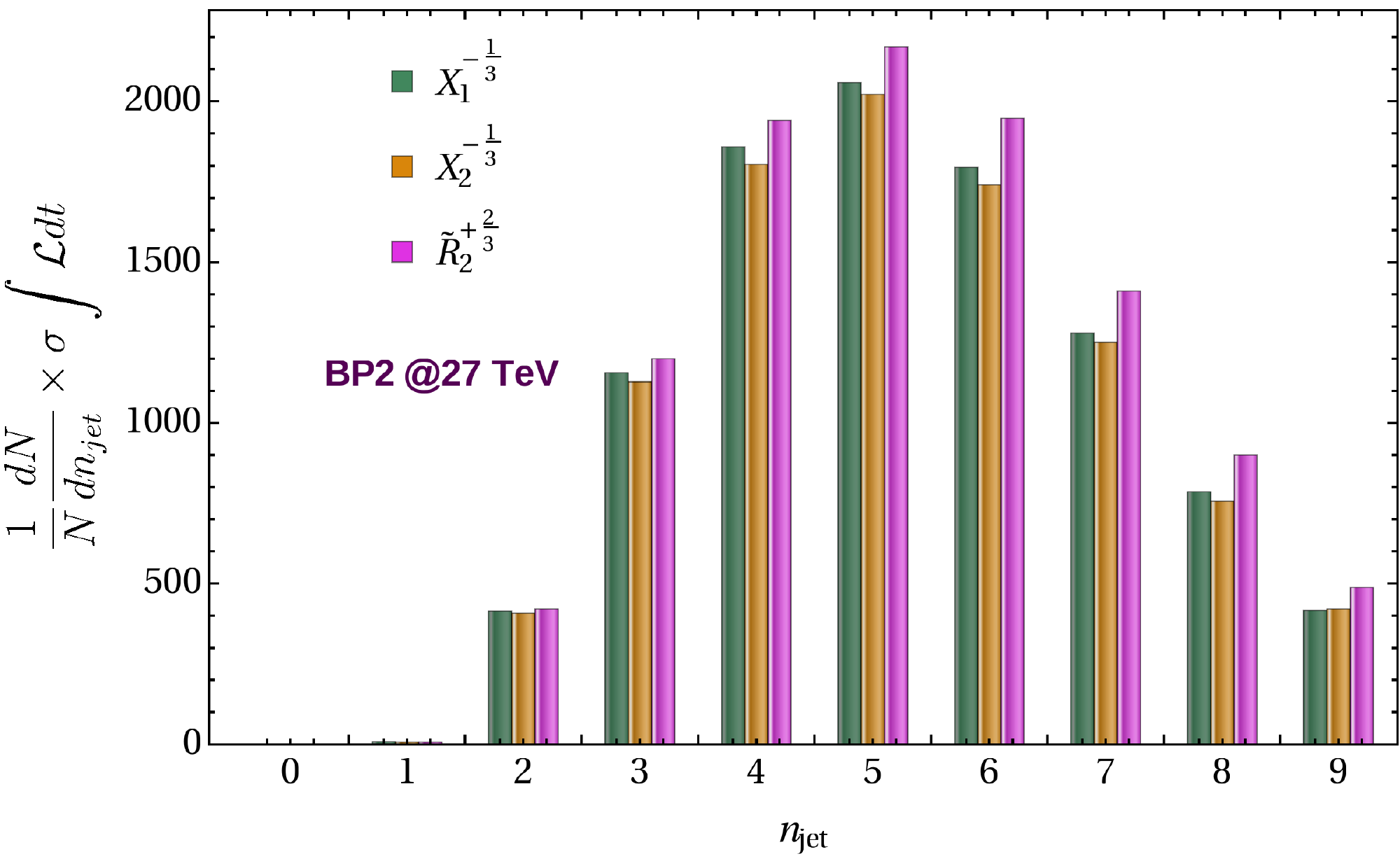}\label{jm2}}
		\subfigure[]{\includegraphics[width=0.33\linewidth]{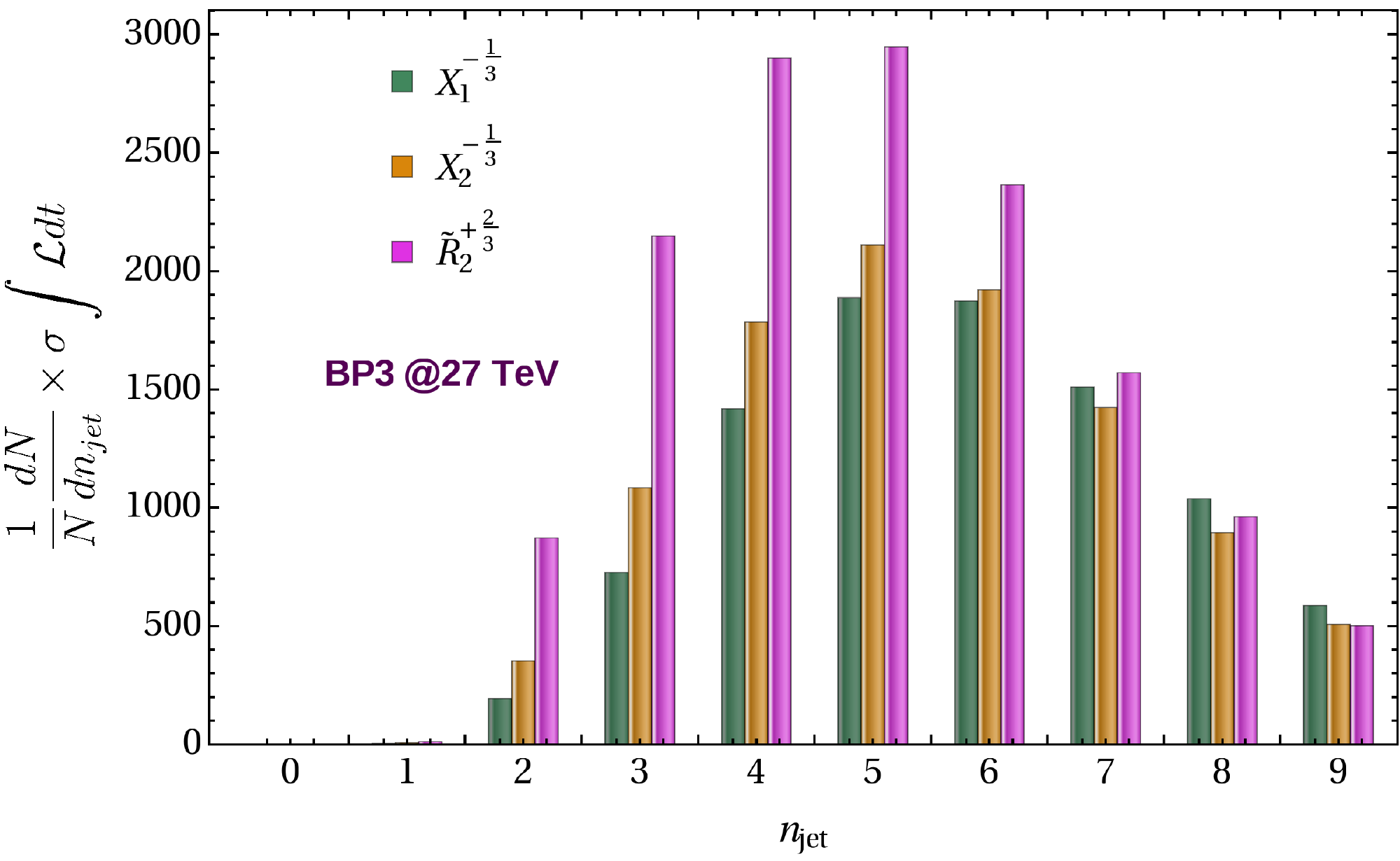}\label{jm3}}
	}
	\caption{Distribution of jet multiplicity coming from pair production of $\rtt$(pink) and $X_{1,2}^{-1/3}$ (green, yellow) for (a) BP1, (b) BP2 and (c) BP3, at the 27 TeV LHC. }
	\label{fig:jetmul}
\end{figure}

\autoref{fig:jetmul} depicts a similar comparative distribution of jet multiplicities for pair productions of $\xo$ (green), $\xt$ (yellow), and $\rtt$ (pink) leptoquarks, at $E_{CM}$ = 27 TeV. Here, irrespective of the benchmark point, we see very similar trends for all three leptoquarks. Although two hadronic jets are always expected from decays of a leptoquark and anti-leptoquark pair, the multiplicities actually peak at $n_{\text{jet}}$=5, owing to the ISR and FSR jets. 

\begin{figure}[h]
%	\makebox[\linewidth][c]{%
		\centering
		\includegraphics[width=0.5\linewidth]{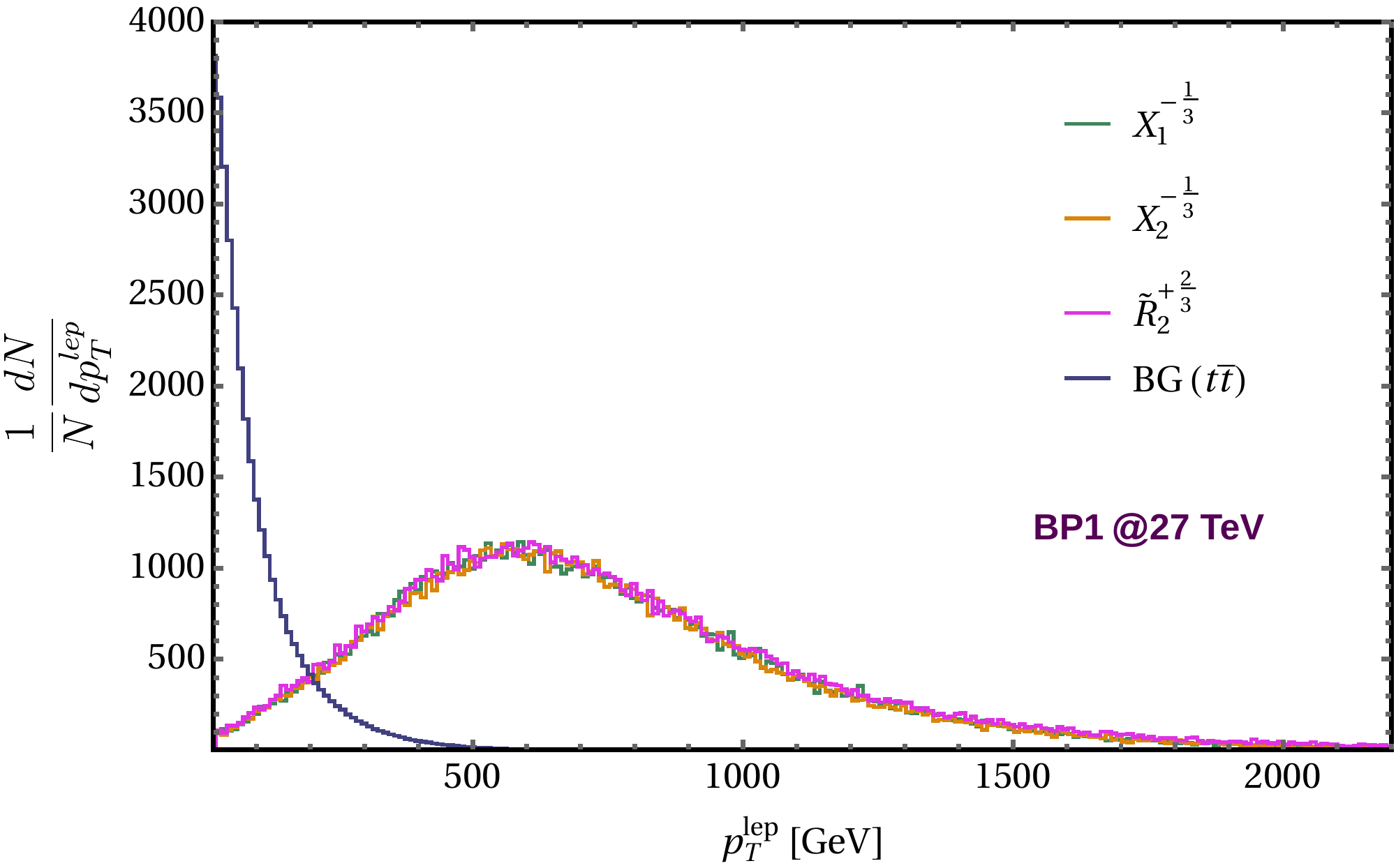}
%		\subfigure[]{\includegraphics[width=0.5\linewidth]{./plots/bp1_jpt_2.pdf}\label{jpt1}}

	\caption{The $p_{T}$ distribution of the hardest lepton from the leptoquark pair productions, compared with that from the $t\bar{t}$ SM background.}
	\label{fig:lepjetpt}
\end{figure}

In \autoref{fig:lepjetpt}, we portray the transverse momentum distributions of the hardest lepton emanating from each of the pair-produced leptoquarks $\xo$ (green), $\xt$ (yellow), and $\rtt$ (pink), at $E_{CM}$=27 TeV for BP1 as well as from the SM background of $t\bar{t}$ (blue). Due to their nearly degenerate masses of $\approx 1.5$ TeV, the $p_T$ distributions for the leptons follow the same pattern for each leptoquark. Wide peaks are observed at around half the mass of the leptoquarks, signifying that the hardest leptons and jets are indeed arising from the decay of $\rtt$, $\xo$ or $\xt$. In case of $t\bar{t}$, the lepton $p_T$ peaks at roughly half of the $W^\pm$ boson mass, because the primary source of leptons in this case are the $W^\pm$ bosons resulting from the $t\to bW^\pm$ decays.

 Now we look for the suitable finalstate topologies that can act as signatures for this doublet-singlet leptoquark extension of the SM. From \autoref{tab:pairprodbr} we notice that, in each benchmark scenario, a finalstate of two light jets along with a pair of oppositely signed dileptons (OSD) is common between all three leptoquark mass eigenstates. However, for $\rtt$ in BP1 and BP2, a finalstate with two $b$-tagged jets and OSD is also highly probable. Similar observation is made for $\xo$ and $\xt$ with a  topology of two $c$-jets along with OSD. Therefore, after carefully observing the expected finalstates and probabilities from \autoref{tab:pairprodbr}, we choose the two following topologies for our model signature:
\begin{equation}
	\text{\textbf{FS1:}}\quad n_{\text{jet}} \geq 2 \,\, \& \,\, n_{\ell^+} = n_{\ell^-} = 1 \,\, \& \,\, n_{b/\tau\text{-jets}} = 0 \,\, \& \,\, p_T^{\ell^+ , \ell^-} \geq 300.0 \, \text{GeV},
	\label{eq:fs1}
\end{equation}
and
\begin{equation}
	\text{\textbf{FS2:}}\quad n_{\text{jet}} \geq 2 \,\, \& \,\, n_{\ell^+} = n_{\ell^-} = 1 \,\, \& \,\, n_{c/\tau\text{-jets}} = 0 \,\, \& \,\, p_T^{\ell^+ , \ell^-} \geq 300.0 \, \text{GeV}.
	\label{eq:fs2}
\end{equation}
Here,  to obtain a cleaner signal by eliminating some background events, a 300 GeV cut on both the oppositely signed leptons' $p_T$ has been implemented which is motivated by \autoref{fig:lepjetpt}. The reason why we demand the number of jets $\geq 2$ rather than exactly equal to 2 is for the fact that, as ISR and FSR increase the jet multiplicity, we do not wish to lose events by just looking at exactly two jets in the finalstate. In the next subsections, we will discuss the event numbers of these finalstates in detail.

\subsubsection{2 jets + 0 $b/\tau$-jets + OSD}
\label{sec:fs1}

\begin{table}[h]
	\begin{adjustwidth}{}{}
		\centering
		\begin{tabular}{|c|c||c|c|c||c|c|c|c|c|}
		\hline
		\multirow{3}{*}{\makecell{$E_{CM}$\\(TeV)}}&\multirow{3}{*}{\makecell{Pair\\produced\\leptoquark}}&\multicolumn{8}{c|}{$\geq$2 jets + 0 $b/\tau$-jets + OSD + $p_T^{\ell^+ , \ell^-} \geq 300.0$ GeV}\\
		\cline{3-10}
		&&\multicolumn{3}{c||}{Signal}&\multicolumn{5}{c|}{Backgrounds}\\
		\cline{3-10}
		&&BP1&BP2&BP3&$t\bar{t}$&$VV$&$VVV$&$t\bar{t}V$&$tVV$\\
		\hline
		\multirow{3}{*}{14}&$\rtt$&155.98&226.12&606.62&&&&&\\
		&$\xo$&445.53&143.65&28.30&246.63&202.17&17.80&2.53&0.0\\
		&$\xt$&241.62&136.48&11.10&&&&&\\
		\hline
		\multicolumn{2}{|c||}{Total}&843.13&506.25&646.02&\multicolumn{5}{c|}{469.13}\\
		\hline
		\multicolumn{2}{|c||}{Significance}&$23.27\sigma$&$16.21\sigma$&$19.34\sigma$&\multicolumn{5}{c|}{}\\
		\cline{1-5}
		\multicolumn{2}{|c||}{$\mathcal{L}_{5\sigma} \, (\text{fb}^{-1})$}&138.45&285.43&200.40&\multicolumn{5}{c|}{}\\
		\hline\hline
		\multirow{3}{*}{27}&$\rtt$&1235.47&1807.68&4944.14&&&&&\\
		&$\xo$&3279.64&1330.15&264.12&526.31&458.52&50.42&3.44&2.70\\
		&$\xt$&2085.87&1200.95&100.66&&&&&\\
		\hline
		\multicolumn{2}{|c||}{Total}&6600.98&4338.78&5308.92&\multicolumn{5}{c|}{1041.39}\\
		\hline
		\multicolumn{2}{|c||}{Significance}&$75.51\sigma$&$59.15\sigma$&$66.62\sigma$&\multicolumn{5}{c|}{}\\
		\cline{1-5}
		\multicolumn{2}{|c||}{$\mathcal{L}_{5\sigma} \, (\text{fb}^{-1})$}&4.97&7.14&5.63&\multicolumn{5}{c|}{}\\
		\hline\hline
		\multirow{3}{*}{100}&$\rtt$&7405.74&10931.30&24369.30&&&&&\\
		&$\xo$&15897.10&8695.07&1832.78&1172.99&731.67&111.14&8.39&4.43\\
		&$\xt$&12929.40&7802.14&680.35&&&&&\\
		\hline
		\multicolumn{2}{|c||}{Total}&36232.24&27428.51&26882.43&\multicolumn{5}{c|}{2028.62}\\
		\hline
		\multicolumn{2}{|c||}{Significance}&$185.23\sigma$&$159.81\sigma$&$158.10\sigma$&\multicolumn{5}{c|}{}\\
		\cline{1-5}
		\multicolumn{2}{|c||}{$\mathcal{L}_{5\sigma} \, (\text{fb}^{-1})$}&0.07&0.10&0.10&\multicolumn{5}{c|}{}\\
		\hline\hline
	\end{tabular}
	\end{adjustwidth}
\caption{The number of signal and background events for the topology FS1 in \autoref{eq:fs1} for the pair production of the three leptoquarks at the LHC/FCC, across the three benchmark points. The numbers are presented for centre-of-mass energies of 14, 27, and 100 TeV, with integrated luminosities of 3000, 1000 and 100 $\text{fb}^{-1}$ for each $E_{CM}$, respectively. The signal significances and the required luminosities for a 5$\sigma$ signal strength $\mathcal{L}_{5\sigma}$ are also presented.}
\label{tab:fs1}
\end{table}

In \autoref{tab:fs1} we present the number of events obtained in the finalstate FS1 given by \autoref{eq:fs1}, for our three leptoquark pair production processes and the dominant SM backgrounds, at three different centre-of-mass energies of 14, 27 and 100 TeV, at the LHC/FCC. In each of three benchmark points, all the three leptoquark mass eigenstates lead to two light jets and two oppositely signed leptons, albeit with different probabilities. Additionally, in BP2 and BP3, the $\xot$ leptoquarks decay to $c$-jets, with probabilities of $\sim10\%$ in BP2 and $\sim 2 \%$ in BP2, for 2 $c$-jets + OSD. We also veto out the events with any $b$- or $\tau$-tagged jets for this finalstate. These observations are reflected in the observed number of events for the three benchmark scenarios. Contributing to availability of OSD and light jets, $t\bar t$ and di-boson act as dominant SM background. As for the signal, BP1 receives the most contribution from $\xot$, BP2 has $\rtt, \xot$ contributing equivalently, and BP3 sees the majority of events from $\rtt$. At 14 TeV, the highest number of events in BP1 is observed for $\xo$, owing to a larger cross-section (\autoref{tab:pairprod}) and a $\sim68\%$ probability of leading to two lights jets and OSD, as seen in \autoref{tab:pairprodbr}. On the other hand, $\xt$, having the same probability but less cross-section, gives us the second largest number of events. Due to the absence of $c$-jets as well as a less probability ($\sim14\%$) of $2j$+OSD, the $\rtt$ pair production events in this finalstate are the lowest, but they still contribute $\sim18\%$ to the total number of events. A significance of $23.27\sigma$ is achieved at an integrated luminosity of 3000 $\text{fb}^{-1}$ for BP1. In BP2, the $\rtt$ pair production leads to the largest number of events, as a result of the $\sim40\%$ probability of $2j$ + OSD. Lesser number of events are observed in case of $\xot$ individually, while  showing a combined $\sim55\%$ probability of achieving the FS1 in this BP2. The signal significance for the total number of events is found to be $16.21\sigma$ at 3000 $\text{fb}^{-1}$ of integrated luminosity. In BP3, as a consequence of $\sim100\%$ probability of $2j$+OSD, the numbers for $\rtt$ dominates this finalstate. The tiny probabilities of $\sim8\%$ for $\xo$ and $\sim3\%$ for $\xt$ leads to very small event numbers in this finalstate, constituting $\sim7\%$ of the total number. However, the signal significance of $19.34\sigma$ is promising for this BP3 as well, with an integrated luminosity of 3000 $\text{fb}^{-1}$. Moving to a higher $E_{CM}$ of 27 TeV, similar patterns are observed for the event numbers in case of each pair-produced leptoquark as those at 14 TeV. Here, at an integrated luminosity of 1000 $\text{fb}^{-1}$, $\geq 59\sigma$ significances are achieved for all three BPs. Finally, at the highest $E_{CM}$ of 100 TeV, $\sim158\sigma$ or larger significances are obtained at 100 $\text{fb}^{-1}$ of integrated luminosity. From the \autoref{tab:fs1}, it is also clear that, in this finalstate the model can be probed with $5\sigma$ significance at $< 300\, \text{fb}^{-1}$ luminosities for 14 TeV, with the lowest being $\sim 138 \,\text{fb}^{-1}$ for BP1. With this observation, one can predict a $5\sigma$ probe of the BP1 scenario in this FS1 at the end of the current LHC run 3 of 13.6 TeV centre-of-mass energy, with the projected integrated luminosity of $\sim400$ \fbi \cite{Fartoukh:2790409}. For 27 and 100 TeV, the $5\sigma$ probe of this model can be achieved with very early stage data.

\subsubsection{2 jets + 0 $c/\tau$-jets + OSD}
\label{sec:fs2}

\begin{table}[h]
	\begin{adjustwidth}{}{}
		\centering
		\begin{tabular}{|c|c||c|c|c||c|c|c|c|c|}
			\hline
			\multirow{3}{*}{\makecell{$E_{CM}$\\(TeV)}}&\multirow{3}{*}{\makecell{Pair\\produced\\leptoquark}}&\multicolumn{8}{c|}{$\geq$2 jets + 0 $c/\tau$-jets + OSD + $p_T^{\ell^+ , \ell^-} \geq 300.0$ GeV}\\
			\cline{3-10}
			&&\multicolumn{3}{c||}{Signal}&\multicolumn{5}{c|}{Backgrounds}\\
			\cline{3-10}
			&&BP1&BP2&BP3&$t\bar{t}$&$VV$&$VVV$&$t\bar{t}V$&$tVV$\\
			\hline
			\multirow{3}{*}{14}&$\rtt$&329.61&329.86&567.93&&&&&\\
			&$\xo$&396.86&49.73&12.04&662.83&195.65&20.77&5.06&2.29\\
			&$\xt$&215.19&48.47&5.00&&&&&\\
			\hline
			\multicolumn{2}{|c||}{Total}&941.66&428.06&584.97&\multicolumn{5}{c|}{886.60}\\
			\hline
			\multicolumn{2}{|c||}{Significance}&$22.02\sigma$&$11.81\sigma$&$15.25\sigma$&\multicolumn{5}{c|}{}\\
			\cline{1-5}
			\multicolumn{2}{|c||}{$\mathcal{L}_{5\sigma} \, (\text{fb}^{-1})$}&154.63&538.10&322.53&\multicolumn{5}{c|}{}\\
			\hline\hline
			\multirow{3}{*}{27}&$\rtt$&2634.68&2625.18&4599.96&&&&&\\
			&$\xo$&2874.61&459.54&109.77&1498.83&456.20&70.59&14.64&4.97\\
			&$\xt$&1826.10&405.47&46.81&&&&&\\
			\hline
			\multicolumn{2}{|c||}{Total}&7335.39&3490.19&4756.54&\multicolumn{5}{c|}{2045.24}\\
			\hline
			\multicolumn{2}{|c||}{Significance}&$75.73\sigma$&$46.91\sigma$&$57.67\sigma$&\multicolumn{5}{c|}{}\\
			\cline{1-5}
			\multicolumn{2}{|c||}{$\mathcal{L}_{5\sigma} \, (\text{fb}^{-1})$}&4.36&11.36&7.51&\multicolumn{5}{c|}{}\\
			\hline\hline
			\multirow{3}{*}{100}&$\rtt$&15089.50&15384.40&21893.60&&&&&\\
			&$\xo$&13632.90&2871.55&759.10&3113.17&769.85&138.90&33.61&11.84\\
			&$\xt$&10979.60&2616.28&288.68&&&&&\\
			\hline
			\multicolumn{2}{|c||}{Total}&39702.00&20872.20&22941.38&\multicolumn{5}{c|}{4067.37}\\
			\hline
			\multicolumn{2}{|c||}{Significance}&$189.77\sigma$&$132.17\sigma$&$139.59\sigma$&\multicolumn{5}{c|}{}\\
			\cline{1-5}
			\multicolumn{2}{|c||}{$\mathcal{L}_{5\sigma} \, (\text{fb}^{-1})$}&0.07&0.14&0.13&\multicolumn{5}{c|}{}\\
			\hline\hline
		\end{tabular}
	\end{adjustwidth}
\caption{The number of signal and background events for the topology FS2 in \autoref{eq:fs2} for the pair production of the three leptoquarks at the LHC/FCC, across the three benchmark points. The numbers are presented for centre-of-mass energies of 14, 27, and 100 TeV, with integrated luminosities of 3000, 1000 and 100 $\text{fb}^{-1}$ for each $E_{CM}$, respectively. The signal significances and the required luminosities for a 5$\sigma$ signal strength $\mathcal{L}_{5\sigma}$ are also presented.}
\label{tab:fs2}
\end{table}

While in FS1 we considered a finalstate with $c$-jets to facilitate higher numbers for $\xot$, in this finalstate (FS2) given by \autoref{eq:fs2}, we veto out events with $c$-jets instead of $b$-jets, keeping in mind that 2 $b$-jets + OSD can be obtained from $\rtt$ for BP1 and BP2. The number of signal and background events are tabulated in \autoref{tab:fs2}, for the three leptoquarks across the benchmark scenarios, at the centre-of-mass energies of 14, 27, and 100 TeV at the LHC/FCC.The SM decay chain of $t\bar{t} \to 2b + 2W \to 2b\text{-jets + OSD}$ ensures a very dominant background contribution from $t\bar{t}$. concerning the signal, we see near-equivalent contributions from $\rtt, \xot$ in BP1, while in both BP2 and BP3, $\rtt$ dominates. At 14 TeV for BP1, we observe an enhancement of events for $\rtt$ compared to FS1, as a result of bringing the $b$-jets into the picture, as suggested by \autoref{tab:pairprodbr}. However, $\xot$ shows similar numbers as FS1, indifferent to the consideration of $b$- or $c$-jets alike, owing to no decay mode of $\xot$ directly leading to a $b/c$-jet and a charged lepton (\autoref{bp1decay}). In BP2 and BP3, the numbers are dominated by $\rtt$, both due to the consideration of $b$-jets and the rejection of $c$-jets, similarly predicted by \autoref{tab:pairprodbr}. The number of events for $\xo$ and $\xt$ together contribute $\sim23\%$ for BP2, and $\sim3\%$ for BP3. The combined number of events for the three leptoquarks lead to the feasibility of probing the model with $22.02\sigma$, $11.81\sigma$, and $15.25\sigma$, respectively for BP1, BP2 and BP3, at the 14 TeV LHC with an integrated luminosity of 3000 $\text{fb}^{-1}$. Moving to a higher centre-of-mass energy of 27 TeV, the signal significances are even more improved, with $75.73\sigma, \, 46.91\sigma$, and $57.67\sigma$, for BP1-BP3 respectively, considering an integrated luminosity of 1000 $\text{fb}^{-1}$. The simulation at 100 TeV FCC improves these numbers even more drastically, where at 100 $\text{fb}^{-1}$ luminosity, the model can be probed with significances of $189.77\sigma,\, 132.17\sigma$, and $139.59\sigma$, respectively for the three BPs. Similar to FS1, a $5\sigma$ probe of the model can be achieved at the 14 TeV LHC with $< 550 \,\text{fb}^{-1}$ luminosity for each benchmark point. Similar to the case of FS1, the numbers predict a possible 5$\sigma$ probe of BP1 at the end of the LHC run 3 of $E_{CM} = 13.6$ TeV, with $\sim400$ \fbi of projected integrated luminosity. In the higher energies of 27 and 100 TeV, this $5\sigma$ probe is predicted to be obtained with much earlier data. 

Thus, the finalstate topologies described by \autoref{eq:fs1} and \autoref{eq:fs2} contain the potential of achieving $5\sigma$ signatures for the $\widetilde{R}_2$ and $S_1$ extension of the SM, with rich phenomenology encompassing the three chosen benchmark scenarios. These finalstates are considered in such a way that, for each benchmark point, some entries of the Yukawa couplings contribute towards obtaining them for each pair of $\rtt$ and $\xot$ that are produced at the LHC/FCC. While FS1 and FS2 here are probes of this Beyond the Standard Model (BSM) extension as a whole, we also intend to study the possibilities of distinguishing between the pure doublet $\rtt$ and the mixed states $\xot$, as well as between the mixed states $\xo$ and $\xt$ themselves. In the next sections, we will discuss these possibilities one by one.

\section{Distinguishing between $\rtt$ and $\xot$}
\label{sec:rtvx12}

For the three chosen benchmark scenarios in this work, the three leptoquarks $\rtt, \, \xo$ and $\xt$ are nearly degenerate, with masses $\approx 1.5$ TeV. However, $\rtt$ being a pure doublet and $\xot$ being mixed states have different combinations of the Yukawa couplings dictating their dominant decay modes. In particular, $\rtt$ only couples to a down-type quark and a charged lepton via $\y_2$, which can be seen by expanding the model Lagrangian in \autoref{eq:Lag_SD}. On the other hand, $\xot$ can couple to an up-type quark and a charged lepton, or a down-type quark and a neutrino, via different combinations of $\y_{1}^{L,R}$ and $\y_2$. Nonetheless, both the pure doublet and the mixed states can lead to finalstates with non-tagged light jets and a pair of oppositely signed leptons that are not identified as electrons or muons. Now, a distinction can be made between $\rtt$ and $\xot$ by virtue of their charge difference, by studying the charge distribution of the jets emanating from them, similar to the work done in Refs. \cite{Krohn:2012fg,CMS:2017yer,Tokar:2017syr,Bandyopadhyay:2020jez,Bandyopadhyay:2020wfv}. However, in this article, we wish to look at a few finalstate topologies that will give us event numbers that are heavily dominated by either $\rtt$ or $\xot$. While we witnessed such a dominance for $\rtt$ pair-production events in case of BP3 for the FS2 as discussed in \autoref{sec:modsig}, it is important to note that the FS2 events still contained direct consequences of the Yukawa couplings for $\xot$. In this section, we will look at topologies facilitated by the Yukawa couplings for either the pure doublet or the mixed states, so that any small contamination from the either is a resultant of the ISR/FSR activities. For example, a finalstate of two $b$-jets + 2OSe is only allowed for $\rtt$ by the coupling $\y_2$ in BP1 and BP2, but there can be slight contaminations from $\xot$ due to $b$-jets arising from ISR/FSR effects. With this discussion in mind, as well as the information provided by \autoref{tab:pairprodbr}, we can now look for four such finalstates with which we will try to discern $\rtt$ and $\xot$.

\subsection{2 $b$-jets + 2 OS$e$}
\label{sec:fs3}

In both BP1 and BP2, the pure doublet leptoquark $\rtt$, when pair-produced, can lead to a finalstate of two $b$-tagged jets and two oppositely signed electrons (\autoref{tab:pairprodbr}), from the $\rtt \to be^+$ decay happening via the $(\y_2)^{31}$ element. Similar to \autoref{sec:modsig}, we can put a cut of $p_T^{e^+ , e^-} \geq 300.0$ GeV, to remove a large portion of the SM background for a cleaner signal. The complete finalstate is given as follows:
\begin{equation}
	\text{\textbf{FS3:}} \quad n_{\text{\textit{b}-jet}} \geq 2 \,\, \& \,\, n_{e^+} = n_{e^-} = 1 \,\, \& \,\, n_{\text{light}/c/\tau\text{-jets}} = 0 \,\, \& \,\, p_T^{e^+ , e^-} \geq 300.0 \, \text{GeV}.
	\label{eq:fs3}
\end{equation}

\begin{table}[h!]
	\begin{adjustwidth}{}{}
		\centering
	\begin{tabular}{|c|c||c|c|c||c|c|c|c|c|}
		\hline
		\multirow{3}{*}{\makecell{$E_{CM}$\\(TeV)}}&\multirow{3}{*}{\makecell{Pair\\produced\\leptoquark}}&\multicolumn{8}{c|}{$\geq$2 $b$-jets + 0 light/$c/\tau$-jets + 2 OS$e$ + $p_T^{e^+ , e^-} \geq 300.0$ GeV}\\
		\cline{3-10}
		&&\multicolumn{3}{c||}{Signal}&\multicolumn{5}{c|}{Backgrounds}\\
		\cline{3-10}
		&&BP1&BP2&BP3&$t\bar{t}$&$VV$&$VVV$&$t\bar{t}V$&$tVV$\\
		\hline
		\multirow{3}{*}{14}&$\rtt$&72.11&49.79&0.03&&&&&\\
		&$\xo$&1.44&0.0&0.0&33.39&0.0&0.0&0.0&0.0\\
		&$\xt$&0.71&0.0&0.0&&&&&\\
		\hline
		\multicolumn{2}{|c||}{Total}&74.26&49.79&0.03&\multicolumn{5}{c|}{}\\
		\hline
		\multicolumn{2}{|c||}{Significance}&$7.16\sigma$&$5.46\sigma$&$0.005\sigma$&\multicolumn{5}{c|}{}\\
		\cline{1-5}
		\multicolumn{2}{|c||}{$\mathcal{L}_{5\sigma} \, (\text{fb}^{-1})$}&1464.17&2516.68&$\gg$3000&\multicolumn{5}{c|}{}\\
		\hline\hline
		\multirow{3}{*}{27}&$\rtt$&578.60&400.79&0.15&&&&&\\
		&$\xo$&13.08&0.0&0.0&106.78&0.0&0.0&1.71&0.0\\
		&$\xt$&7.37&0.0&0.0&&&&&\\
		\hline
		\multicolumn{2}{|c||}{Total}&599.05&400.79&0.15&\multicolumn{5}{c|}{108.49}\\
		\hline
		\multicolumn{2}{|c||}{Significance}&$22.52\sigma$&$17.76\sigma$&$0.01\sigma$&\multicolumn{5}{c|}{}\\
		\cline{1-5}
		\multicolumn{2}{|c||}{$\mathcal{L}_{5\sigma} \, (\text{fb}^{-1})$}&49.29&79.26&$\gg$1000&\multicolumn{5}{c|}{}\\
		\hline\hline
		\multirow{3}{*}{100}&$\rtt$&3123.94&2274.82&0.88&&&&&\\
		&$\xo$&88.35&0.0&0.0&355.07&2.12&0.0&3.36&0.0\\
		&$\xt$&81.14&0.0&0.0&&&&&\\
		\hline
		\multicolumn{2}{|c||}{Total}&3293.43&2274.82&0.88&\multicolumn{5}{c|}{360.55}\\
		\hline
		\multicolumn{2}{|c||}{Significance}&$54.48\sigma$&$44.31\sigma$&$0.05\sigma$&\multicolumn{5}{c|}{}\\
		\cline{1-5}
		\multicolumn{2}{|c||}{$\mathcal{L}_{5\sigma} \, (\text{fb}^{-1})$}&0.84&1.27&$\gg100$&\multicolumn{5}{c|}{}\\
		\hline\hline
	\end{tabular}
	\end{adjustwidth}
\caption{The number of signal and background events for the topology FS3 in \autoref{eq:fs3} for the pair production of the three leptoquarks at the LHC/FCC, across the three benchmark points. The numbers are presented for centre-of-mass energies of 14, 27, and 100 TeV, with integrated luminosities of 3000, 1000 and 100 $\text{fb}^{-1}$ for each $E_{CM}$, respectively. The signal significances and the required luminosities for a 5$\sigma$ signal strength $\mathcal{L}_{5\sigma}$ are also presented.}
\label{tab:fs3}
\end{table}

\autoref{tab:fs3} displays the signal and background event numbers for this finalstate, simulated from pair productions of each leptoquark under consideration at the LHC/FCC with three centre-of-mass energies of 14, 27 and 100 TeV. The SM background is very small in this case and it mainly comes from $t\bar t$, contributing to the $b$-jet pair. The first thing we notice here is that, for BP3, we get negligible number of events due the the absence of this finalstate there, as shown in \autoref{tab:pairprodbr}. However, BP1 and BP2 give us very interesting results. At 14 TeV in BP1, a large majority of the events pertain to $\rtt$, with a $\sim3\%$ contamination from $\xot$ combined. For BP2, only $\rtt$ contributes to this finalstate, and zero contamination is observed from the mixed states. This is due to the absence of di-electron finalstates for $\xot$ pair production in the BP2 scenario. The FS3 in these benchmark points are shown to be probed with significances of $7.16\sigma$ and $5.46\sigma$, respectively for BP1 and BP2, with 3000 $\text{fb}^{-1}$ of integrated luminosity. This distinction of $\rtt$ from $\xot$ can be observed with $5\sigma$ significance, at integrated luminosities of $1464.17$ and $2516.68$ $\text{fb}^{-1}$, which are within the limits of the HL-LHC. The observed SM background here is very minimal due to the stringent cuts on the jet and lepton identities, with most of it being contributed by $t\bar{t}$. The same $\rtt$ dominance is observed at $E_{CM}$ = 27 TeV, with the signal significance now increasing to $22.52\sigma$ in BP1, and $17.76\sigma$ in BP2, with an integrated luminosity of 1000 $\text{fb}^{-1}$. At 100 TeV centre-of-mass energy, this distinction can be observed with a significance of $54.48\sigma$ in BP1, and $44.31\sigma$ in BP2, with 100 $\text{fb}^{-1}$ luminosity. The required $5\sigma$ significance can be obtained with $<80 \,\text{fb}^{-1}$ luminosity at 27 TeV, while at 100 TeV it comes down to $<2 \,\text{fb}^{-1}$, for both BP1 and BP2. The FS3 topology can thus act as a probe of the Yukawa coupling $(\y_2)^{31}$ itself, while being able to distinguish the $\rtt$ pair production signal from $\xot$.

\subsection{1 light jet + 1 $c$-jet + 2 OS$\mu$}
\label{sec:fs4}

In case of BP2, \autoref{tab:pairprodbr} predicts a significant probability of di-muons in the finalstates for $\xot$ pair production, while for $\rtt$ this probability is seen to be zero. On the other hand, BP3 sees a 100\% probability of having oppositely signed muon pairs in the finalstates from $\rtt$ pair production, with a negligibly small probability in case of $\xot$. Considering the $\sim20\%$ effective branching for a finalstate involving one light and one $c$-tagged jet alongside a pair of oppositely signed muons (OS$\mu$) from $\xot$ at BP2, we decide to study the events of the following finalstate as a means of distinguishing between the pure doublet and the mixed states:
\begin{equation}
	\text{\textbf{FS4:}} \quad n_{\text{light jet}} \geq 1 \,\, \& \,\, n_{\text{\textit{c}-jet}} \geq 1 \,\, \& \,\, n_{\mu^+} = n_{\mu^-} = 1 \,\, \& \,\, n_{b/\tau\text{-jets}} = 0 \,\, \& \,\, p_T^{\mu^+ , \mu^-} \geq 300.0 \, \text{GeV}.
	\label{eq:fs4}
\end{equation}

\begin{table}[h]
	\begin{adjustwidth}{}{}
		\centering
		\begin{tabular}{|c|c||c|c|c||c|c|c|c|c|}
			\hline
			\multirow{3}{*}{\makecell{$E_{CM}$\\(TeV)}}&\multirow{3}{*}{\makecell{Pair\\produced\\leptoquark}}&\multicolumn{8}{c|}{$\geq$1 light jets + $\geq$1 $c$-jets+ 0 $b/\tau$-jets + 2 OS$\mu$ + $p_T^{\mu^+ , \mu^-} \geq 300.0$ GeV}\\
			\cline{3-10}
			&&\multicolumn{3}{c||}{Signal}&\multicolumn{5}{c|}{Backgrounds}\\
			\cline{3-10}
			&&BP1&BP2&BP3&$t\bar{t}$&$VV$&$VVV$&$t\bar{t}V$&$tVV$\\
			\hline
			\multirow{3}{*}{14}&$\rtt$&0.0&0.0&48.76&&&&&\\
			&$\xo$&0.0&94.35&1.26&5.13&3.25&0.0&0.0&0.0\\
			&$\xt$&0.01&88.46&0.44&&&&&\\
			\hline
			\multicolumn{2}{|c||}{Total}&0.01&182.81&50.46&\multicolumn{5}{c|}{8.38}\\
			\hline
			\multicolumn{2}{|c||}{Significance}&$0.003\sigma$&$13.22\sigma$&$6.57\sigma$&\multicolumn{5}{c|}{}\\
			\cline{1-5}
			\multicolumn{2}{|c||}{$\mathcal{L}_{5\sigma} \, (\text{fb}^{-1})$}&$\gg$3000&429.06&1733.16&\multicolumn{5}{c|}{}\\
			\hline\hline
			\multirow{3}{*}{27}&$\rtt$&0.0&0.0&473.73&&&&&\\
			&$\xo$&0.06&875.87&13.53&38.13&13.89&9.43&0.0&1.63\\
			&$\xt$&0.11&800.11&4.41&&&&&\\
			\hline
			\multicolumn{2}{|c||}{Total}&0.17&1675.98&491.67&\multicolumn{5}{c|}{63.08}\\
			\hline
			\multicolumn{2}{|c||}{Significance}&$0.02\sigma$&$40.18\sigma$&$20.87\sigma$&\multicolumn{5}{c|}{}\\
			\cline{1-5}
			\multicolumn{2}{|c||}{$\mathcal{L}_{5\sigma} \, (\text{fb}^{-1})$}&$\gg$1000&15.48&57.37&\multicolumn{5}{c|}{}\\
			\hline\hline
			\multirow{3}{*}{100}&$\rtt$&0.77&0.0&3069.96&&&&&\\
			&$\xo$&0.89&5883.05&109.21&63.40&46.65&22.21&0.0&4.44\\
			&$\xt$&1.54&5235.57&47.59&&&&&\\
			\hline
			\multicolumn{2}{|c||}{Total}&3.2&11118.62&3226.76&\multicolumn{5}{c|}{136.70}\\
			\hline
			\multicolumn{2}{|c||}{Significance}&$0.27\sigma$&$104.80\sigma$&$55.63\sigma$&\multicolumn{5}{c|}{}\\
			\cline{1-5}
			\multicolumn{2}{|c||}{$\mathcal{L}_{5\sigma} \, (\text{fb}^{-1})$}&$\gg100$&0.23&0.81&\multicolumn{5}{c|}{}\\
			\hline\hline
		\end{tabular}
	\end{adjustwidth}
\caption{The number of signal and background events for the topology FS4 in \autoref{eq:fs4} for the pair production of the three leptoquarks at the LHC/FCC, across the three benchmark points. The numbers are presented for centre-of-mass energies of 14, 27, and 100 TeV, with integrated luminosities of 3000, 1000 and 100 $\text{fb}^{-1}$ for each $E_{CM}$, respectively. The signal significances and the required luminosities for a 5$\sigma$ signal strength $\mathcal{L}_{5\sigma}$ are also presented.}
\label{tab:fs4}
\end{table}

In \autoref{tab:fs4}, the signal and background event numbers are presented for the pair production of the three leptoquarks at 14, 27 and 100 TeV LHC/FCC energies, pertaining to the finalstate FS4 described in \autoref{eq:fs4}. The applied cuts and vetoes keep the SM background contribution minimal. This time, the insignificant event numbers are seen for BP1 owing to the unavailability of OS$\mu$s in any finalstate, as seen from \autoref{tab:pairprodbr}. As a consequence of the $\sim32\%$ branching ratios of both $\xot$ in $c\mu$ and $u\mu$ decay modes, as well as the zero probability of OS$\mu$ pairs from $\rtt$, the event numbers for BP2 are very similar for $\xo$ and $\xt$, while zero events are observed from $\rtt$ pair production. The situation gets reversed in BP3, where the total number of events are dominated by $\rtt$, with $\sim4\%$ contamination from $\xot$ pair productions combined. Here, even though the exact finalstate of FS4 is not predicted for $\rtt$ in \autoref{tab:pairprodbr}, the 100\% probability of obtaining OS$\mu$ pairs lead to a significant number of events here, with the additional required $c$-jet provided by ISR/FSR effects. Thus, in the BP2 scenario, events in this FS4 topology pertain only to the mixed leptoquark states $\xot$, governed by the Yukawa coupling combinations $(\yr)_{12}+(\widetilde\y_{1}^L)_{12}$ and $(\yr)_{23}+(\widetilde\y_{1}^L)_{23}$ (\autoref{bp2decay}). However, in the BP3 scenario, FS3 is a probe of the pure doublet state $\rtt$, where the required jet and di-muon finalstate is provided by the Yukawa couplings $(\y_2)_{12}$ and $(\y_2)_{22}$ (\autoref{bp3decay}). These distinctions are achieved with significances of $13.22\sigma$ for BP2 and $6.57\sigma$ for BP3, at the 14 TeV LHC with an integrated luminosity of 3000 $\text{fb}^{-1}$. A 5$\sigma$ probe can be achieved with HL-LHC luminosities of 429.06 and 1733.16 $\text{fb}^{-1}$, respectively for BP2 ad BP3. At the higher centre-of-mass energy of 27 TeV, the same discerning signatures are obtained with $40.18\sigma$ and $20.87\sigma$ significances, respectively for BP2 and BP3, with 1000 $\text{fb}^{-1}$ luminosity. At 100 TeV centre-of-mass energy and 100 $\text{fb}^{-1}$ luminosity, these significances increase to $104.8\sigma$ for BP2, and $55.63\sigma$ for BP3. In both of these higher centre-of-mass energies, early data can provide the required $5\sigma$ significance (< $60 \,\text{fb}^{-1}$ for 27 TeV, < $1 \,\text{fb}^{-1}$ for 100 TeV). Thus, unlike FS3, this topology can point towards either the mixed states or the pure doublet state, considering the benchmark scenario.

\subsection{1 light jet + 1 $b$-jet + 2 OS$e$}
\label{sec:fs5}

In case of both BP1 and BP2, \autoref{tab:pairprodbr} shows us the strong presence of a finalstate containing one light jet, one $b$-jet and a pair of oppositely signed electrons (OS$e$), that come from the $\rtt$ pair production, but are absent in case of $\xot$. This presents us with another opportunity of distinguishing $\rtt$ from $\xot$ from pair production events. The complete finalstate topology is given as follows:
\begin{equation}
	\text{\textbf{FS5:}} \quad n_{\text{light jet}} \geq 1 \,\, \& \,\, n_{\text{\textit{b}-jet}} \geq 1 \,\, \& \,\, n_{e^+} = n_{e^-} = 1 \,\, \& \,\, n_{c/\tau\text{-jets}} = 0 \,\, \& \,\, p_T^{e^+ , e^-} \geq 300.0 \, \text{GeV}.
	\label{eq:fs5}
\end{equation}

\begin{table}[h!]
	\begin{adjustwidth}{}{}
		\centering
	\begin{tabular}{|c|c||c|c|c||c|c|c|c|c|}
		\hline
		\multirow{3}{*}{\makecell{$E_{CM}$\\(TeV)}}&\multirow{3}{*}{\makecell{Pair\\produced\\leptoquark}}&\multicolumn{8}{c|}{$\geq$1 light jets + $\geq$1 $b$-jets+ 0 $c/\tau$-jets + 2 OS$e$ + $p_T^{e^+ , e^-} \geq 300.0$ GeV}\\
		\cline{3-10}
		&&\multicolumn{3}{c||}{Signal}&\multicolumn{5}{c|}{Backgrounds}\\
		\cline{3-10}
		&&BP1&BP2&BP3&$t\bar{t}$&$VV$&$VVV$&$t\bar{t}V$&$tVV$\\
		\hline
		\multirow{3}{*}{14}&$\rtt$&184.31&122.80&0.06&&&&&\\
		&$\xo$&3.44&0.0&0.0&184.98&0.0&1.97&0.0&0.45\\
		&$\xt$&2.05&0.0&0.0&&&&&\\
		\hline
		\multicolumn{2}{|c||}{Total}&189.80&122.80&0.06&\multicolumn{5}{c|}{187.40}\\
		\hline
		\multicolumn{2}{|c||}{Significance}&$9.77\sigma$&$6.97\sigma$&$0.005\sigma$&\multicolumn{5}{c|}{}\\
		\cline{1-5}
		\multicolumn{2}{|c||}{$\mathcal{L}_{5\sigma} \, (\text{fb}^{-1})$}&785.32&1542.82&$\gg$3000&\multicolumn{5}{c|}{}\\
		\hline\hline
		\multirow{3}{*}{27}&$\rtt$&1515.37&1001.96&0.3&&&&&\\
		&$\xo$&34.54&0.0&0.0&556.82&16.20&5.03&3.60&3.61\\
		&$\xt$&23.48&0.0&0.0&&&&&\\
		\hline
		\multicolumn{2}{|c||}{Total}&1573.39&1001.96&0.3&\multicolumn{5}{c|}{585.26}\\
		\hline
		\multicolumn{2}{|c||}{Significance}&$33.82\sigma$&$25.11\sigma$&$0.01\sigma$&\multicolumn{5}{c|}{}\\
		\cline{1-5}
		\multicolumn{2}{|c||}{$\mathcal{L}_{5\sigma} \, (\text{fb}^{-1})$}&21.85&39.65&$\gg$1000&\multicolumn{5}{c|}{}\\
		\hline\hline
		\multirow{3}{*}{100}&$\rtt$&8600.97&5913.14&0.88&&&&&\\
		&$\xo$&239.17&0.0&0.0&1014.47&42.40&20.37&15.12&7.39\\
		&$\xt$&217.16&0.0&0.78&&&&&\\
		\hline
		\multicolumn{2}{|c||}{Total}&9057.30&5913.14&1.66&\multicolumn{5}{c|}{1099.75}\\
		\hline
		\multicolumn{2}{|c||}{Significance}&$89.87\sigma$&$70.61\sigma$&$0.05\sigma$&\multicolumn{5}{c|}{}\\
		\cline{1-5}
		\multicolumn{2}{|c||}{$\mathcal{L}_{5\sigma} \, (\text{fb}^{-1})$}&0.31&0.50&$\gg100$&\multicolumn{5}{c|}{}\\
		\hline\hline
	\end{tabular}
	\end{adjustwidth}
\caption{The number of signal and background events for the topology FS5 in \autoref{eq:fs5} for the pair production of the three leptoquarks at the LHC/FCC, across the three benchmark points. The numbers are presented for centre-of-mass energies of 14, 27, and 100 TeV, with integrated luminosities of 3000, 1000 and 100 $\text{fb}^{-1}$ for each $E_{CM}$, respectively. The signal significances and the required luminosities for a 5$\sigma$ signal strength $\mathcal{L}_{5\sigma}$ are also presented.}
\label{tab:fs5}
\end{table}

\autoref{tab:fs5} depicts the number of signal and background events for FS5, coming from the pair production of all the three leptoquarks at 14, 27 and 100 TeV centre-of-mass energies at the LHC/FCC. As expected, for BP3 the signal event numbers are insignificant. BP1 shows a heavy dominance of $\rtt$ events, while a tiny $\sim4\%$ combined contamination from $\xot$ pair production is observed, due to the high probability of di-electron events (\autoref{tab:pairprodbr}). In BP2 however, $\rtt$ pair production is responsible for all the events in this topology, as no di-electron finalstates are facilitated by the $\y_1^{L,R}$ for $\xot$ in this scenario. The pattern remains consistent through all the three centre-of-mass energies, with small backgrounds mostly arising from $t\bar{t}$, owing to more $b$-jet-tagged events. Similar to all the previous finalstates, very promising signal significances are obtained at each centre-of-mass energy for both BP1 and BP2. At 14 TeV, the pure doublet $\rtt$ stands apart from the mixed states with a significance of $9.77\sigma$ for BP1, and $6.97\sigma$ for BP2, with a luminosity of 3000 $\text{fb}^{-1}$. the required $5\sigma$ significance is achieved at < 1600 $\text{fb}^{-1}$ luminosity for both BP1 and BP2 in this case. At 27 TeV energy with 1000 $\text{fb}^{-1}$ luminosity, these signals are enhanced to $33.82\sigma$ for BP1 and $25.11\sigma$ for BP2, requiring < 40 $\text{fb}^{-1}$ luminosities to probe them up to $5\sigma$ significance. Finally, the 100 TeV FCC simulation at 100 $\text{fb}^{-1}$ luminosity gives us $89.87\sigma$ significance for BP1 and $70.61\sigma$ in case of BP2. Here, < 1 $\text{fb}^{-1}$ luminosity is enough to probe this discerning signature with a $5\sigma$ significance. In both BP1 and BP2, the FS5 topology mainly arises from the combination of doublet Yukawa couplings $(\y_2)_{21} + (\y_2)_{31}$ (\autoref{bp3decay}).

\subsection{2 light-jets + $\ptmiss$}
\label{sec:fs6}

The neutrinos that emerge from any SM or BSM particle decay, remain ``invisible" at the LHC detectors. Due to the conservation of total $p_T$, these invisible neutrinos account for some missing transverse momentum $\ptmiss$. From our model Lagrangian it is observed that, the component of $\widetilde{R}_2$ with electric charge -1/3, as well as $S_1^{-1/3}$, can couple to a down-type quark and a neutrino via $\y_2$ and $\y_1^{L}$ couplings, respectively. Especially for BP2 and BP3, \autoref{tab:pairprodbr} shows us the probabilities of obtaining a pair of light jets along with some $\ptmiss$ that is carried by the neutrinos directly emanating from $\xot$ leptoquarks. Having a heavy ($\sim 1.5$ TeV) particle as the mother, the $p_T$ of these neutrinos are expected to be large, similar to the case with the leptons as seen in Figure \ref{fig:lepjetpt}.

\begin{figure}[h]
	\centering
	\includegraphics[width=0.6\linewidth]{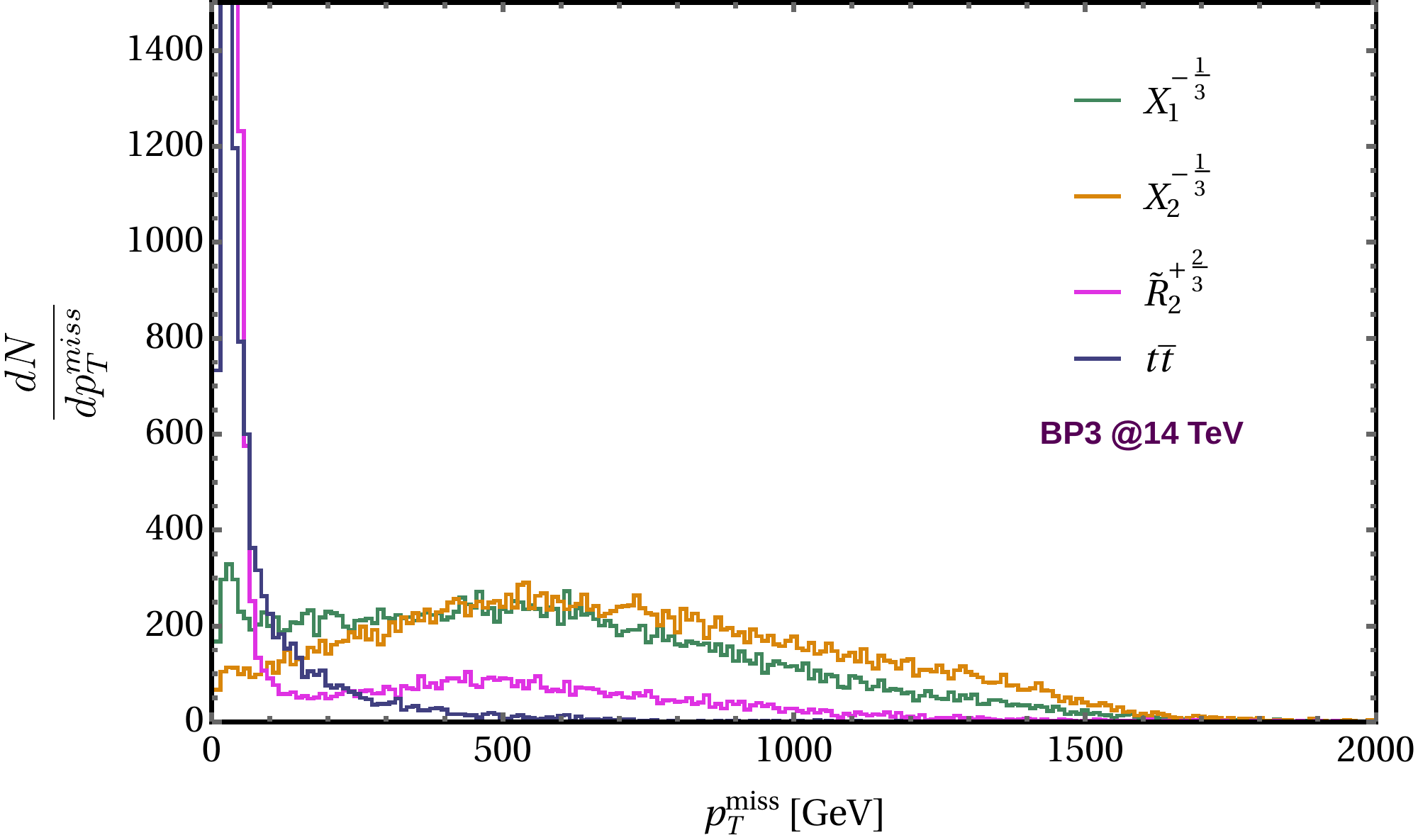}
	\caption{Distributions of $\ptmiss$ in a $2j + \ptmiss$ finalstate, from pair production of the three leptoquarks at 14 TeV LHC, for BP1. The same distribution is also shown for the $t\bar{t}$ SM background.}
	\label{fig:2jptmiss}
\end{figure}

In \autoref{fig:2jptmiss}, we show the missing transverse momentum distributions in a finalstate with two light jets and $\ptmiss$, for the pair productions of $\rtt$ (pink), $\xo$ (green), and $\xt$ (yellow), for BP3 at the 14 TeV LHC. The same distribution arising from the SM background of $t\bar{t}$ (dominant) is also shown (dark blue). As expected, a wide peak is observed around $\sim600$ GeV for both $\xot$ leptoquark cases, accounting for the neutrinos that they decay into. However, owing to the lesser probability of this state for $\xt$ (\autoref{tab:pairprodbr}), another sharp peak is seen at around $\sim50$ GeV. This peak signifies the neutrinos that come from other SM processes, such as decays of the $\tau$ from $\xot \to u \tau^-$ modes. In case of $\xo$, this earlier peak is negligible due to higher probability of obtaining $2j + \ptmiss$. The distributions for $\rtt$ almost mimics that of $t\bar{t}$, but with a longer tail, with the only significant peak occurring at around $\sim50$ GeV. This allows us to put an advanced cut of $\ptmiss \geq 500$ GeV in order to eliminate more of the SM background contribution, leading us to the finalstate:
\begin{equation}
	\text{\textbf{FS6:}} \quad n_{\text{light jet}} \geq 2 \,\,  \& \,\, n_{\text{lep}} = 0 \,\, \& \,\, n_{b/c/\tau\text{-jets}} = 0 \,\, \& \,\,  \ptmiss \geq 500.0\,\text{GeV}.
	\label{eq:fs6}
\end{equation}

\begin{table}[h!]
	\begin{adjustwidth}{}{}
		\centering
	\begin{tabular}{|c|c||c|c|c||c|c|c|c|c|}
		\hline
		\multirow{3}{*}{\makecell{$E_{CM}$\\(TeV)}}&\multirow{3}{*}{\makecell{Pair\\produced\\leptoquark}}&\multicolumn{8}{c|}{$\geq$2 light jets + 0 leptons + 0 $b/c/\tau$-jets + $ \ptmiss \geq 500.0$ GeV}\\
		\cline{3-10}
		&&\multicolumn{3}{c||}{Signal}&\multicolumn{5}{c|}{Backgrounds}\\
		\cline{3-10}
		&&BP1&BP2&BP3&$t\bar{t}$&$VV$&$VVV$&$t\bar{t}V$&$tVV$\\
		\hline
		\multirow{3}{*}{14}&$\rtt$&3.04&5.31&8.16&&&&&\\
		&$\xo$&20.05&28.89&51.70&236.35&45.65&5.93&7.63&0.23\\
		&$\xt$&11.91&31.53&108.83&&&&&\\
		\hline
		\multicolumn{2}{|c||}{Total}&35.00&65.73&168.69&\multicolumn{5}{c|}{295.80}\\
		\hline
		\multicolumn{2}{|c||}{Significance}&$1.92\sigma$&$3.46\sigma$&$7.82\sigma$&\multicolumn{5}{c|}{}\\
		\cline{1-5}
		\multicolumn{2}{|c||}{$\mathcal{L}_{5\sigma} \, (\text{fb}^{-1})$}&$\gg$3000&$\gg$3000&1224.22&\multicolumn{5}{c|}{}\\
		\hline\hline
		\multirow{3}{*}{27}&$\rtt$&26.85&47.78&67.63&&&&&\\
		&$\xo$&158.23&303.62&519.61&1666.64&272.10&57.14&31.00&3.61\\
		&$\xt$&116.03&302.95&1088.45&&&&&\\
		\hline
		\multicolumn{2}{|c||}{Total}&301.11&654.35&1675.69&\multicolumn{5}{c|}{2030.49}\\
		\hline
		\multicolumn{2}{|c||}{Significance}&$6.23\sigma$&$12.62\sigma$&$27.52\sigma$&\multicolumn{5}{c|}{}\\
		\cline{1-5}
		\multicolumn{2}{|c||}{$\mathcal{L}_{5\sigma} \, (\text{fb}^{-1})$}&642.90&156.76&32.99&\multicolumn{5}{c|}{}\\
		\hline\hline
		\multirow{3}{*}{100}&$\rtt$&209.45&329.19&399.46&&&&&\\
		&$\xo$&900.47&2011.79&3688.64&5202.36&816.50&142.61&157.95&59.20\\
		&$\xt$&770.51&2114.71&7315.32&&&&&\\
		\hline
		\multicolumn{2}{|c||}{Total}&1880.43&4455.69&11403.42&\multicolumn{5}{c|}{6378.63}\\
		\hline
		\multicolumn{2}{|c||}{Significance}&$20.69\sigma$&$42.81\sigma$&$85.51\sigma$&\multicolumn{5}{c|}{}\\
		\cline{1-5}
		\multicolumn{2}{|c||}{$\mathcal{L}_{5\sigma} \, (\text{fb}^{-1})$}&5.83&1.36&0.34&\multicolumn{5}{c|}{}\\
		\hline\hline
	\end{tabular}
	\end{adjustwidth}
\caption{The number of signal and background events for the topology FS6 in \autoref{eq:fs6} for the pair production of the three leptoquarks at the LHC/FCC, across the three benchmark points. The numbers are presented for centre-of-mass energies of 14, 27, and 100 TeV, with integrated luminosities of 3000, 1000 and 100 $\text{fb}^{-1}$ for each $E_{CM}$, respectively. The signal significances and the required luminosities for a 5$\sigma$ signal strength $\mathcal{L}_{5\sigma}$ are also presented.}
\label{tab:fs6}
\end{table}

In \autoref{tab:fs6} we present the number of events in the FS6 topology for the leptoquark pair production signals as well as the SM backgrounds, at 14, 27 and 100 TeV centre-of-mass energies of the LHC/FCC. While in each benchmark scenario the majority of events are seen to be coming from $\xot$, the $\rtt$ pair production events account for about $11\%$ in BP1, and $10\%$ in BP2, which are not very negligible. This happens due to the lesser or no probability of obtaining a $2j + \ptmiss$ finalstate from $\xot$ in BP1 and BP2, as seen in \autoref{tab:pairprodbr}. However, the situation is interesting in case of BP3. Owing to the higher probabilities of $2j+\ptmiss$ for $\xot$, the events pertaining to $\rtt$ contribute < 5\% to the FS6 for BP3. Again, dominated by the comparatively large doublet Yukawa coupling of $(\y_2)_{12} \approx 0.876$, the numbers for $\xt$ (containing $\sim66\%$ doublet) are almost twice of those for $\xo$. This approximate 1:2 ratio is maintained by $\xo$ and $\xt$, irrespective of the centre-of-mass energy. Considering the tiny contamination from $\rtt$ pair production events, for BP3 this finalstate topology can act as a distinguisher in favour of the mixed leptoquarks $\xot$, against the pure doublet. At the 14 TeV LHC, this probe is achieved with a significance of $7.82\sigma$, for 3000 $\text{fb}^{-1}$ luminosity. The $5\sigma$ signal strength is achievable at $1224.22 \text{fb}^{-1}$ luminosity. Moving to the higher centre-of-mass energy of 27 TeV with a luminosity of 1000 $\text{fb}^{-1}$, the signal significance increases to $27.52\sigma$, while at 100 TeV FCC with 100 $\text{fb}^{-1}$ of luminosity, the FS6 shows a $85.51\sigma$ strength of signal. In both of these higher-energy LHC/FCC, the $5\sigma$ probe is shown to be obtained with very early data. 

The study of these four finalstate event numbers can thus provide a way to discern the singlet and doublet leptoquarks in different benchmark scenarios of the Yukawa coupling structures. The difference in decay modes between singlet and doublet states as shown in \autoref{tab:pairprodbr} can also help us reconstruct the invariant mass of these leptoquarks, which we will discuss in detail in the next subsection.

\subsection{Invariant mass distributions}
\label{sec:imr}

One peculiar characteristic of a leptoquark, by definition, is that it can decay to a quark and a lepton, after production. The quark eventually leads to a hadronic jet at the LHC, and the invariant mass of the jet-lepton pair can be used to reconstruct the leptoquark resonance. However, the resolutions of high-momentum jets and leptons are reported to be low, hampering a high-precision reconstruction of the leptoquark mass peaks for the time being. The CMS detector currently reports $\sim5\%$ resolution for $p_T^{\text{jet}} \geq 200.0$ GeV \cite{CMS:2016lmd}. While a resolution of 3 GeV is achieved for muons at the $Z$-boson peak, the resolutions for muons in general with $p_T^{\mu} \geq 200.0$ GeV may vary from $\sim3\%$ to $\sim5\%$ \cite{CMS:2019ied}. This leads to the current reports of CMS searches for leptoquarks using 100 GeV invariant mass bins for a jet-lepton pair \cite{CMS:2018ncu, CMS:2018lab}. Nevertheless, in this work we are using a more optimistic scenario where the bin widths for such distributions are taken as 10 GeV. 

Across our three benchmark scenarios, the leptoquark masses remain the same, with the values of 1.502 TeV, 1.499 TeV, and 1.506 TeV, respectively for $\rtt, \, \xo$ and $\xt$. Such nearly degenerate mass resonances are very challenging to distinguish in the same finalstate, due to the high resolution that is required. However, each of the BPs provide us with a finalstate that is different for the pure state $\rtt$ with respect to the mixed states $\xot$ as seen in \autoref{tab:pairprodbr}, and we will be focusing on these finalstates in this subsection. For BP1, we can look for the invariant mass distribution of one $b$-jet and one electron, in a finalstate of two $b$-jets and a pair of OS$e$, without any light, $c$- or $\tau$-tagged jets, in which we expect to obtain a clean peak for $\rtt$, considering the SM backgrounds as well as the combined events from $\rtt$ and $\xot$ pair production. For BP2, a distinct peak for $\rtt$ is expected from a two light jets and two OS$e$ in the finalstate, whereas for BP3 we will be looking at a finalstate of two light jets and two OS$\mu$. It is important to note that, for each of these cases, different combinations of jet-lepton are possible, and only the correct combination will lead to the mass peak. Additionally, tagging the exact charge of the lepton can help us obtain the peak specific to $\rtt$ or $\widetilde{R}_2^{-2/3}$. In each case, a cut of $p_T^{\ell} \geq$ 300.0 GeV is imposed, in order to filter out the majority of backgrounds.

\begin{figure}[h]
%	\makebox[\linewidth][c]{%
		\centering
		\subfigure[]{\includegraphics[width=0.45\linewidth]{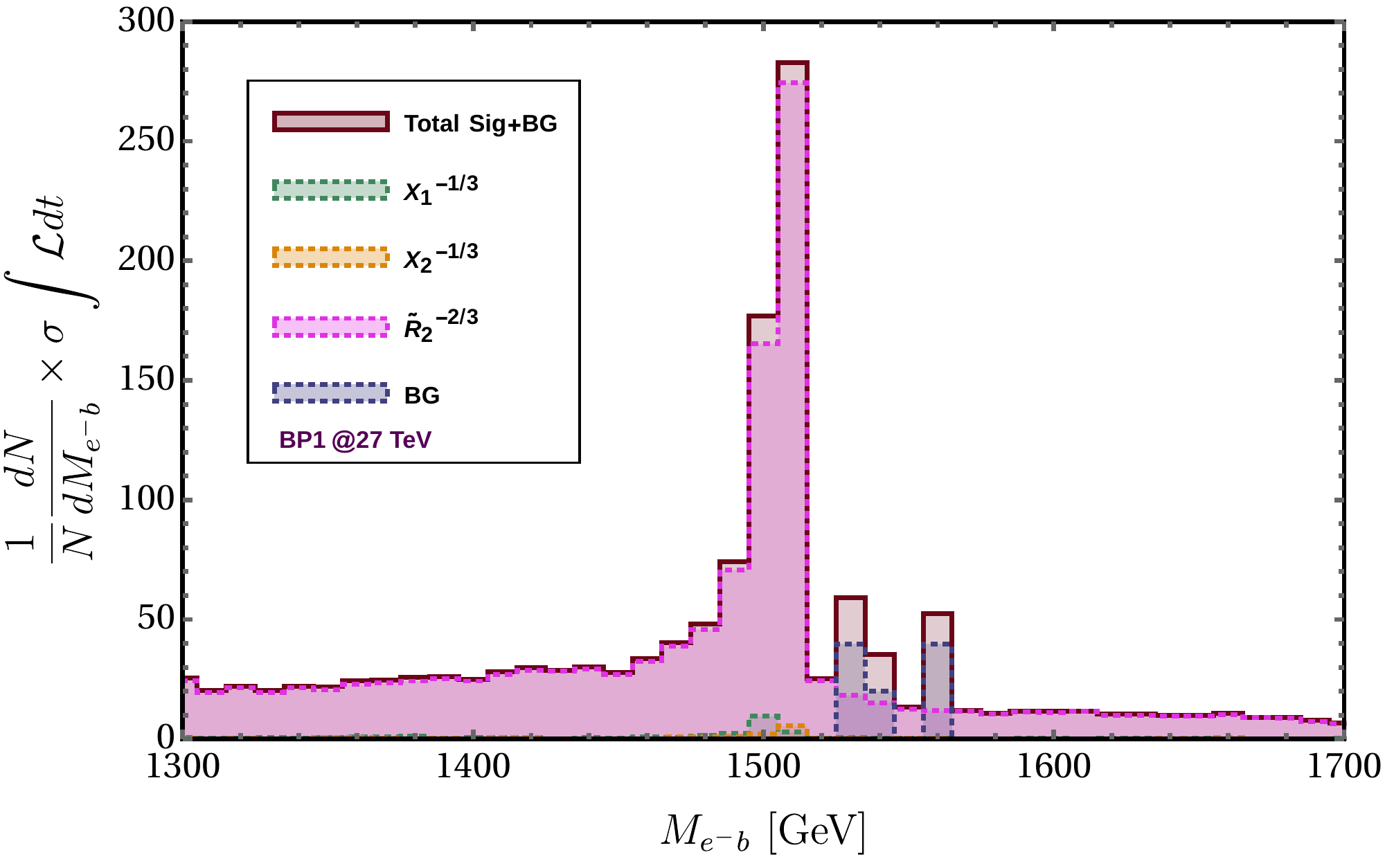}\label{r2m1}}
		\hfil
		\subfigure[]{\includegraphics[width=0.45\linewidth]{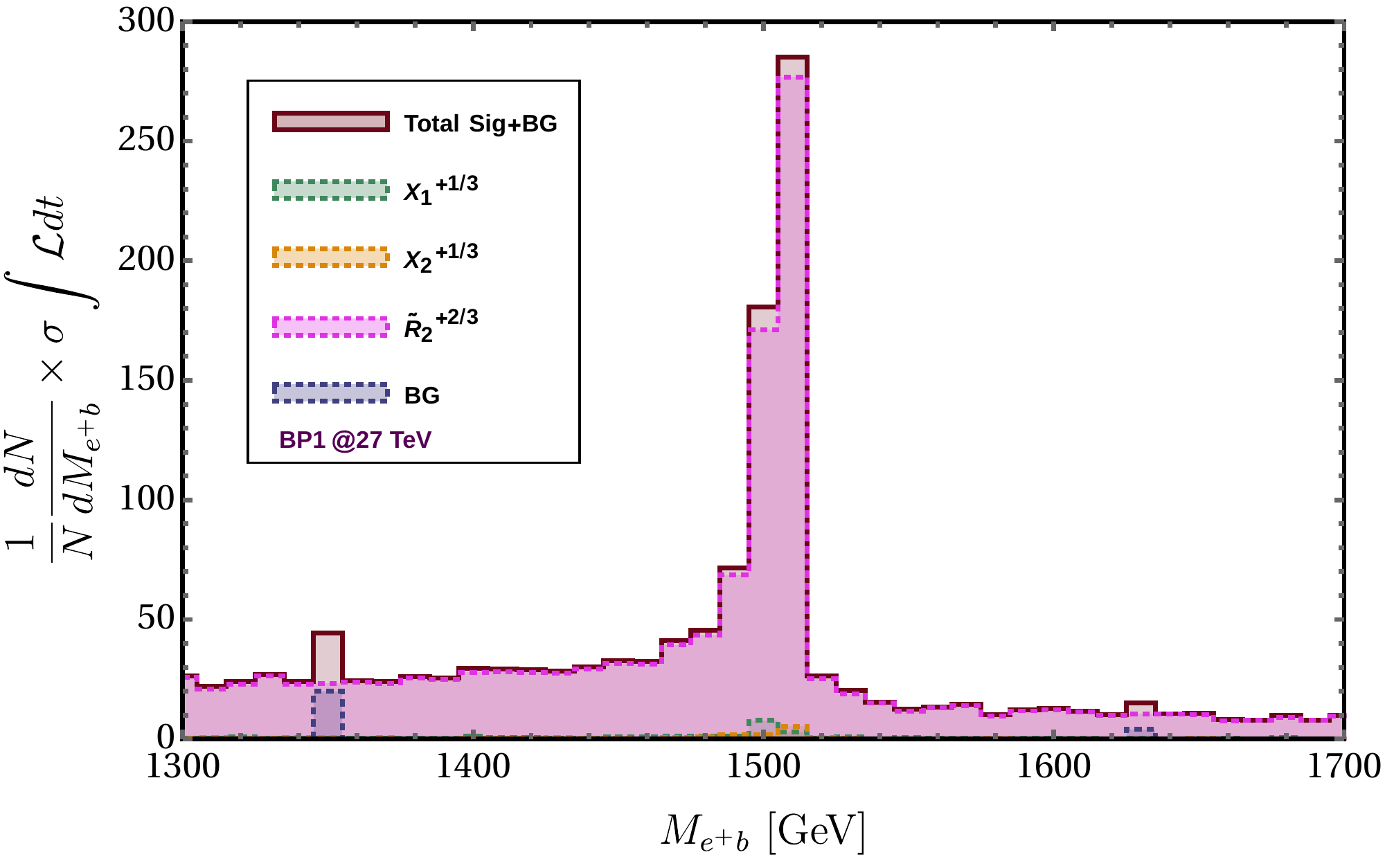}\label{r2p1}}
		
		\subfigure[]{\includegraphics[width=0.45\linewidth]{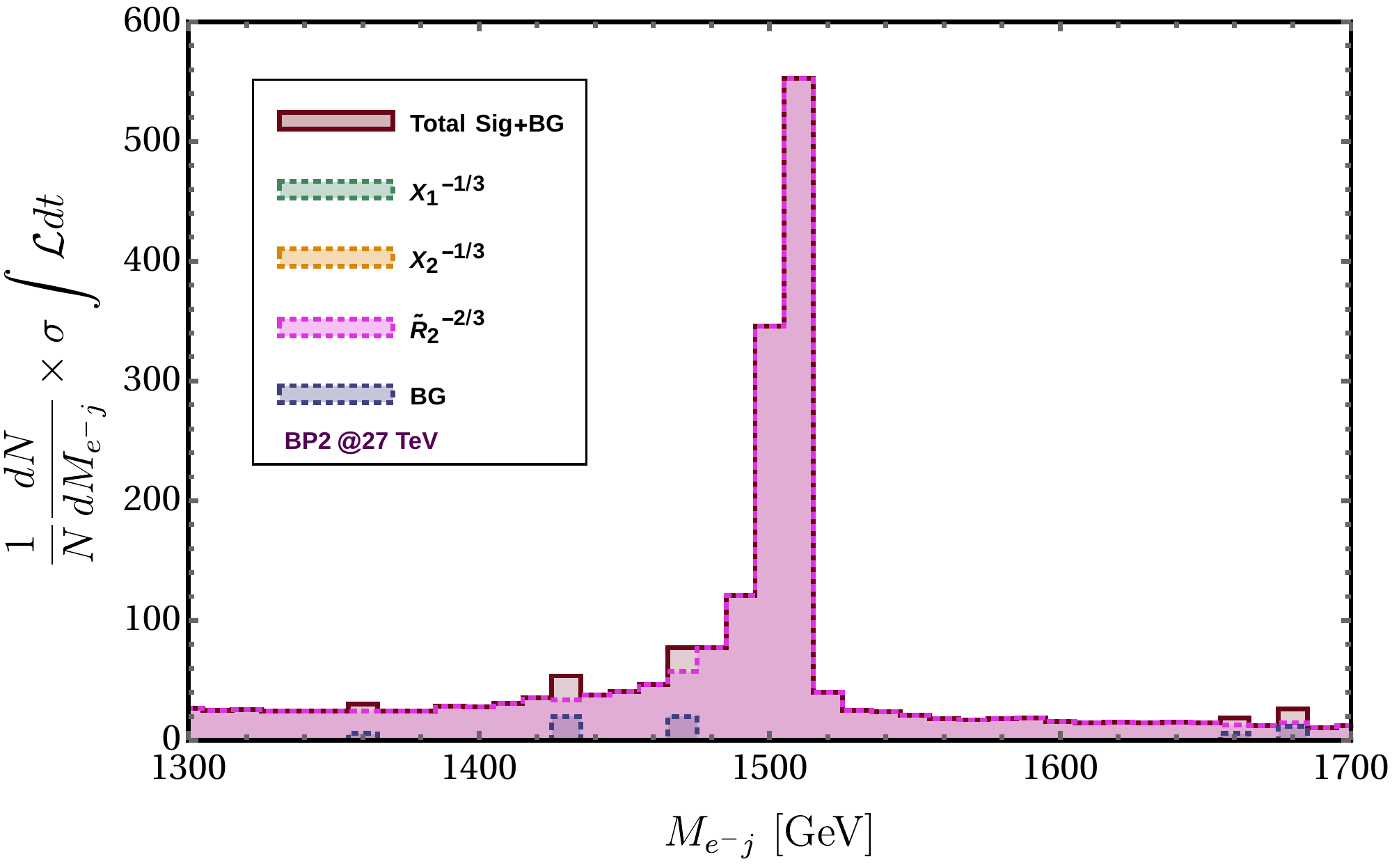}\label{r2m2}}
		\hfil
		\subfigure[]{\includegraphics[width=0.45\linewidth]{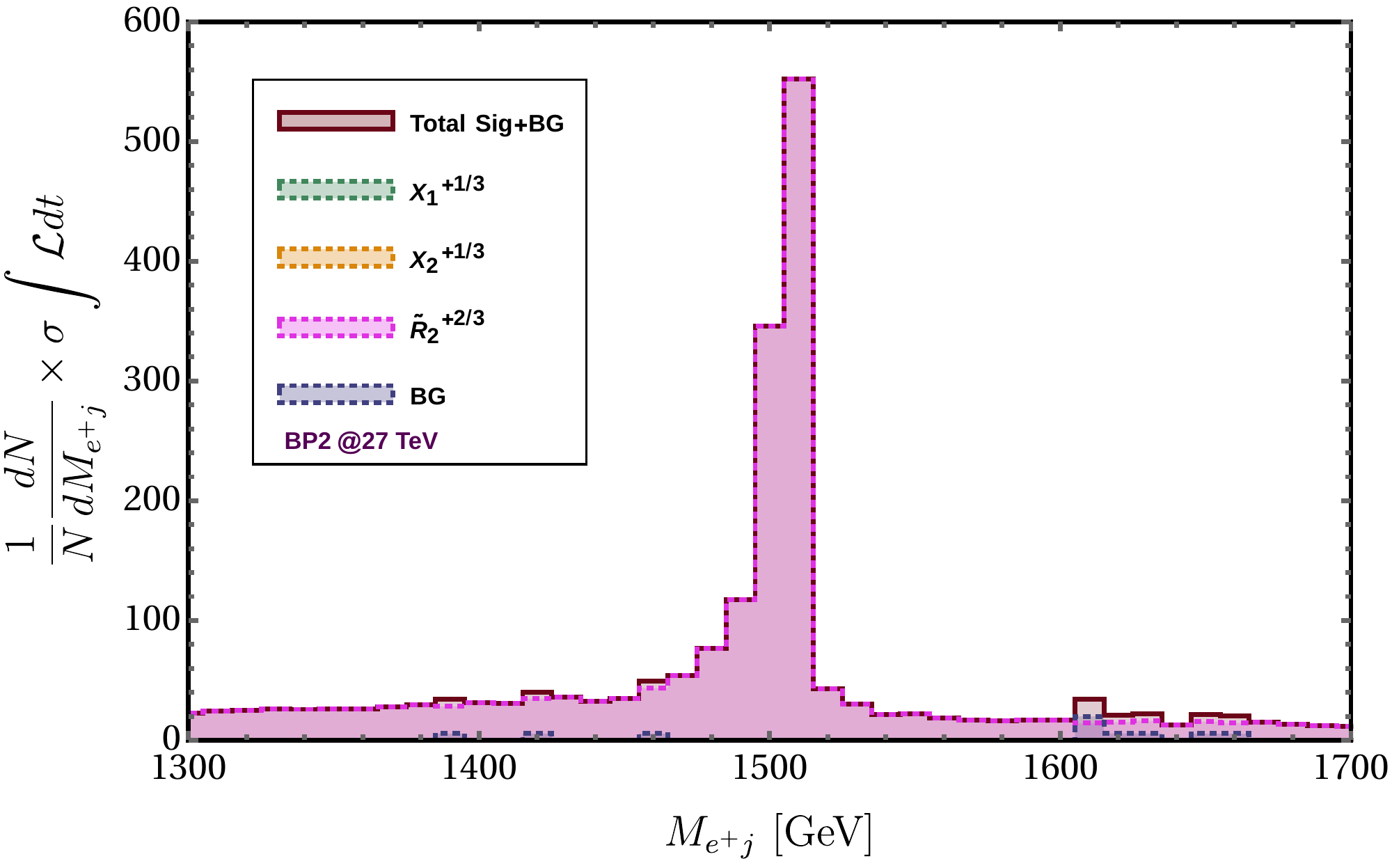}\label{r2p2}}
		
		\subfigure[]{\includegraphics[width=0.45\linewidth]{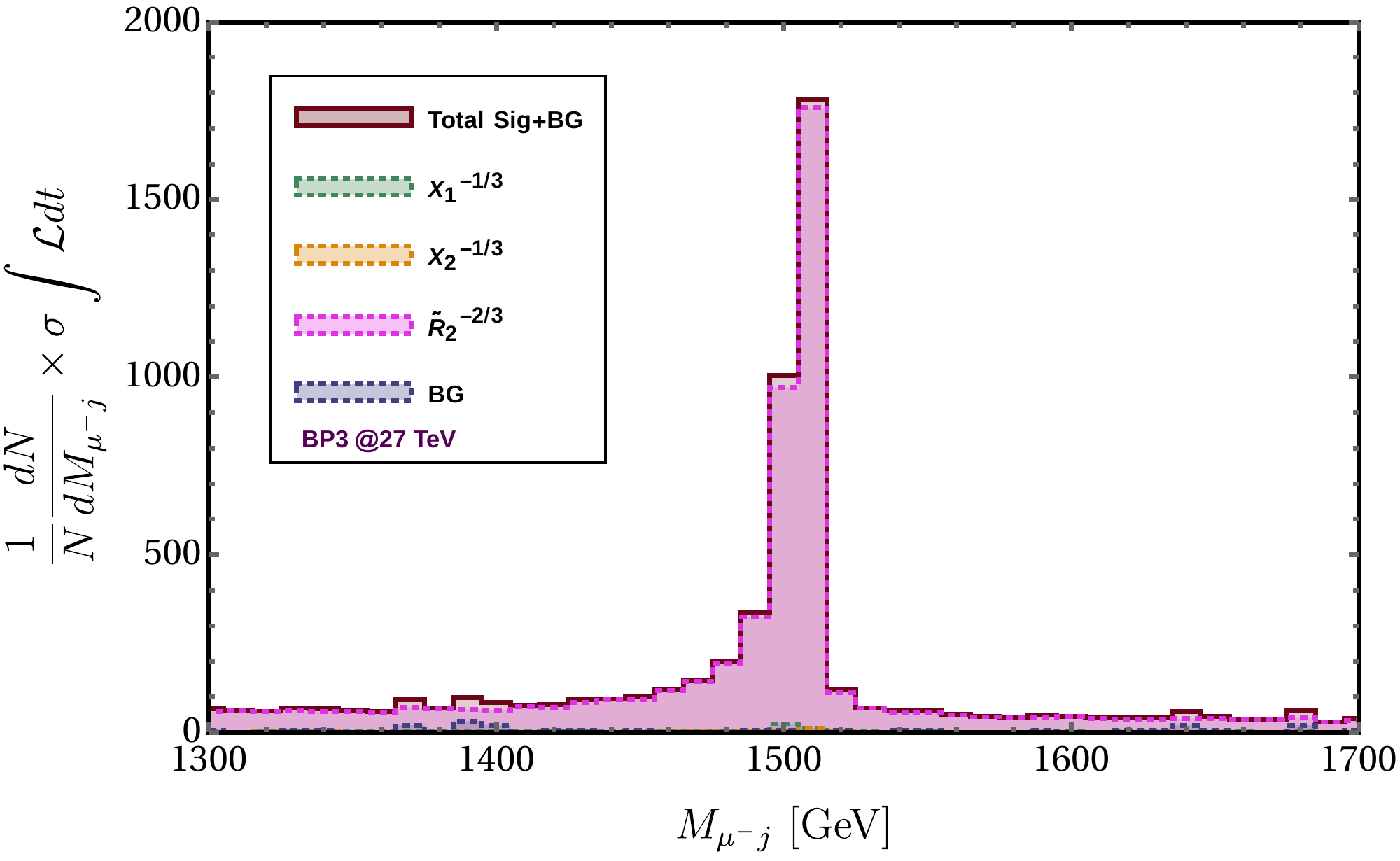}\label{r2m3}}
		\hfil
		\subfigure[]{\includegraphics[width=0.45\linewidth]{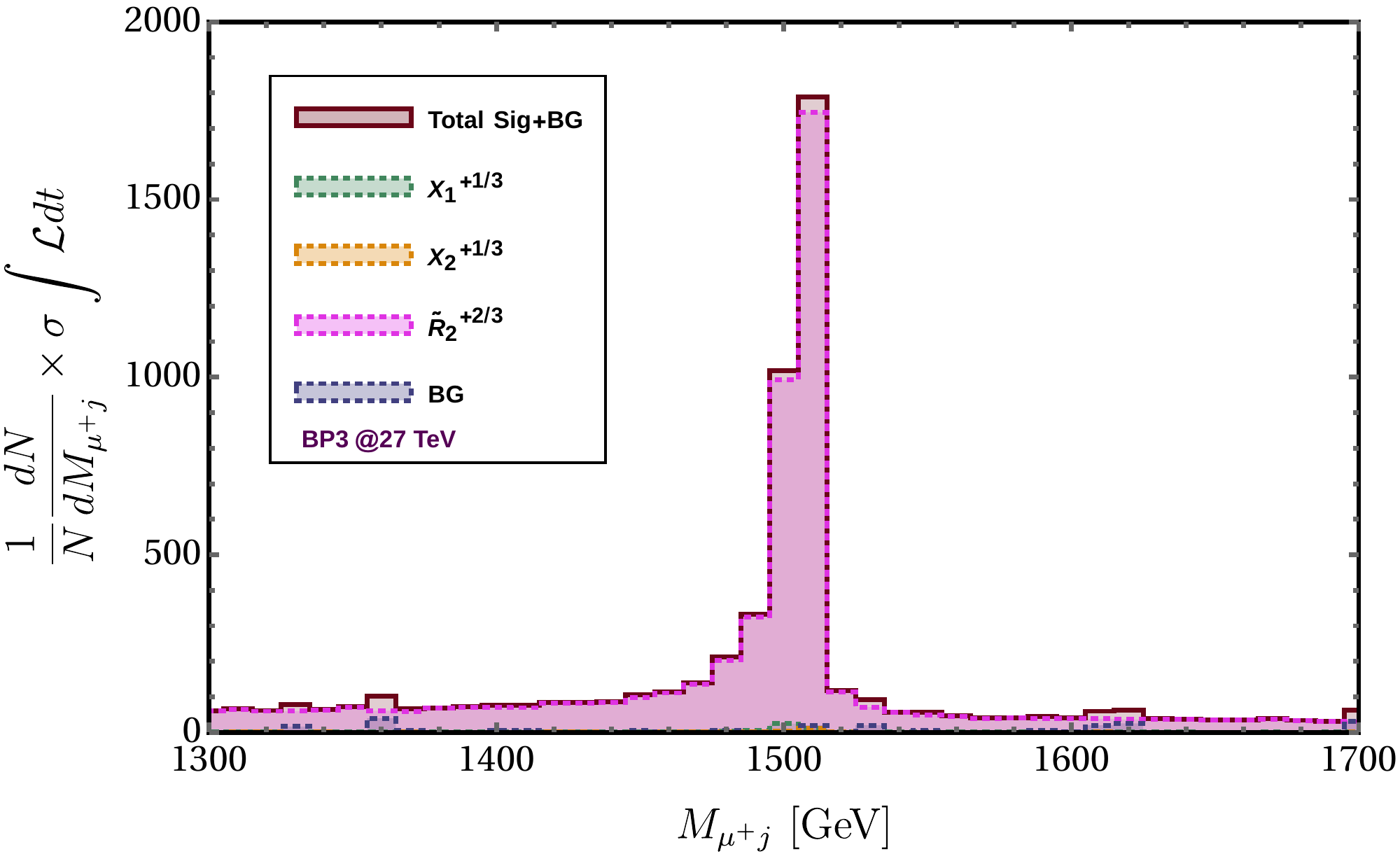}\label{r2p3}}
%	}
	
	\caption{Invariant mass distributions of (a) $e^- b$ and (b) $e^+ b$ in BP1, (c)$e^- j$ and (d) $e^+ j$ in BP2, (e) $\mu^- j$ and (f) $\mu^+ j$ in BP3, from the pair production of leptoquarks at the 27 TeV LHC. }
	\label{fig:imr2}
\end{figure}

In \autoref{fig:imr2} we display the invariant mass distributions from the channels discussed above, simulated at the 27 TeV LHC with 1000 $\text{fb}^{-1}$ of integrated luminosity. In each panel, the dark red distributions show the combined events for $\rtt$ and $\xot$ pair production, summed with the total SM background events in that finalstate. The dotted distributions show the events for each individual pair production of leptoquarks $\rtt$ (pink), $\xo$ (green), and $\xt$ (yellow), along with the SM background (dark blue). \autoref{r2m1} shows the invariant mass distribution of one $e^-$ and one $b$-tagged jet ($M_{e^- b}$) in a 2 $b$-jet + 2 OS$e$ finalstate, obtained from BP1. The absence of $b$-jet events in the finalstates of $\xot$ in this scenario, as well as the demand of high-$p_T$ electrons, almost nullify the contributions from $\xot$ and the SM backgrounds, and the peak obtained at $\sim1500$ GeV can be pinpointed to the pure doublet $\widetilde{R}_2^{-2/3}$. Similarly, in \autoref{r2p1}, the invariant mass of one $e^+$ and one $b$-jet ($M_{e^+ b}$) gives the peak for $\rtt$. In \autoref{r2m2} we observe the invariant mass distribution of one light jet and one $e^-$ ($M_{e^-j}$), for BP2. Here, the lack of di-electron finalstates in $\xot$ pair production leads to almost zero contamination from the mixed states, and the peak at $\sim1500$ GeV again consists almost entirely of events pointing to $\widetilde{R}_2^{-2/3}$. One again, demanding $e^+$ instead of $e^-$ in \autoref{r2p2} leads to a clean peak for $\rtt$. A similar story is observed in \autoref{r2m3}, where the invariant mass distribution of one $\mu^-$ and one light jet ($M_{\mu^- j}$) for BP3 leads to a clean peak at $\sim1500$ GeV for $\widetilde{R}_2^{-2/3}$, with next to zero contribution from $\xot$ and the SM backgrounds, while $M_{\mu^+ j}$ peak consists almost entirely of events from $\rtt$, as seen in \autoref{r2p3}.

To summarize, the benchmark point-specific invariant mass distribution of one lepton and one jet can lead to a distinct resonance peak of $\rtt$ and its antiparticle, untainted by the SM backgrounds as well as events from $\xot$ pair production. This provides us with another way of distinguishing the pure doublet leptoquark from the mixed states. However, the conversation changes when it comes to distinguishing between the mixed states themselves, which will be detailed in the following section.

\section{Distinguishing between $\xo$ and $\xt$}
\label{sec:x1vx2}

In \autoref{sec:rtvx12}, we performed an analysis of finalstates observed in the pair production events of $\rtt$, $\xo$, and $\xt$, through which we can distinguish the pure state $\rtt$ from the mixed states $\xot$. We also successfully obtained $\rtt$ resonance peaks from different jet-lepton invariant mass distributions, for each of the three benchmark points in consideration. The difference in their electromagnetic charge makes it relatively easier to obtain discerning signatures that favour either $\rtt$ or $\xot$, even though they are nearly degenerate in the mass spectrum. However, when it comes to distinguishing between the mixed states $\xo$ and $\xt$, in context of our benchmark points, it becomes more challenging. Especially in BP1 and BP2, where both $\xo$ and $\xt$ have almost the same branching ratios and probable finalstates from their pair production (\autoref{tab:pairprodbr}), it becomes increasingly difficult to obtain a signature where either one of them have negligible contribution compared to the other. In BP3, due to the difference in branching ratios, we observed an event ratio of $\approx 1:2$ in the finalstate FS6, described in \autoref{tab:fs6}. Regardless, for the purpose of distinguishing, the number differences need to be even larger. The large mixing angle $\theta_{LQ} = -0.618$ radians leads to $\sim66.5\%$ singlet and $\sim33.5\%$ doublet content in $\xo$, with the ratios reversed for $\xt$, as discussed in \autoref{sec:bps}. Therefore, even with comparably large entries in $\y_2$ and $\y_1^{L,R}$, there is no significant dominance observed for any one of them in BP3, for either of the two mixed leptoquarks. In the following subsections, we will be discussing some approaches that are viable in this regard.

\subsection{The leptoquark mixing angle}
The mixing between the $Q=1/3$ component of the doublet leptoquark $\widetilde{R}_2$ and the singlet leptoquark $S_1$ influences the discernability of $\xo$ and $\xt$ significantly. As seen from \autoref{eq:tlq}, for fixed values of $m_1,~m_2,~\alpha_1,~\alpha_2$ and $\alpha_3$, the value of $\tlq$ depends only upon the value of the trilinear coupling $\kappa$. Again, \autoref{eq:mass} shows how $\kappa$ affects the masses of $\xot$. Therefore, with the change in $\kappa$, the mass splitting $\Delta M_{21} = M_2 - M_1 $ also changes, which again affects the distinguishability further.
\begin{figure}[h]
	\makebox[\linewidth][c]{%
		\centering
		\subfigure[]{\includegraphics[width=0.45\linewidth]{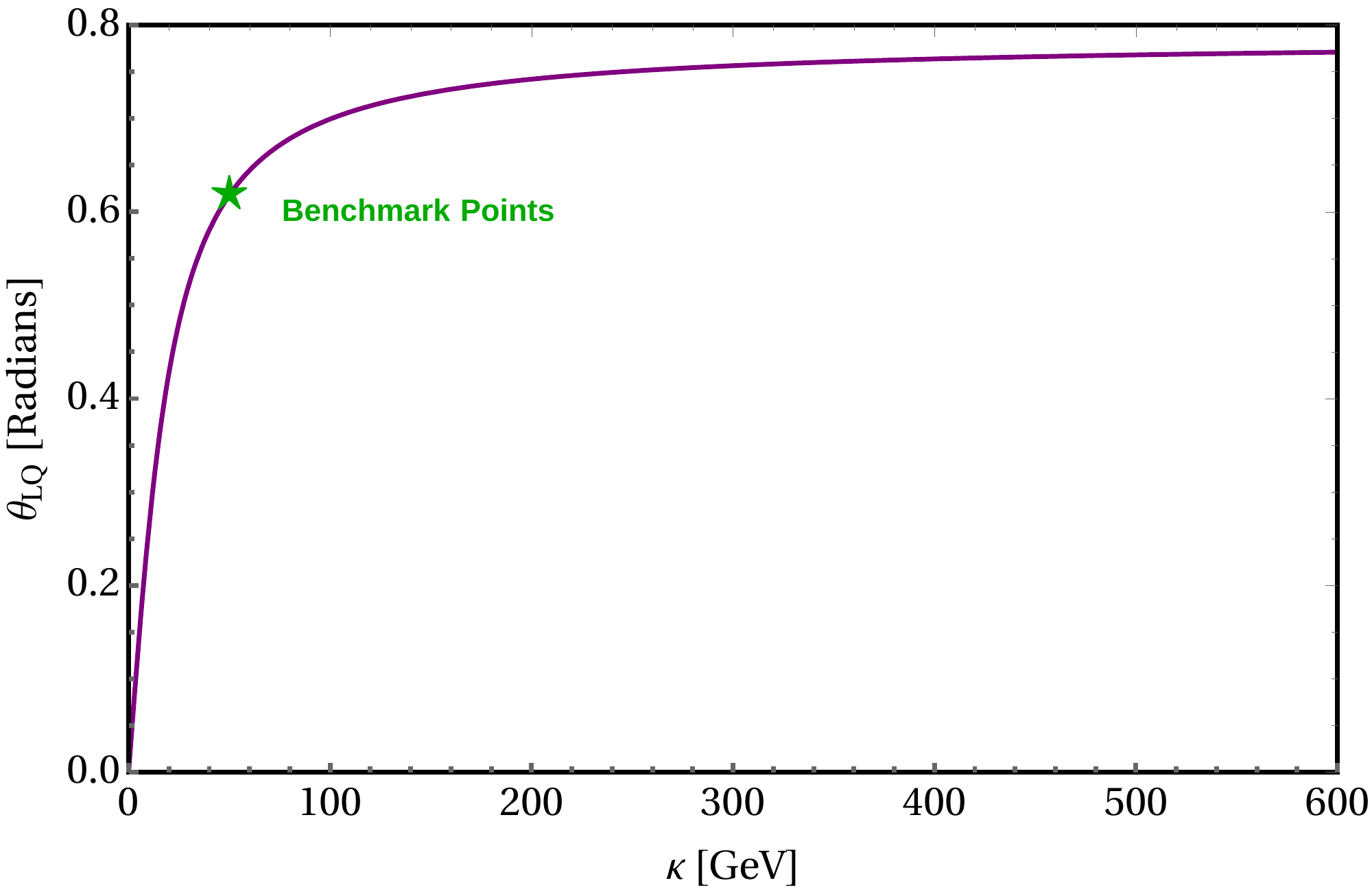}\label{kvtheta}}
		\subfigure[]{\includegraphics[width=0.45\linewidth]{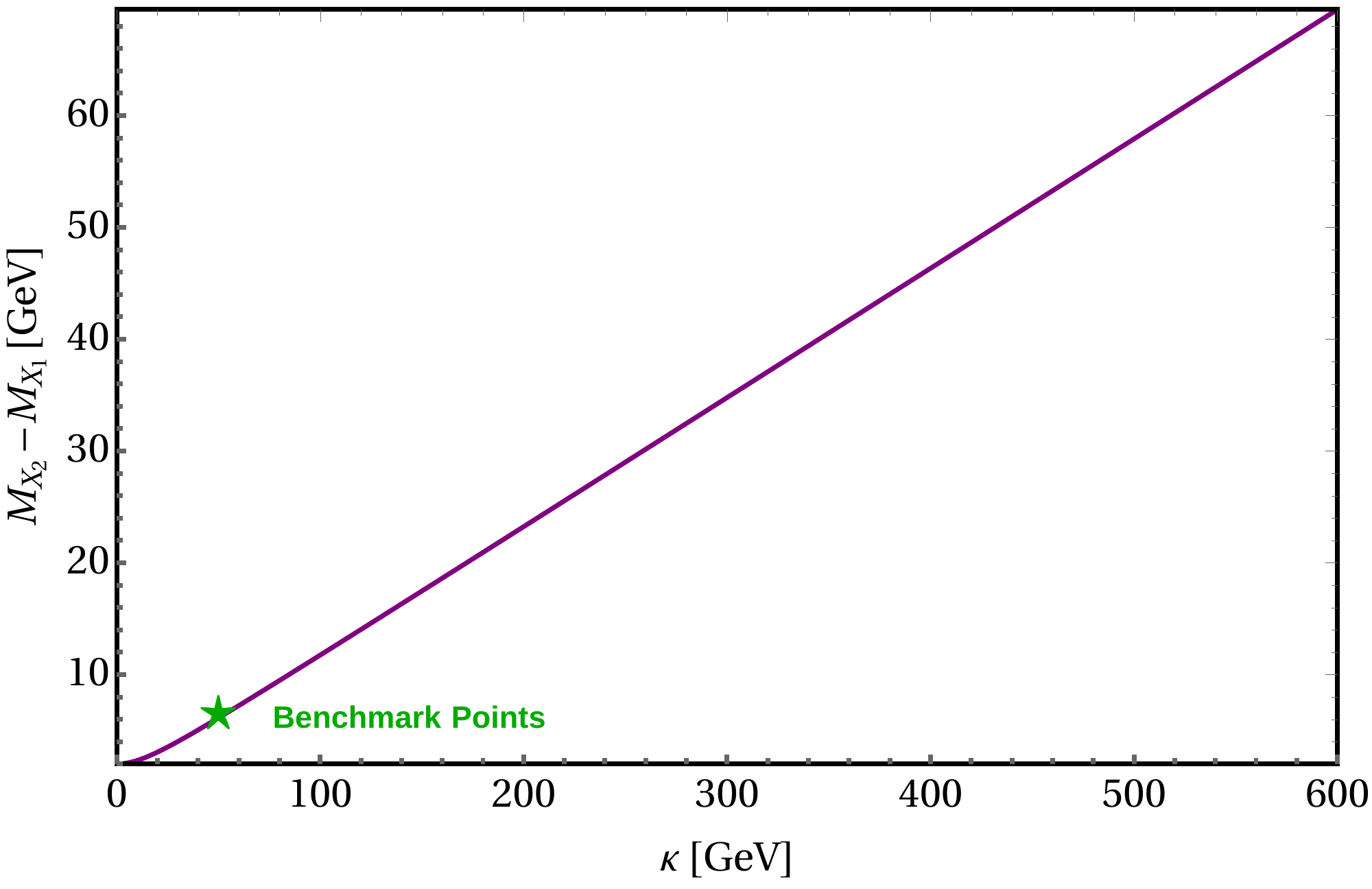}\label{kvmass}}
	}
	
	\caption{Variation of (a) leptoquark mixing angle and (b) $\X_{1,2}$ mass splitting with respect to $\kappa$. The green star corresponds to the benchmark of $\kappa=50$ GeV.}
	\label{fig:kappa_tm}
\end{figure}

\autoref{fig:kappa_tm} illustrates how changes in $\kappa$ affects the mixing angle and the mass splitting, as discussed above. The value of $\theta_{LQ}$ has a fast increase up to $\kappa=100$ GeV, and then it slows down, becoming almost constant over a wide range of $\kappa$, as portrayed by \autoref{kvtheta}. The mass splitting however follows a linear trend, shown in \autoref{kvmass}. In both these plots, our benchmark value of $\kappa=50$ GeV is shown with the green stars. For this value of $\kappa$, we have $\abs{\theta_{LQ}} \approx 0.618$, and $\Delta M_{21} \approx 7$ GeV. If we can determine the mixing angle from a production process at the LHC/FCC, then we are at an advantage when it comes to distinguishing between $\xot$. Theoretically, the difference in doublet content of these two mixed states allow one such production process, which we will discussed in the next subsection.

\subsection{Distinguishing $\xot$ from asymmetric production}
\label{sec:wmed}

\begin{figure}[h!]
	\centering
	%		\tikzfeynmanset{every scalar@@/.style={thick, dashed}, every boson@@/.style={thick, decoration={snake,amplitude=1mm},decorate}, every plain@@/.style={thick}}
	\begin{tikzpicture}
		\begin{feynman}
			\vertex (a1);
			\vertex [above left =1.5cm of a1] (i1){\footnotesize\(\uq\)};
			\vertex [below left =1.5cm of a1] (i2){\footnotesize\(\xbar{\dq}\)};
			\vertex [right =1.2cm of a1] (a2);
			\vertex [above right =1.5cm of a2] (i3){\footnotesize\(\rtt\)};
			\vertex [below right =1.5cm of a2] (i4){\footnotesize\(\xbar\X_{1,2}^{+1/3}\)};
			%			\vertex [above right =1cm of i3] (h1){\footnotesize\(T^0\)};
			%			\vertex [right =1cm of i3] (h2);
			%			\vertex [below =1cm of i4] (h3){\footnotesize\(\textcolor{white}{d}\)}; %dummy vertex%
			%			\vertex [above right =0.5cm of h2] (w1){\footnotesize\(l\)};
			%			\vertex [below right =0.5cm of h2] (w2){\footnotesize\(\bar{\nu}\)};
			%			\vertex [below =1cm of a1] (t){\footnotesize\(\textbf{ITM}\)};
			
			\diagram { (i1)--[fermion](a1)--[boson, edge label ={\footnotesize\(W^+\)}](a2), (i2)--[fermion](a1), (a2)--[scalar](i3), (a2)--[scalar](i4)
			};
		\end{feynman}
	\end{tikzpicture}
	
	\caption{Associated production of $\rtt$ with $\xbar\X_{1,2}^{+1/3}$ via an $s$-channel $W^+$ boson at the LHC.}
	\label{fig:wp}
\end{figure}
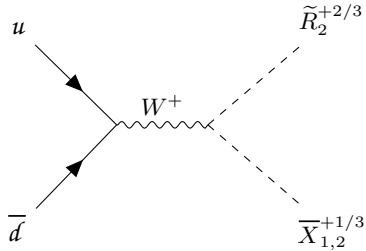

While our previous discussion dealt with the phenomenology of leptoquark pair production, in this subsection we present a possibility of distinguishing between the mixed leptoquark states with the help of the asymmetric production process of $q\bar{q'} \to \rtt \xbar\X_{1,2}^{+1/3}$. This production is facilitated by the $s$-channel exchange of a $W^\pm$ gauge boson, as shown in \autoref{fig:wp}. As only the doublet components of $\xot$ couple to the $W^\pm$ boson, the parameter $\kappa$ and subsequently the mixing angle $\theta_{LQ}$ plays a role in these production cross-sections. 

\begin{figure}[h]
	\makebox[\linewidth][c]{%
		\centering
		\subfigure[]{\includegraphics[width=0.35\linewidth]{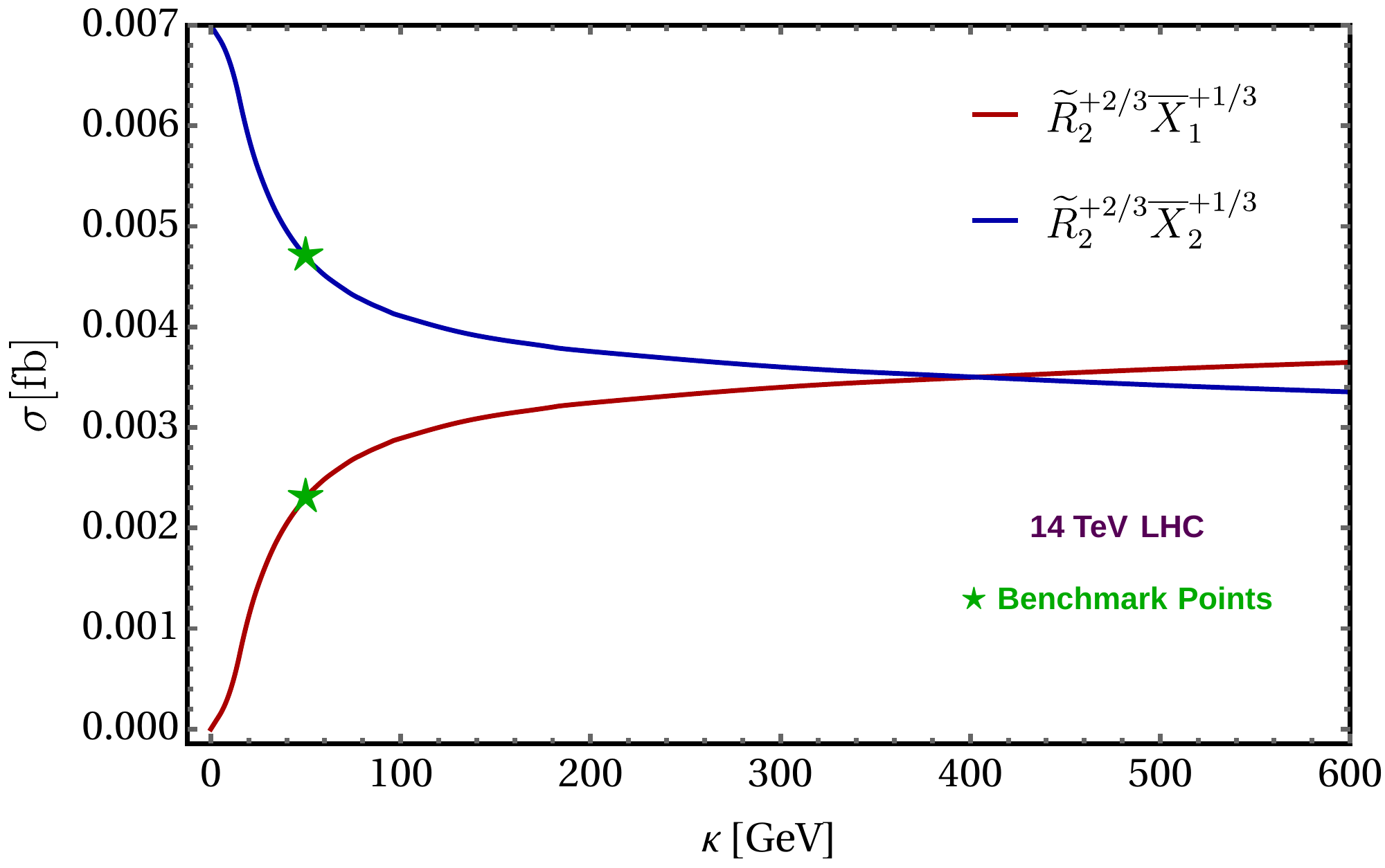}\label{k14}}
		\subfigure[]{\includegraphics[width=0.35\linewidth]{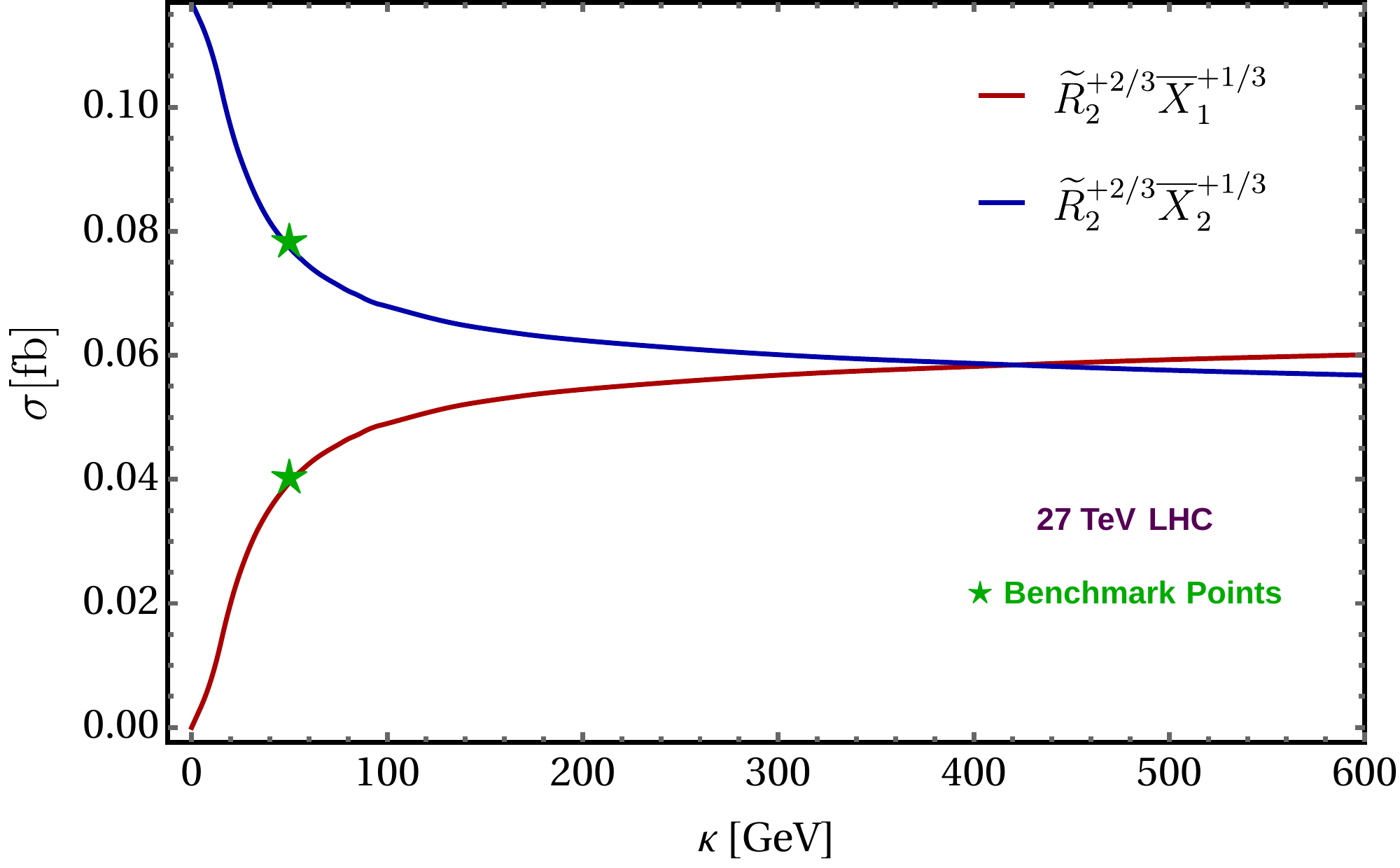}\label{k27}}
		\subfigure[]{\includegraphics[width=0.35\linewidth]{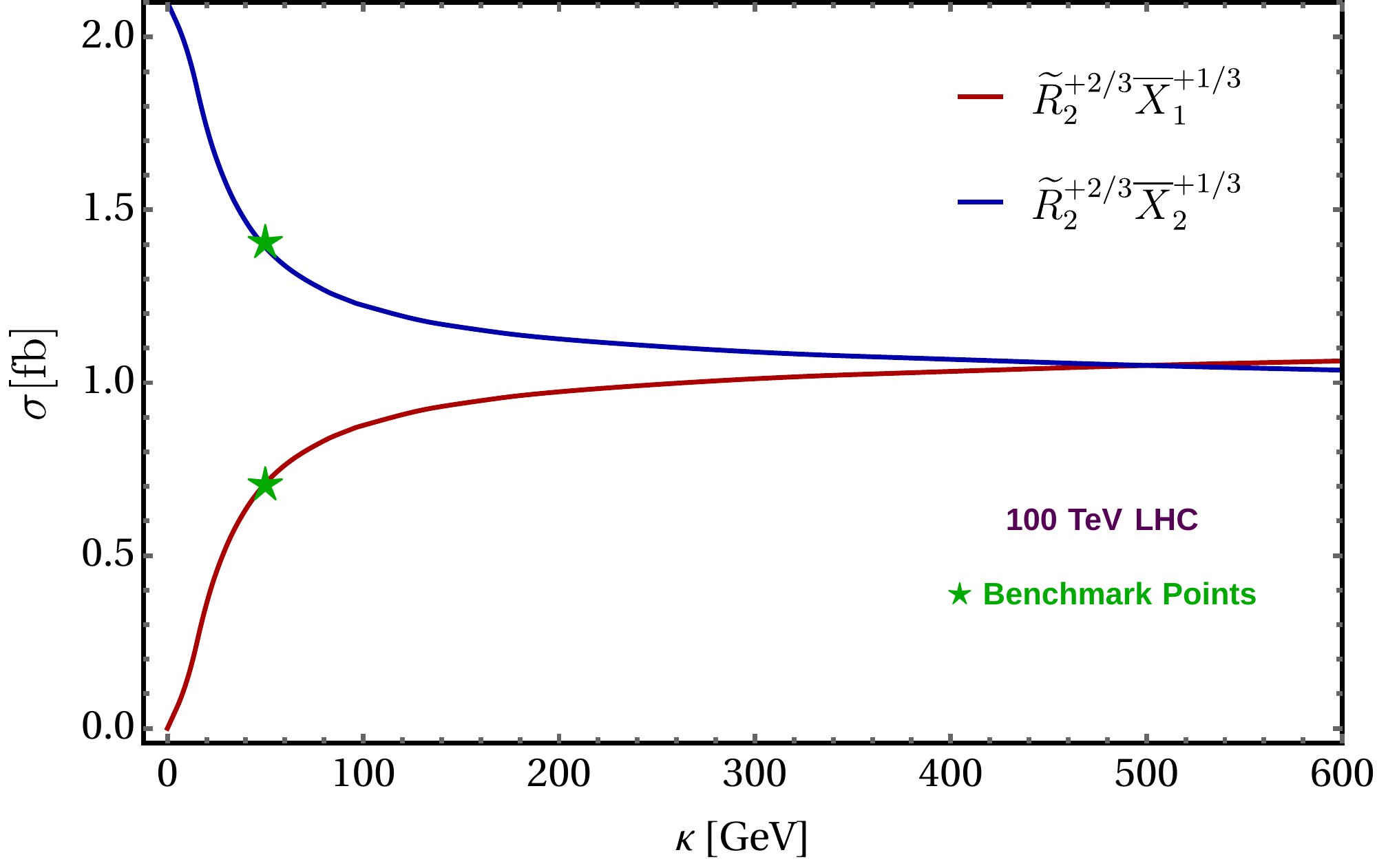}\label{k100}}
	}
	
	\makebox[\linewidth][c]{%
		\centering
		\subfigure[]{\includegraphics[width=0.35\linewidth]{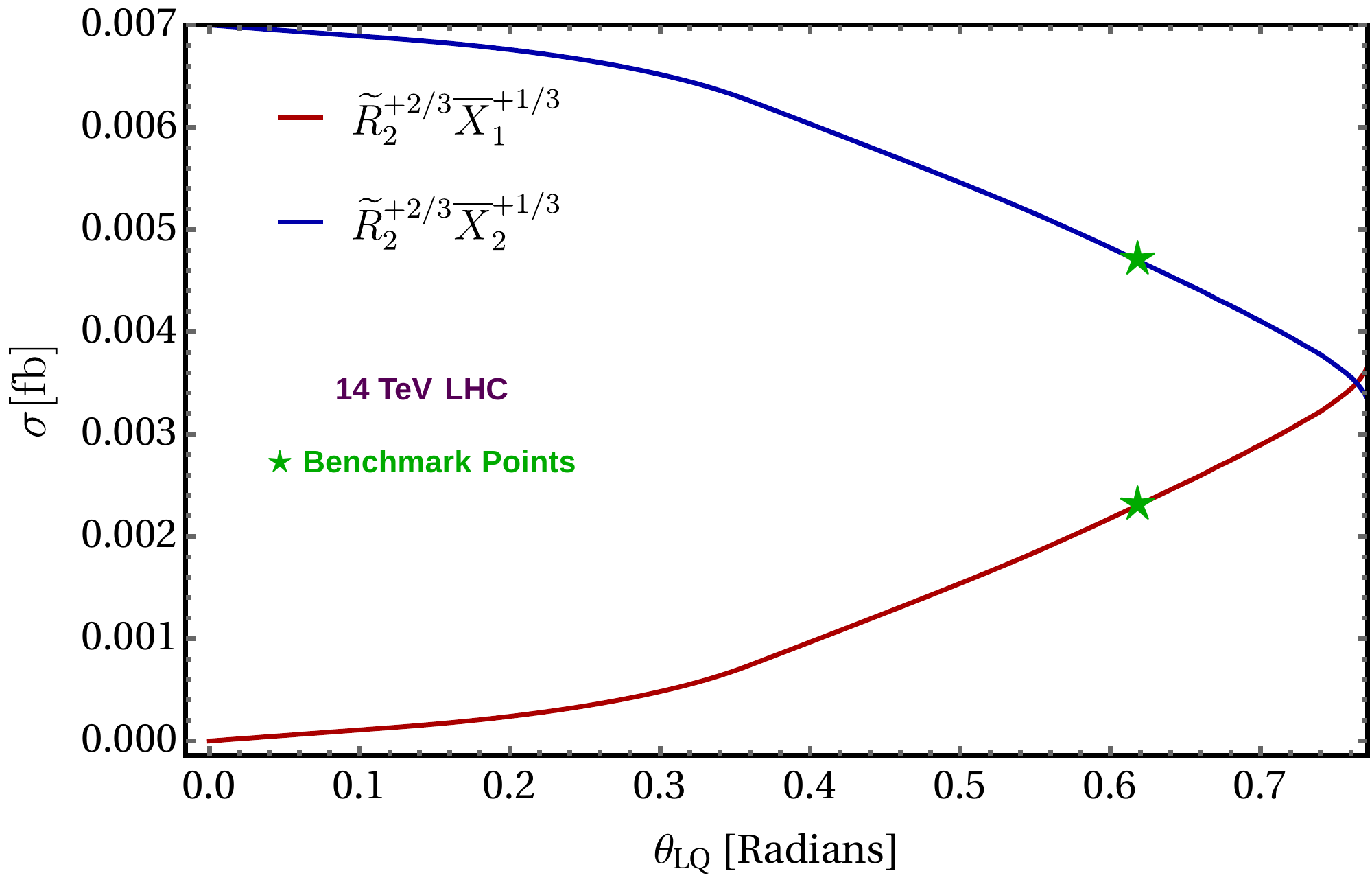}\label{t14}}
		\subfigure[]{\includegraphics[width=0.35\linewidth]{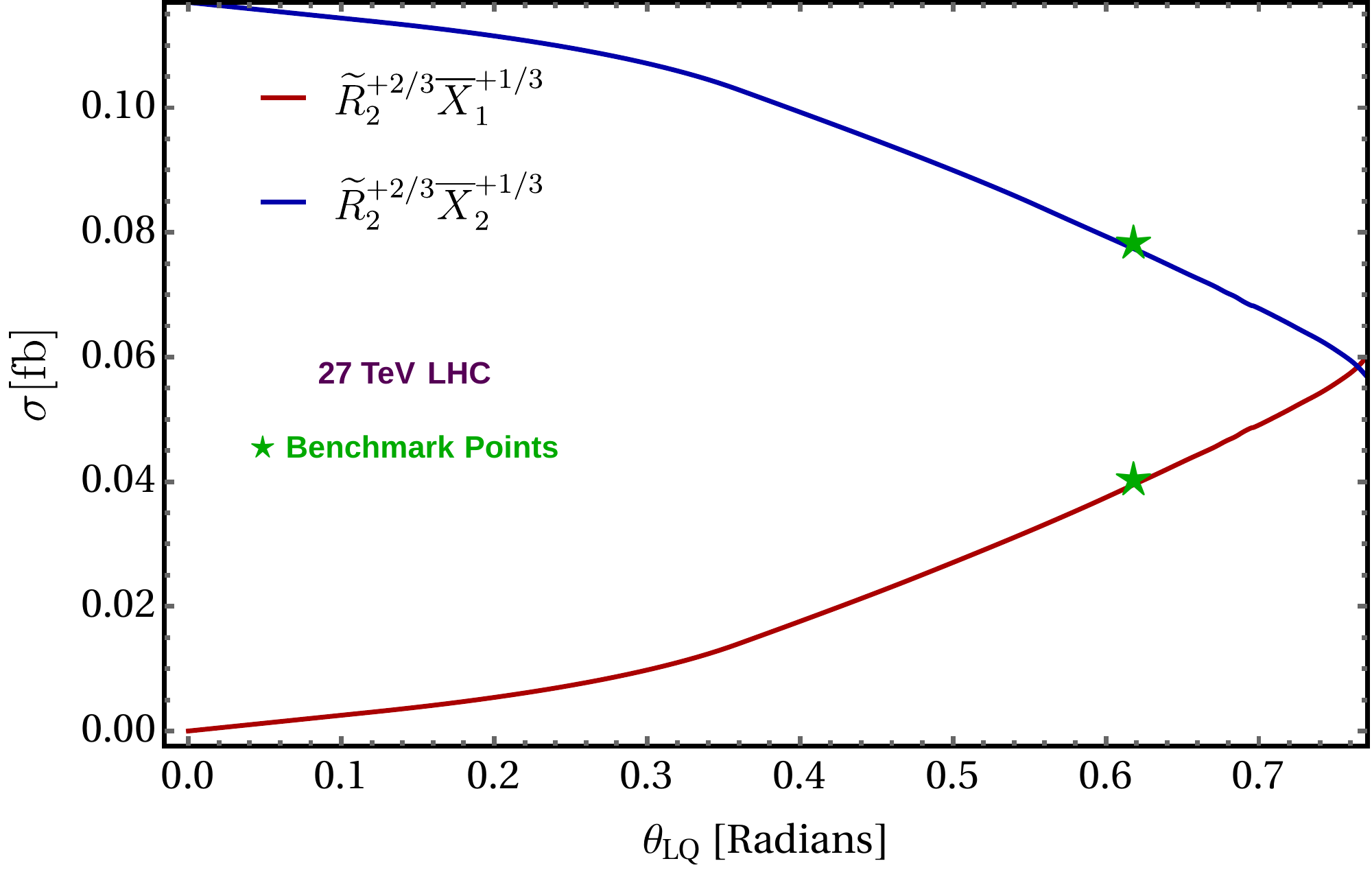}\label{t27}}
		\subfigure[]{\includegraphics[width=0.35\linewidth]{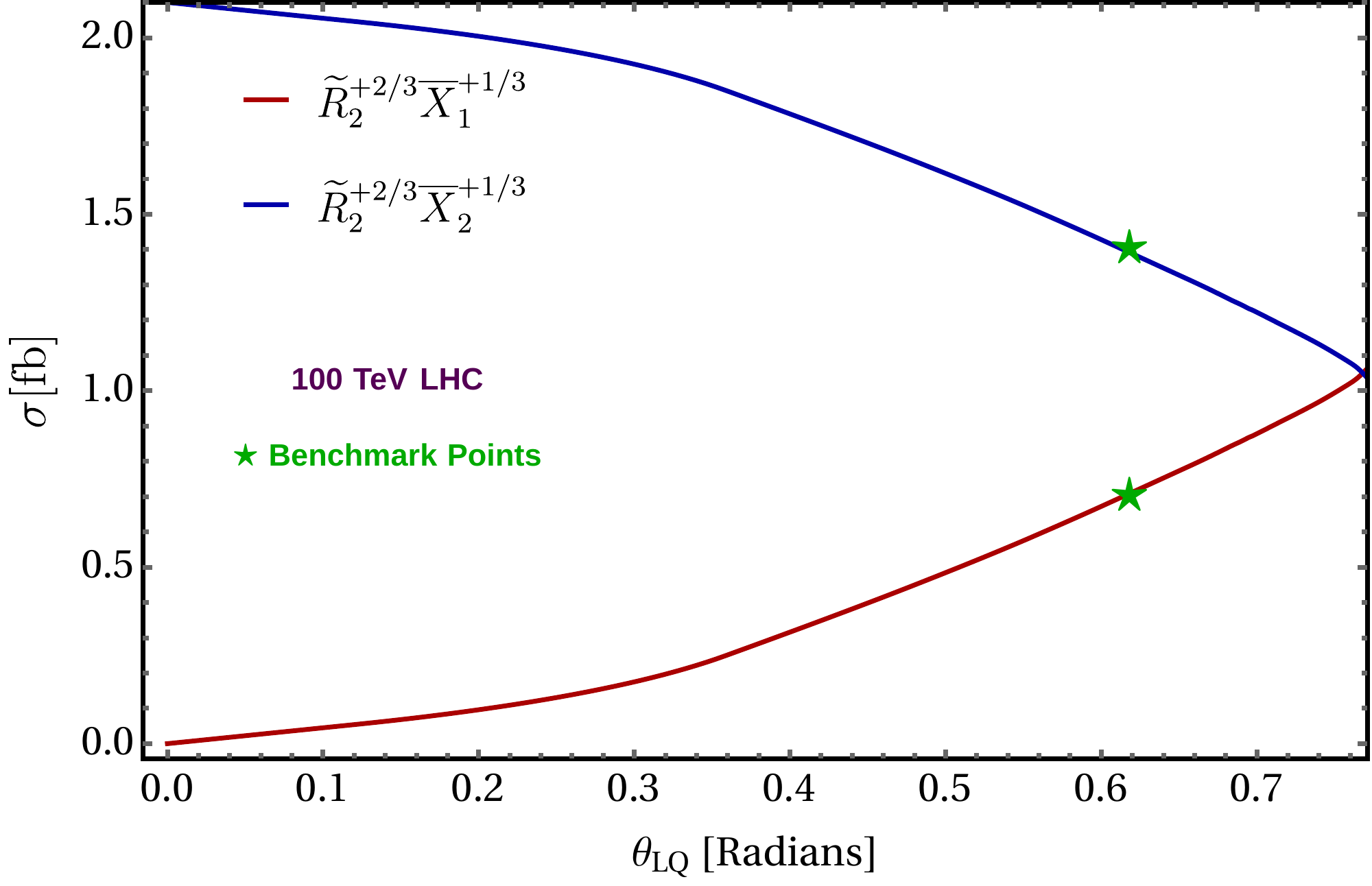}\label{t100}}	
	}

	\caption{The variation of $\sigma_{\widetilde{R}_2^{+2/3} \xbar\X_{1,2}^{+1/3}}$ with respect to $\kappa$ (Figures \ref{k14}-\ref{k100}), and with respect to $\theta_{LQ}$ (Figures \ref{t14}-\ref{t100}) for three different $E_{CM}$s of 14, 27 and 100 TeV. In each plot, the red line corresponds to $\X_1$ and the blue line to $\X_2$. The green star shows our choice of $\kappa=50$ GeV and $\theta_{LQ} = 0.617998$ for the BPs.}
	\label{fig:kthetavcs}
\end{figure}

Figures \ref{k14}-\ref{k100} show the $\kappa$-dependence of the cross-sections at  energies of 14, 27 and 100 TeV, respectively, where the red line corresponds to $\rtt \overline{\X}_{1}^{+1/3}$ and the blue line to $\rtt \overline{\X}_{2}^{+1/3}$ production processes. We see that, when $\kappa = 0$ GeV, the cross-section involving $\xo$ remains zero, as there is zero doublet component in this. With the increase in $\kappa$, the $\xo$ cross-section increases while the $\xt$ cross-section comes down. Figures \ref{t14}-\ref{t100} show the same variation of the two production cross-sections, with respect to the leptoquark mixing angle $\theta_{LQ}$. From both these plots, we observe that even though $M_{2} > M_{1}$, the production cross-section of $\widetilde{R}_2^{+2/3} \xbar\X_{2}^{+1/3}$ stays higher than that of $\widetilde{R}_2^{+2/3} \xbar\X_{1}^{+1/3}$ upto a certain value of $\kappa$ or $\theta_{LQ}$. The point at which these two cross-sections overlap, is different for different centre-of-mass energies. Before reaching the intersection point, the difference in cross-section is dominated by the amount of doublet/singlet percentage in the leptoquark. At the intersection point, the effect of the mass gap cancels out that of $\tlq$. As the mass splitting increases, the availability of phase space becomes the dominant factor, and so after that point the production process involving the lighter of the two mixed leptoquarks has higher cross-section than that of the heavier leptoquark. An important observation here is that, the point of intersection corresponds to a smaller $\kappa$ (or $\theta_{LQ}$) for a lesser $E_{CM}$. The crossover happens at $\kappa$ ($\theta_{LQ}$) values of $\sim \, 400$ GeV (0.7636 rad.), $\sim$ 420 GeV (0.7647 rad.) and $\sim 500$ GeV ( 0.7680 rad.), for centre-of-mass energies of 14, 27 and 100 TeV, respectively. These threshold values increase with $E_{CM}$ due to the enhancement in availability of the phase space with $E_{CM}$. Therefore, at higher energies, the effect of mixing angle can be observed for a longer range compared to lower energies. Within this range, the measured values of cross-sections at the same finalstate can point towards the leptoquark with either more singlet content, or more doublet.

\begin{table}[h]
	\centering
	\renewcommand{\arraystretch}{1}
	\begin{tabular}{|M{0.8cm}|M{1.6cm}|M{1.6cm}|M{1.6cm}|M{1.6cm}|M{1.6cm}|M{1.6cm}|}
		\hline
		\multirow{3}{*}{BPs}&\multicolumn{6}{c|}{$pp\to W^\pm \to LQ_{1} \overline{LQ_{2}}$ at different $E_{CM}$s}\\
		\cline{2-7}
		&\multicolumn{3}{c|}{$\sigma_{\widetilde{R}_2^{+2/3} \xbar\X_{1}^{+1/3}}$ (fb)}&\multicolumn{3}{c|}{$\sigma_{\widetilde{R}_2^{+2/3} \xbar\X_{2}^{+1/3}}$ (fb)}\\
		\cline{2-7}
		&14 TeV & 27 TeV & 100 TeV 	&14 TeV & 27 TeV & 100 TeV 	\\
		\hline
		BP 1-3 & 0.002 & 0.039&0.71& 0.004 & 0.078 & 1.40\\
		\hline
		%		BP2 &&&&&&\\
		%		\hline
		%		BP3 & &&&&&\\
		%		\hline
		%			BP2 &0.39&10.28&657.6&0.35&10.33&660.9&0.35&10.08&650.5\\
		%			\hline
	\end{tabular}
	
	\caption{Associated production cross-sections of $\rtt \xbar\X_{1,2}^{+1/3}$ at LO via $s$-channel $W^+$-boson exchange for all three BPs, and three different $E_{CM}$s of 14, 27 and 100 TeV at the LHC.}
	\label{tab:assoc}
\end{table}

In \autoref{tab:assoc} the LO cross-sections for the production of $\rtt \overline X_1^{+1/3}$ and $\rtt \overline X_2^{+1/3}$ are presented for three centre-of-mass energies of 14, 27, and 100 TeV, at the LHC/FCC. These cross-sections are independent of the choice of Yukawa couplings, and so they remain the same in each case for all the three BPs. Naturally, as $\xt$ contain twice as much doublet percentage as $\xo$ ($\sim 66.5\%$ and $\sim 33.5\%$), the value of $\sigma_{\widetilde{R}_2^{+2/3} \xbar\X_{2}^{+1/3}}$ is almost two times that of $\sigma_{\widetilde{R}_2^{+2/3} \xbar\X_{1}^{+1/3}}$, in each centre-of-mass energy. This difference in cross-sections between $\xo$ and $\xt$ can act as a probe of the mixing angle $\theta_{LQ}$ itself. The cross-sections are, however, very low as the masses of the leptoquarks are high and a massive gauge boson mediates the process. Such cross-sections will not lead us to a number of observed events that is large enough to draw conclusive remarks from it, from an experimental perspective. At the HL-LHC limit of 3000 $\text{fb}^{-1}$ luminosity, the production processes at 14 TeV will lead to a total of less than fifteen observed events, for both $\xo$ and $\xt$. At 27 TeV, with 1000 $\text{fb}^{-1}$ luminosity, the observed events still remain low, with less than eighty events for both the leptoquarks. At the 100 TeV FCC, the cross-sections predict the observation of less than hundred and ten events for $\xo$ and $\xt$, with 100 $\text{fb}^{-1}$ of integrated luminosity. These observed event numbers are too small for a proper analysis, and with advanced kinematic and phase space cuts, they will decrease even further. Thus, while being theoretically sound, this production mode is not experimentally viable, in probing the leptoquark mixing angle, as well as in distinguishing between $\xo$ and $\xt$, and one needs to wait for higher energy and higher integrated luminosity.

\subsection{Distinguishing $\xot$ from pair production}
The experimental impracticality of the asymmetric pair production process in discerning between the mixed leptoquarks leads us back to the discussion on the pair production processes. In the preceding analysis we have observed that, for the chosen set of mass, mixing and Yukawa couplings, obtaining a clear signature that pinpoints towards either $\xo$ or $\xt$ is extremely challenging from the pair production processes. In the following subsections, a few possible approaches for resolving such a near-degenerate pair of mixed leptoquarks are presented.

\subsubsection{Invariant mass distribution}
\label{sec:imx1x2}

In \autoref{sec:imr}, we have already discussed the challenges when it comes to resolving reconstructed invariant mass peaks of leptoquark resonances. As our benchmark points allow different finalstates for $\rtt$ and $\xot$, we could perform reconstructions of the $\rtt$ leptoquark from different finalstates involving two jets and two leptons, with negligible contaminations from the SM backgrounds as well as from $\xot$, as shown in \autoref{fig:imr2} The situation becomes tricky when we want to do the same for $\xo$ and $\xt$ individually. In our BPs, the mass splitting between these two states is just 7 GeV, which require high-precision resolution of the reconstructed peaks. However, as mentioned in \autoref{sec:imr}, this precision cannot be obtained with the current LHC resolutions of high-momentum jets and leptons that need to be used for the reconstruction. To properly resolve between two states separated by a mass of 7 GeV, we would require $\sim2$ GeV of bin width, which is far too optimistic in the current context. Nevertheless, just like for $\rtt$, we present some invariant mass distributions at the 27 TeV LHC with 10 GeV bin widths, to explain the situation better. 

\begin{figure}[h]
%	\makebox[\linewidth][c]{%
		\centering
		\subfigure[]{\includegraphics[width=0.45\linewidth]{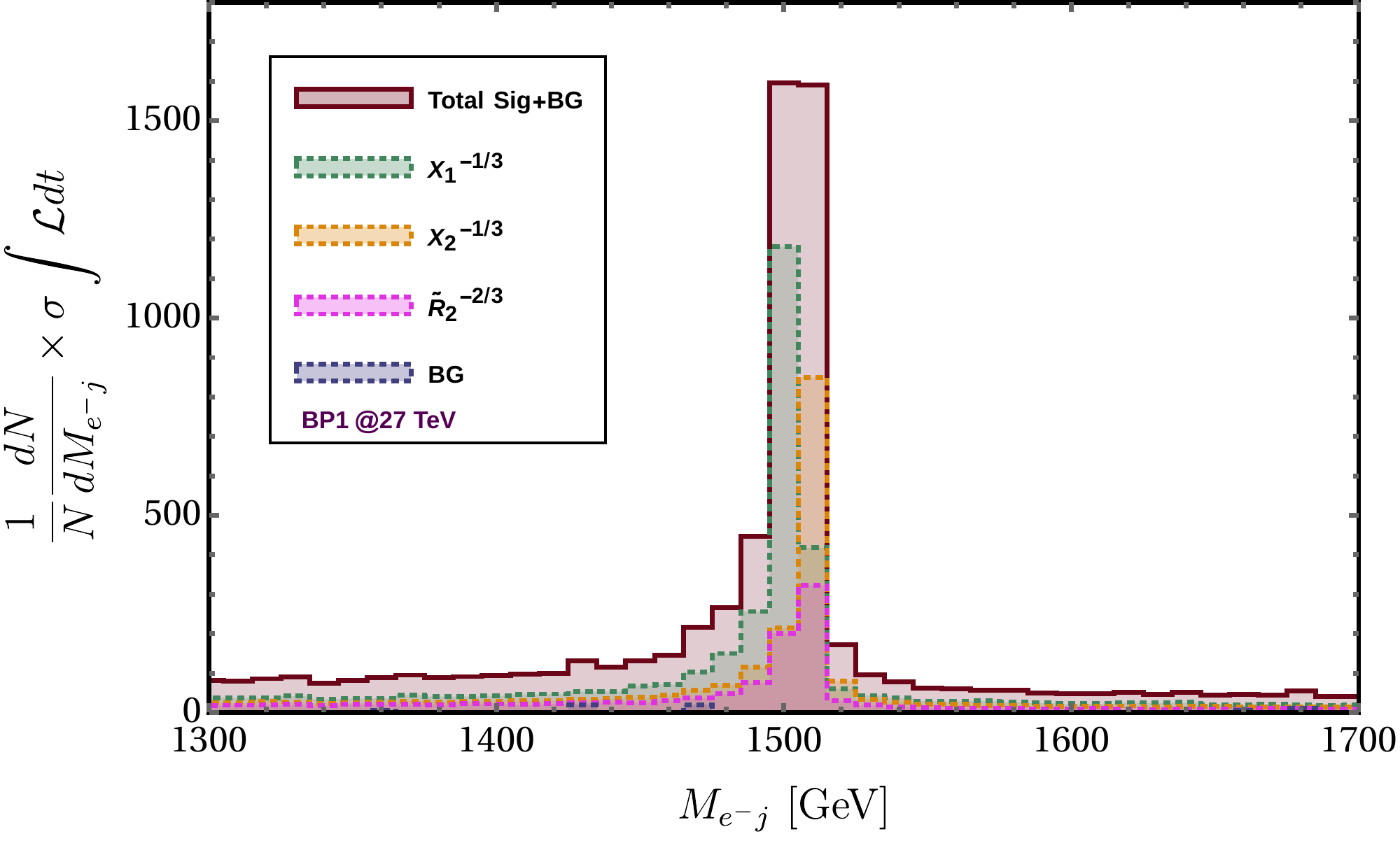}\label{x1m1}}
		\hfil
		\subfigure[]{\includegraphics[width=0.45\linewidth]{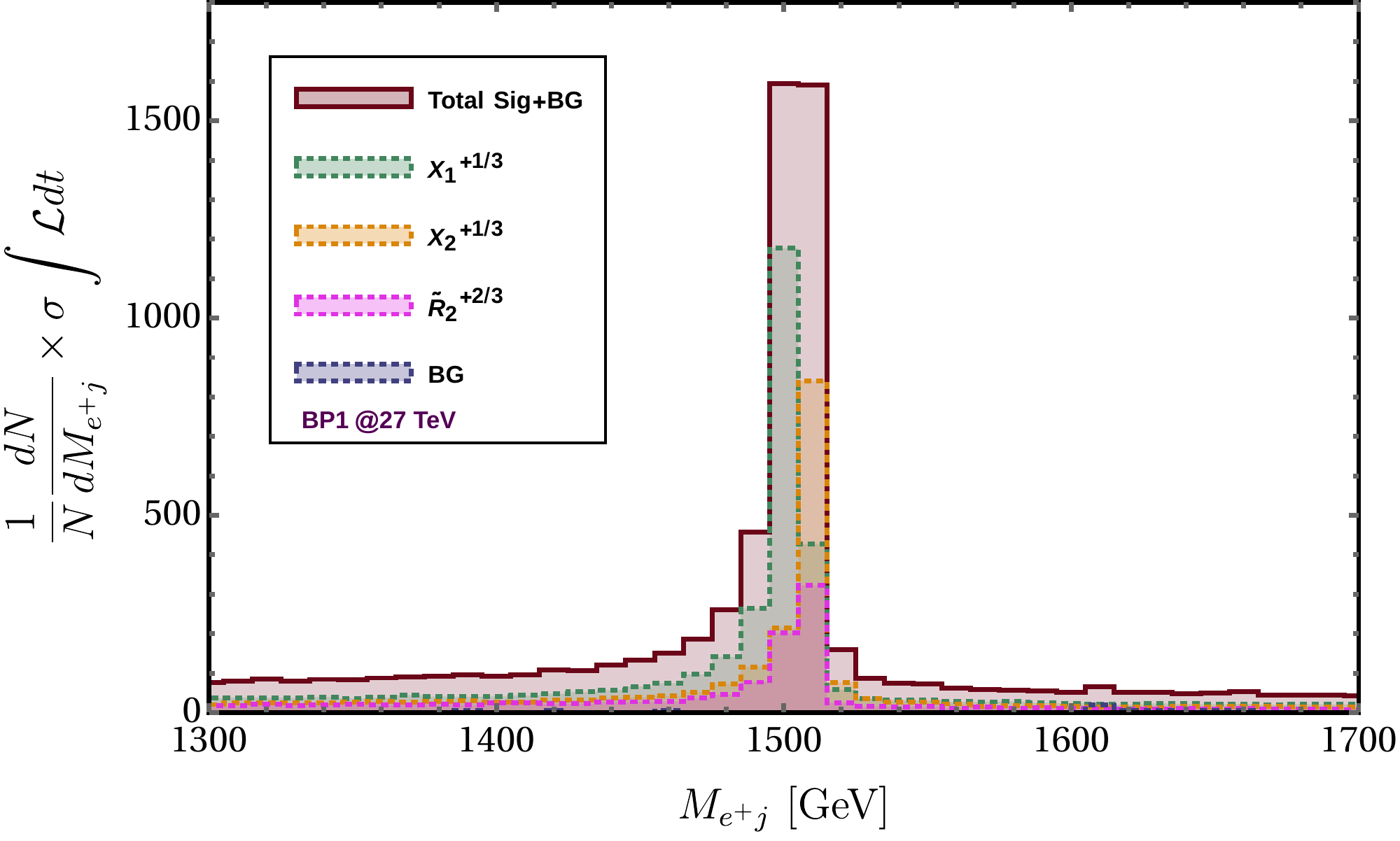}\label{x1p1}}

		\subfigure[]{\includegraphics[width=0.45\linewidth]{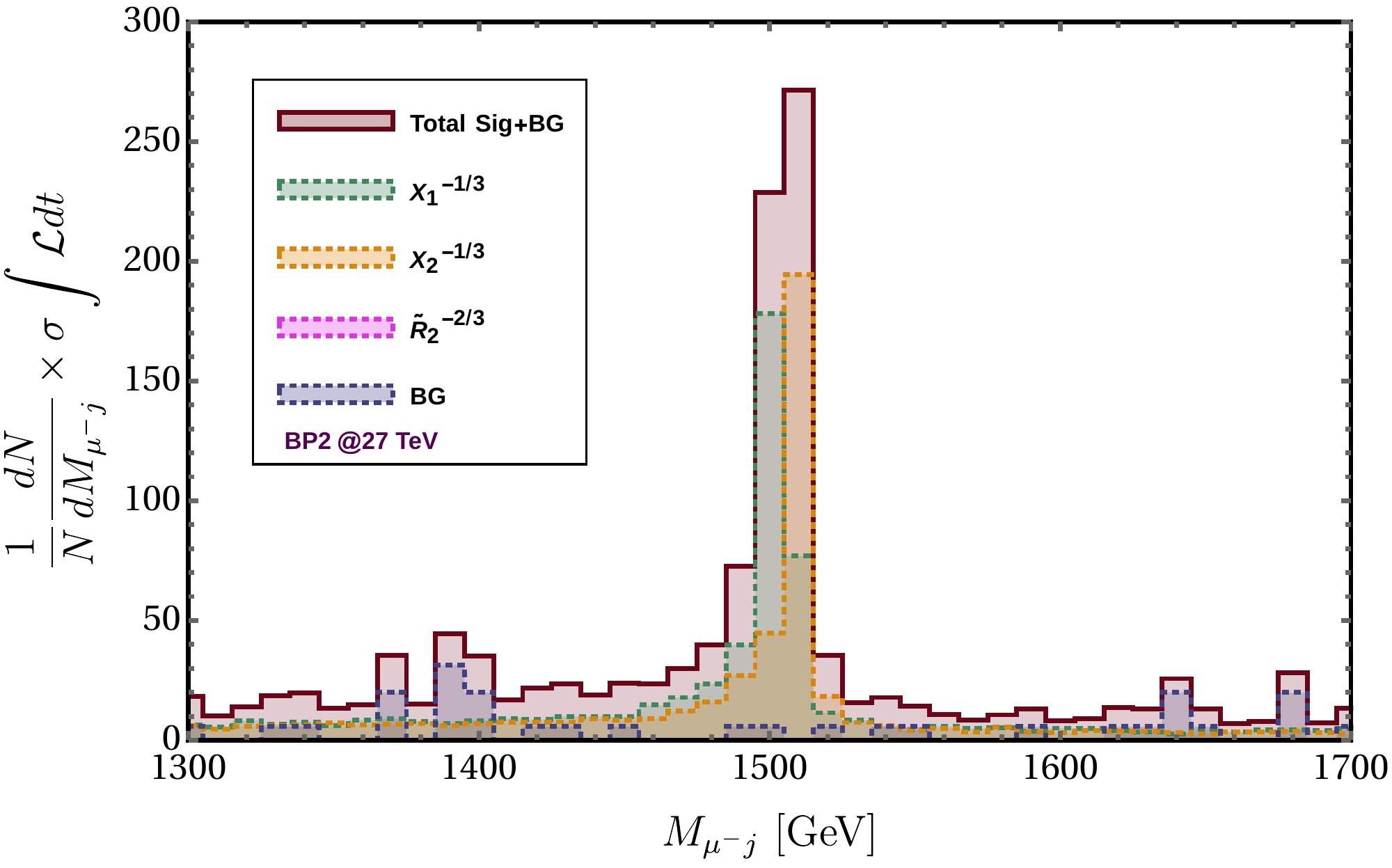}\label{x1m2}}
		\hfil
		\subfigure[]{\includegraphics[width=0.45\linewidth]{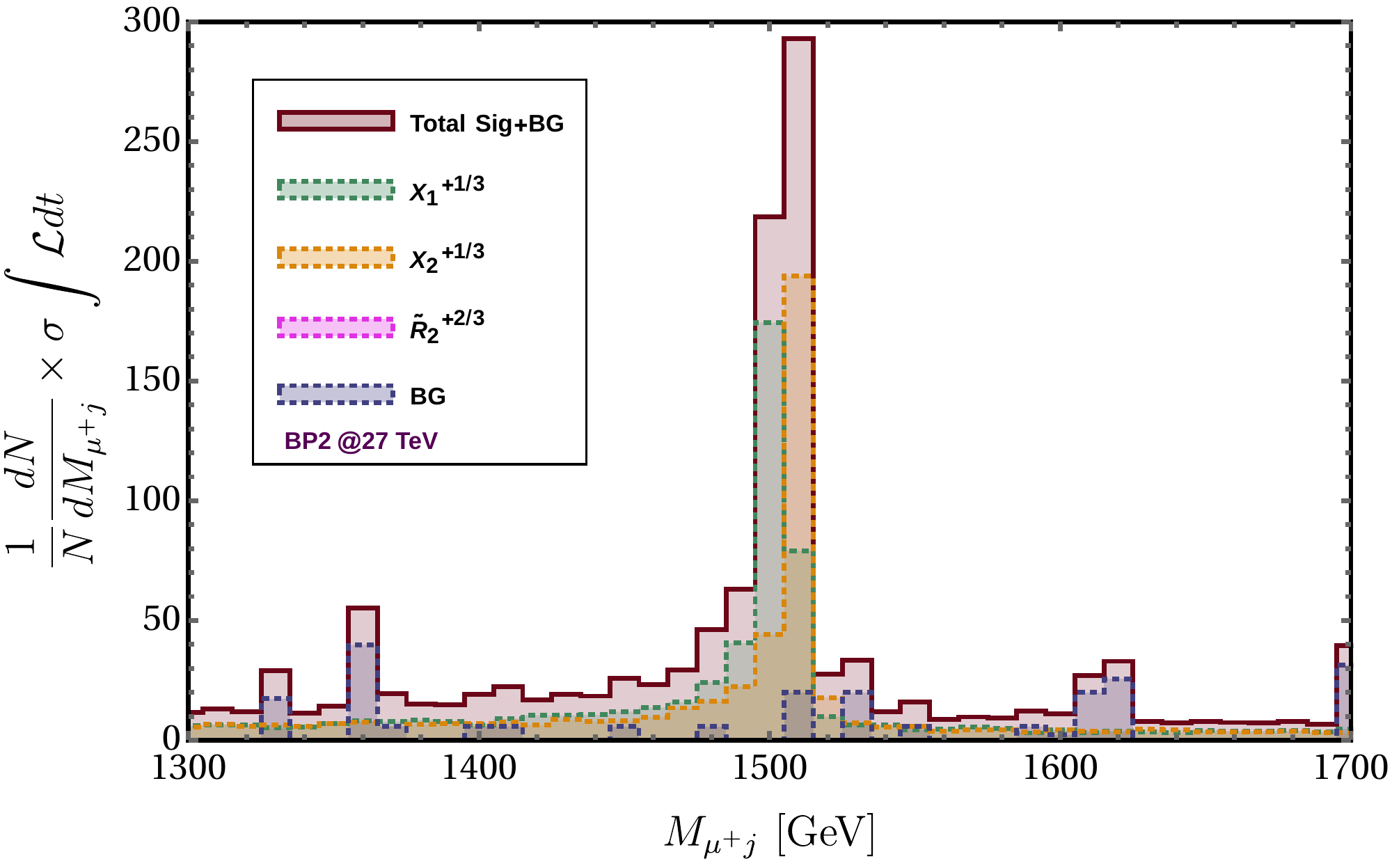}\label{x1p2}}
		
		\subfigure[]{\includegraphics[width=0.45\linewidth]{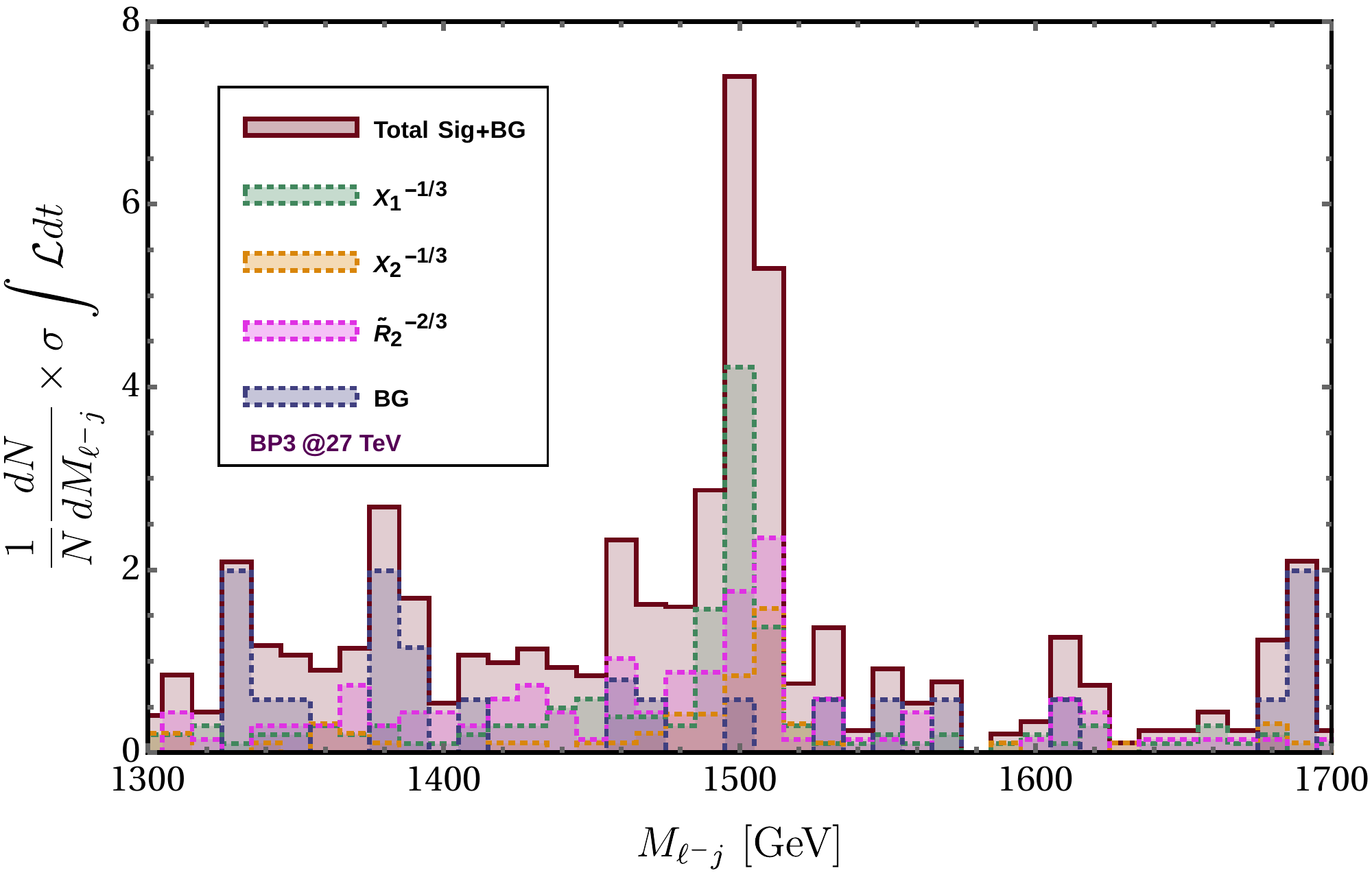}\label{x1m3}}
		\hfil
		\subfigure[]{\includegraphics[width=0.45\linewidth]{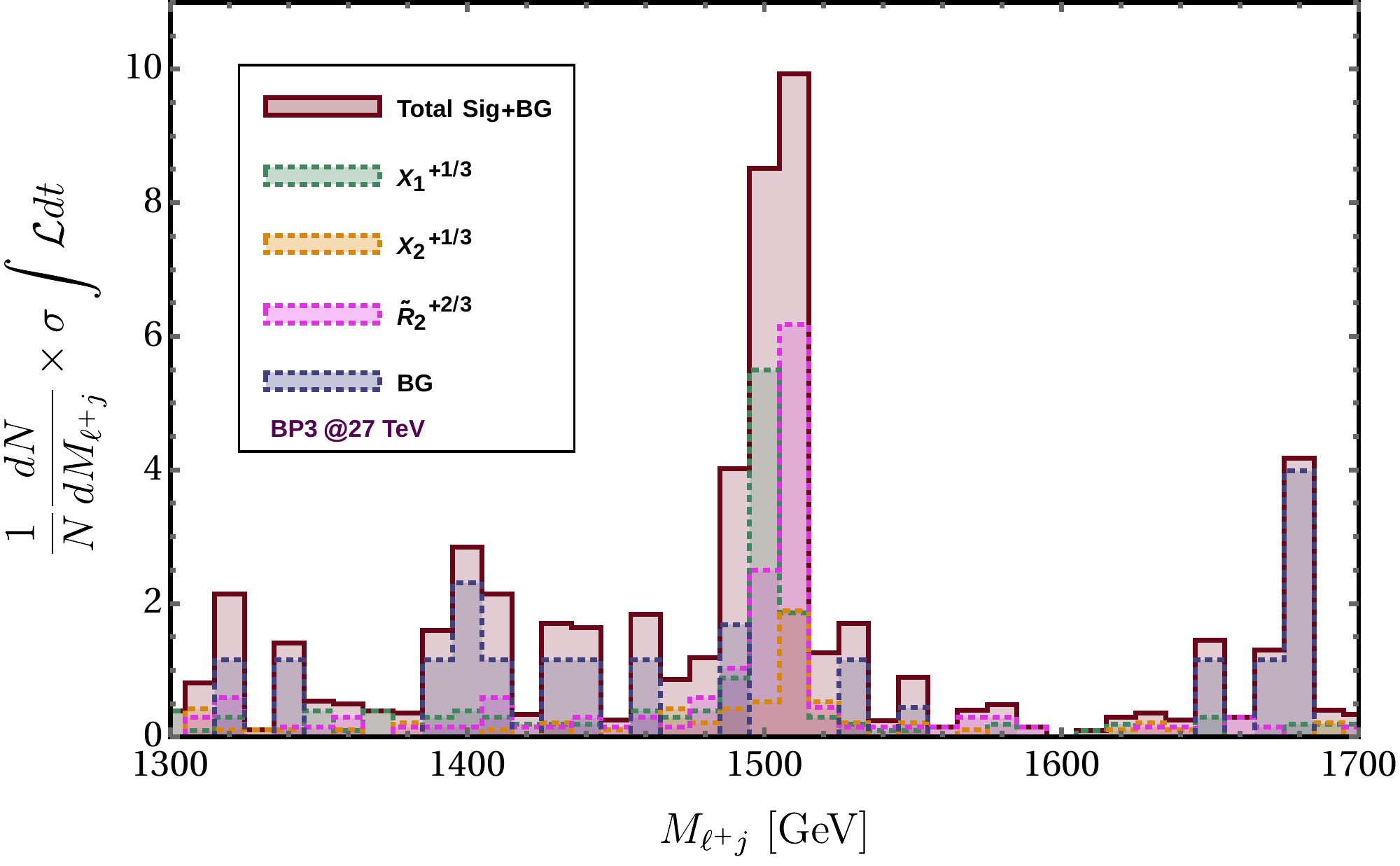}\label{x1p3}}
%	}
	
	\caption{Invariant mass distributions of (a) $e^-j$ and (b) $e^+j$ in BP1, (c) $\mu^- j$ and (d) $\mu^+ j$ in BP2, (e) $\ell^- j$ and (f) $\ell^+ j$ in BP3, from the pair production of leptoquarks at the 27 TeV LHC. }
	\label{fig:imx1x2}
\end{figure}
\autoref{fig:imx1x2} shows a few invariant mass distributions of jet-lepton pairs that can reconstruct the mixed leptoquark mass peaks. In each case, the dark red distribution signifies the total signal events of the three leptoquark pair production processes, combined with the SM background. The dotted lines represent the individual contributions from $\rtt$ (pink), $\xo$ (green), and $\xt$ (yellow), and the SM background (dark blue). In \autoref{x1m1} we show the invariant mass distribution of one light jet and one $e^-$ ($M_{e^-j}$), obtained in a 2$j$+ 2 OS$e$ finalstate in case of BP1.  Here, as expected, events from both $\xo$ and $\xt$ contribute to the total observed peak at $\sim1.5$ TeV, with a small contribution also coming from $\widetilde{R}_2^{-2/3}$. This obtained peak can not pinpoint to either $\xo$ or $\xt$. \autoref{x1p1} shows the distribution of $M_{e^+j}$, with dominant contribution to the peaks from $X_{1,2}^{+1/3}$. Again, \autoref{x1m2} displays the invariant mass distribution of one light jet and one $\mu^-$ ($M_{\mu^-j}$) in a  2$j$+ 2 OS$\mu$ finalstate, pertaining to the leptoquark pair productions in BP2. here, the contribution of $\widetilde{R}_2^{-2/3}$ is nullified due to the absence of muons in its decays. However, similar to BP1, the peak has dominant contributions from both $\xo$ and $\xt$, which cannot be resolved. The same outcome is seen from the $M_{\mu^+j}$ distribution in \autoref{x1p2}, the peak of which has contributions from both $X_{1,2}^{+1/3}$. For BP3, the probabilities of obtaining a finalstate with two jets and two leptons is very rare for $\xot$ pair production. We evaluate the invariant mass of one light jet and one electron or muon (denoted as $\ell^\pm$) from the 2$j$ + 1$e^\pm$ + 1$\mu^\mp$ finalstate as described in \autoref{tab:pairprodbr}. The distributions of $M_{\ell^- j}$ and $M_{\ell^+ j}$ are presented in \autoref{x1m3} and \autoref{x1p3} respectively. Here, the number of events are very low for each leptoquark, and owing to the high abundance of events with two light jets and at least one muon, the $\rtt$ contribution is even more than that of $\xt$. 

If one wishes to utilize this invariant mass peak in distinguishing $\xo$ and $\xt$, one possible way is to have an increase in the mass splitting. This can be done by setting the bare mass terms $m_1$ and $m_2$ in the Lagrangian in \autoref{eq:Lag_SD} to be unequal, unlike in our consideration where $m_1 = m_2 = 1500$ GeV. A difference in $m_1$ and $m_2$ can also lead to a much smaller value of $\tlq$, thus increasing the singlet and doublet percentages in $\xo$ abd $\xt$, respectively. The phenomenology will then change for such a parameter set, and it will also affect the neutrino data and the LFV branching fractions under consideration. In this work we are not considering such a non-degenerate scenario. 

Nevertheless, with the rapid enhancement in machine learning algorithms, the resolution of jets and leptons are becoming more precise at the LHC. The top quark mass, for example, is now measured with a precision of $< 1$ GeV \cite{CMS:2021jnp}, while the resolution for muons at the $Z$-mass peak is $< 3$ GeV \cite{CMS:2019ied}. The SM Higgs boson mass has also been measured with a $< 1$ GeV resolution \cite{CMS:2020xrn, CMS:2017dib}. We need to wait for such dedicated high-resolution analysis being available for high-mass leptoquark studies, so that the invariant mass peaks of $\xot$ are distinguishable.

\subsubsection{finalstate modes}
\label{sec:dcymd}

Considering for the time being that the value of $\kappa$ is negligibly small or zero, the mixing between the leptoquarks also vanish, as seen in \autoref{fig:kappa_tm}. Then, the decays of $\rtt$ and $\widetilde{R}_2^{-1/3}$ are purely governed by $\y_2$, while the $S_1^{-1/3}$ leptoquark decays only via $\y_{1}^{L,R}$, as seen from \autoref{eq:Lag_SD}. This allows $S_1^{-1/3}$ to decay into an up-type quark and a charged lepton, while $\widetilde{R}_2^{-1/3}$ can decay into a down-type quark and a neutrino. Meanwhile, $\rtt$ can decay only to a down-type quark and a charged lepton. Hints of such a case is already found in \autoref{bp1decay}, where, due to the inertness of the small $\y_2$, the decays of $\xo$ and $\xt$ happen into $ue^-$ and $c\tau^-$ for a combined 87\% of the time, while $\rtt$ decays 100\% into $be^+$ and $se^+$.

Assuming all three leptoquark states to be pure, irrespective of the Yukawa coupling structures, the generic finalstates that can be obtained from their pair production are as follows:
\begin{align}
	\rtt \widetilde{R}_2^{-2/3} \to 2\text{-jets + OSD}; \quad \widetilde{R}_2^{-1/3}\widetilde{R}_2^{+1/3} \to 2\text{-jets} + \ptmiss;\\
	S_1^{-1/3} S_1^{+1/3} \to 2\text{-jets + OSD} \,\, \& \,\, 2\text{-jets} + \ptmiss.\quad\quad\quad
\end{align}
We can expect an abundance of di-jet and di-lepton events in case of the singlet pair production, while the $Q=-1/3$ doublet component will lead to di-jet events with some large missing transverse momenta. For smaller values of $\tlq$, an analysis of events in either of these finalstates can lead to the heavy domination of $\xo$ or $\xt$ pair production. Even with cases like BP3 where the entries of $\y_{1}^{L,R}$ and $\y_2$ are of comparable order, distinct signatures can be achieved if the points are chosen with a tiny mixing angle. Again, each one of $\widetilde{R}_2^{-1/3}$ and $S_1^{1/3}$ emanate differently charged quarks for a fixed lepton charge. This can also be crucial for differentiating between $\xo$ and $\xt$, which we will discuss in the next subsection.

\subsubsection{Jet charge}

\begin{figure}[h!]
	\centering
	%		\tikzfeynmanset{every scalar@@/.style={thick, dashed}, every boson@@/.style={thick, decoration={snake,amplitude=1mm},decorate}, every plain@@/.style={thick}}
	\begin{tikzpicture}
		\begin{feynman}
			\vertex[blob,small] (a1){};
			\vertex [above left =2cm of a1] (i1){\footnotesize\(p\)};
			\vertex [below left =2cm of a1] (i2){\footnotesize\(p\)};
%			\vertex [right =1.2cm of a1] (a2);
			\vertex [above right =1.5cm of a1] (i3);
			\vertex [below right =1.5cm of a1] (i4);
			\vertex [above right =1.5cm of i3] (l1) {\footnotesize\(e^+\)};
			\vertex [right =1cm of i3] (j1){\footnotesize\(b/s\)};
			\vertex [below right =1.5cm of i4] (l2) {\footnotesize\(e^-\)};
			\vertex [right =1cm of i4] (j2){\footnotesize\(\bar{b}/\bar{s}\)};
			\vertex [above left = -0.1cm of i3] (d1) {\footnotesize{\bl{Leg 1}}};
			\vertex [below left = -0.1cm of i4] (d2) {\footnotesize{\bl{Leg 2}}};

			\diagram { (i1)--(a1), (i2)--(a1), (a1)--[scalar, edge label' = \footnotesize$\rtt$](i3), (a1)--[scalar](i4), (i3)--(l1), (i3)--(j1), (a1)--[scalar, edge label = \footnotesize$\widetilde{R}_2^{-2/3}$](i4), (i4)--(l2), (i4)--(j2)
			};
		
		\draw [decoration={brace}, decorate] (l1.north east) -- (j1.south east)
		node [pos=0.5, right] {\(M_{e^+ j}\)};
		\draw [decoration={brace}, decorate] (j2.north east) -- (l2.south east)
		node [pos=0.5, right] {\(M_{e^- j}\)};
		\end{feynman}
	\end{tikzpicture}
\hfil
\begin{tikzpicture}
	\begin{feynman}
		\vertex[blob,small] (a1){};
		\vertex [above left =2cm of a1] (i1){\footnotesize\(p\)};
		\vertex [below left =2cm of a1] (i2){\footnotesize\(p\)};
		%			\vertex [right =1.2cm of a1] (a2);
		\vertex [above right =1.5cm of a1] (i3);
		\vertex [below right =1.5cm of a1] (i4);
		\vertex [above right =1.5cm of i3] (l1) {\footnotesize\(e^+\)};
		\vertex [right =1.25cm of i3] (j1){\footnotesize\(\bar{u}\)};
		\vertex [below right =1.5cm of i4] (l2) {\footnotesize\(e^-\)};
		\vertex [right =1.25cm of i4] (j2){\footnotesize\(u\)};
		\vertex [above left = -0.1cm of i3] (d1) {\footnotesize{\bl{Leg 1}}};
		\vertex [below left = -0.1cm of i4] (d2) {\footnotesize{\bl{Leg 2}}};
		%			\vertex [above right =1cm of i3] (h1){\footnotesize\(T^0\)};
		%			\vertex [right =1cm of i3] (h2);
		%			\vertex [below =1cm of i4] (h3){\footnotesize\(\textcolor{white}{d}\)}; %dummy vertex%
		%			\vertex [above right =0.5cm of h2] (w1){\footnotesize\(l\)};
		%			\vertex [below right =0.5cm of h2] (w2){\footnotesize\(\bar{\nu}\)};
		%			\vertex [below =1cm of a1] (t){\footnotesize\(\textbf{ITM}\)};
		
		\diagram { (i1)--(a1), (i2)--(a1), (a1)--[scalar, edge label' = \footnotesize$S_1^{+1/3}$](i3), (a1)--[scalar](i4), (i3)--(l1), (i3)--(j1), (a1)--[scalar, edge label = \footnotesize$S_1^{-1/3}$](i4), (i4)--(l2), (i4)--(j2)
		};
		
		\draw [decoration={brace}, decorate] (l1.north east) -- (j1.south east)
		node [pos=0.5, right] {\(M_{e^+ j}\)};
		\draw [decoration={brace}, decorate] (j2.north east) -- (l2.south east)
		node [pos=0.5, right] {\(M_{e^- j}\)};
	\end{feynman}
\end{tikzpicture}
	
	\caption{Cartoon diagrams in context of jet charge analysis, for pair production of $\rtt$ (left) and $S_1^{-1/3}$ (right) in case of BP1.}
	\label{fig:jc}
\end{figure}
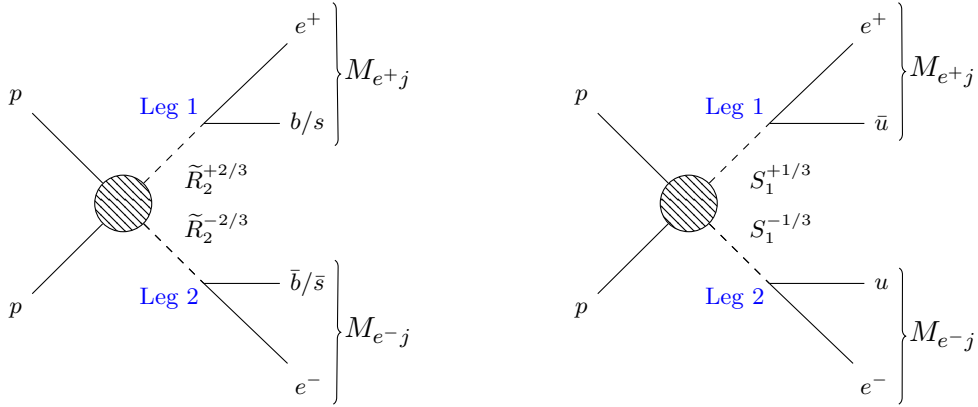

In our analysis of segregating between $\rtt$ and $\xot$ in \autoref{sec:rtvx12}, the utility of jet charge reconstruction was mentioned. Continuing our assumption of negligible mixing where $\xo$ and $\xt$ are essentially $S_1^{-1/3}$ and $\widetilde{R}_2^{-1/3}$, we can tag a lepton with its identity and charge specified, and then identify the jet that is produced with it from the same leg of the pair production diagram \cite{Krohn:2012fg,CMS:2017yer,Tokar:2017syr,Bandyopadhyay:2020jez,Bandyopadhyay:2020wfv}. The correct jet can be identified by demanding the jet-lepton invariant mass to be within a $\pm 5$ GeV window of the leptoquark resonance, with the help of the discussion in \autoref{sec:imx1x2} and \autoref{sec:imr}. After identifying the jet, we can study the charge distribution of it, to determine whether it originates from an up-type quark (coming from $\rtt$) or a down-type quark (coming from $S_1^{-1/3}$), which can eventually lead us to correctly identifying the responsible leptoquark, as shown in \autoref{fig:jc}. For example, in case of BP1, the \autoref{fig:jc} shows the 2-jet + 2OS$e$ finalstates obtainable from $\rtt$ and $S_1^{-1/3}$ pair production. \autoref{bp1decay} shows how from the pair production of $\widetilde{R}_2^{+2/3}$, tagging an $e^-$ automatically ensures the jet coming from Leg 2 to be of $\bar{s}$ or $\bar{b}$ flavour, in the correct invariant mass pairing. Whereas, in case of $S_1^{-1/3}$, the tagged $e^-$ from the invariant mass pair will point to a $u$-flavoured jet from Leg 2. Considering a scenario of negligible $\tlq$, $\widetilde{R}_2^{-1/3}$ will not even have that required electron at the first place. Hence, from the combined analysis of pair production of the three leptoquarks, we can observe two different gaussian distributions, which may overlap for some part. Now, if $\widetilde{R}_2$ did not exist in the model, we would be seeing only one distribution pertaining to the $u$-flavoured jets. Similarly, absence of $S_1$ will show us the $\bar{s}$ or $\bar{b}$-flavoured distribution. As the mixing angle increases, the number of events in the $\xo$ jet charge distribution decrease, due to the enhancement of $\widetilde{R}_2^{-1/3}$ percentage in it. This deviation from the ideal case can be utilized to estimate the mixing angle itself. 
	
	\section{Conclusion}
	\label{sec:conc}
	In this work we generate Majorana neutrinio masses via one-loop Weinberg operator involving 
	$\widetilde{R}_2$ and $S_1$ leptoquarks, which are in $SU(2)$ doublet and singlet representations, respectively. We further investigate the model parameter space which can satisfy the neutrino mixing angles, anomalous magnetic moment of the muon and electron, as well as various experimental bounds coming from lepton flavour violating processes. Once setting up with these we study various finalstate topologies in probing different mass eigenstates involving $\widetilde{R}_2$ and $S_1$ at the LHC/FCC.
	
	Due to the presence of a trilinear coupling $\kappa$, the component of $\widetilde{R}_2$ with charge $1/3$ mixes with $S_1$ through an angle $\theta_{LQ}$ to give us two mass eigenstates $\X_{1,2}^{-1/3}$, while the $\rtt$ mass eigenstate remains a pure doublet. Thus, we have three physically observable leptoquark states from this model, on which we perform a study at the LHC/FCC. For the purpose of the collider simulations, we choose three sets of Yukawa couplings in three benchmark points with different phenomenological implications. The entries of these couplings are chosen in such a way that they agree to the neutrino mass and oscillation data, the $g-2$ of muon and electron, as well as the experimental bounds on various LFV processes, simultaneously. In each benchmark point, a nearly degenerate mass spectrum is considered for the leptoquarks, with $M_{\rtt} = 1.502$ TeV, $M_{\xo} = 1.499$ TeV, and $M_{\xt} = 1.506$ TeV.
	
	We begin our study with the pair production of each of the three physical leptoquark states. The pair production is mostly dominated by the QCD processes of gluon-gluon fusion and $s$-channel gluon exchange from quark fusions. However, based on the choice of Yukawa couplings, the $t$-channel lepton exchange diagrams can also contribute to the pair production cross-section. The three sets of Yukawa couplings in the three benchmark points show us different phenomenology in each case. The analysis is performed at centre-of-mass energies of 14, 27 and 100 TeV to simulate the present and future LHC/FCC environments.
	
	The first step of the collider analysis (\autoref{sec:modsig}) involves obtaining signatures to probe the model at the LHC/FCC with two different finalstate topologies: 2 light/$c$-jets + OSD (FS1), and 2 light/$b$-jets + OSD (FS2). In both finalstates, 5$\sigma$ probes of the model is achieved across the three benchmark points at the 14 TeV LHC, with $<540 \,\text{fb}^{-1}$ integrated luminosity in each case (\autoref{tab:fs1},\autoref{tab:fs2}). With higher energies of 27 and 100 TeV, the $5\sigma$ significance is shown to be obtained with much earlier data. 
	
	Next, in \autoref{sec:rtvx12}, signatures to distinguish between the pure doublet $\rtt$ and the mixed states $\xot$ are explored from the pair production processes. Four different finalstate topologies are identified (FS3-FS6), with which we can perform such a distinction. Depending on the benchmark points, the obtained signal events in these states are dominated by either $\rtt$, or $\xot$. In each finalstate, discerning signatures are obtained for at least one BP, with $5\sigma$ probes being possible within $\sim2500$ $\text{fb}^{-1}$ luminosity limit at the 14 TeV LHC. One again, at higher centre-of-mass energies, these signatures are obtained at much less luminosities of $<80 \, \text{fb}^{-1}$ at 27 TeV, and $<1.5 \,\text{fb}^{-1}$ at 100 TeV. The $\rtt$ leptoquark is also shown to be distinctly reconstructed from the invariant mass distributions of one jet and one lepton, with the identities varying for different BPs, with negligible contamination from $\xot$ and SM background events. 
	
	Lastly, in \autoref{sec:x1vx2}, we discuss the challenges and possibilities of discerning $\xo$ from $\xt$. Their small mass splitting of $\sim 7$ GeV, same electric charge, and large mixing angle of $\abs{\tlq} = 0.618$ radians lead to complications in obtaining finalstates that are dominated heavily by either one of them. The mixing angle plays a part in the asymmetric production mode of $q\bar q' \to \rtt \xbar\X_{1,2}^{+1/3}$ via $s$-channel $W^\pm$-boson exchange, which can, in theory, allow us to probe $\theta_{LQ}$ and act as a way of distinguishing. However, small values of cross-sections make this probe impractical from the current experimental perspective. Moving back to pair production, the tiny mass gap remains unresolved, even with an optimistic bin width of 10 GeV, and we need to wait for the advancement in the precision of high-momentum jet and lepton resolution. For a smaller mixing angle, the stark difference in the singlet and doublet decay modes can help us obtain distinguishing signatures. A jet charge analysis of these modes can also reflect the effect of the mixing, and act as a probe of $\xo$ or $\xt$.

\acknowledgments

SP sincerely thanks the Council of Scientific and Industrial Research (CSIR), India for funding his research (File no: 09/1001(0082)/2020-EMR-I). SP also acknowledges Chandrima Sen for some valuable discussion regarding {\tt PYTHIA}. PB and AK acknowledge SERB CORE Grant CRG/2018/004971 and MATRICS Grant MTR/2020/000668 for the financial support. A.K also acknowledges the partial support by MCIN/AEI/10.13039/501100011033 Grant No. PID2020-114473GB-I00, and Grant PROMETEO/2021/071 (Generalitat Valenciana). K. G. and Avnish acknowledge SERB Core Research Grant No. CRG/2019/006831. S. P. and P. B. also acknowledge SERB Karyashala titled “Data and Machine Learning at the Large Hadron Collider” (DML@LHC-2022), Grant
No. AV/KAR/2022/0167.

\bibliography{refs_2}

\end{document}